\documentclass[11pt,a4paper]{article}
\usepackage{amsmath, amsfonts,amssymb,mathtools,nicefrac,nccmath,cases}
\usepackage{microtype,booktabs,multirow}
\usepackage[usenames,dvipsnames,table]{xcolor}
\usepackage{colortbl}
\usepackage{tikz}
\usepackage{verbatim}
\usetikzlibrary{shapes,arrows}
\usetikzlibrary{calc,shapes.callouts,shapes.arrows,bending,intersections}
\usetikzlibrary{shapes.geometric}
\usetikzlibrary{arrows.meta,arrows}
\usetikzlibrary{decorations.pathmorphing}
\usetikzlibrary{decorations.markings}
\usetikzlibrary{shapes.multipart}
\usetikzlibrary{tikzmark}
\usepackage{caption}
\usepackage{subcaption}
\usepackage{bm}
\usepackage[nosort]{cite}
\usepackage{colortbl}
\usepackage{amsthm}
\usepackage{soul}

\usepackage{xcolor}
\usepackage{lipsum}

\newcommand\boxedB[1]{{\setlength\fboxsep{6pt}\boxed{#1}}}

\definecolor{babyblue}{rgb}{0.54, 0.81, 0.94}
\usepackage[linktocpage=true]{hyperref}
\hypersetup{colorlinks=true,linkcolor=nicecolor,citecolor=nicecolor,urlcolor=nicecolor}
\definecolor{nicecolor}{rgb}{0.1, 0.3, 0.4}

\definecolor{blue}{rgb}{0.06, 0.3, 0.57}
\definecolor{Gray}{gray}{0.4}

\definecolor{nicecolor}{rgb}{0.1, 0.3, 0.4}
\definecolor{blue}{rgb}{0.06, 0.3, 0.57}
\definecolor{Gray}{gray}{0.4}
\colorlet{tableheadcolor}{gray!15} 
\colorlet{tablerowcolor}{gray!7} 

\def\hybrid{\topmargin -20pt    \oddsidemargin 0pt
	\headheight 0pt \headsep 0pt
	\textwidth 6.5in        
	\textheight 9in         
	\textwidth 6.25in       
	\textheight 9 in       
	\marginparwidth .875in
	\parskip 5pt plus 1pt 
	\jot = 1.5ex
}
\hybrid
\numberwithin{equation}{section}
\numberwithin{table}{section}
\usepackage{float}
\usepackage{tabularx}
\usepackage{multirow}
\usepackage{hhline}
\usepackage{stackengine}

\newcolumntype{D}{>{\centering\arraybackslash}X}
\newcolumntype{L}{>{$}l<{$}}
\newcolumntype{R}{>{$}r<{$}}
\newcolumntype{C}{>{$}c<{$}}
\usepackage{IEEEtrantools}

\newcommand{\beq}{\begin{equation}}  \newcommand{\eeq}{\end{equation}}
\newcommand{\bal}{\begin{aligned}}   \newcommand{\eal}{\end{aligned}}
\newcommand{\bea}{\begin{eqnarray}}  \newcommand{\eea}{\end{eqnarray}}
\def\beqa{\begin{eqnarray}}
\def\eeqa{\end{eqnarray}}

\newcommand{\bmat}{\left(\begin{array}}
\newcommand{\emat}{\end{array}\right)}

\newcommand{\cE}{\mathcal{E}}

\newcommand{\cW}{\mathcal{W}}

\usepackage{cancel}

\newcommand{\be}{\begin{equation}}
\newcommand{\ee}{\end{equation}}

\definecolor{Gray}{gray}{0.95}
\definecolor{darkspringgreen}{rgb}{0.09, 0.45, 0.27}
\definecolor{darkseagreen}{rgb}{0.56, 0.74, 0.56}
\definecolor{darkmouthgreen}{rgb}{0.05, 0.5, 0.06}
\definecolor{darkcyan}{rgb}{0.0, 0.55, 0.55}

\def\d {{\rm d}}
\def\del          {\partial}
\def\delbar       {\bar\partial}
\def\ii           {{\rm i}}

\def\tr           {\mathop{\rm tr}}
\def\Re           {{\rm Re\hskip0.1em}}
\def\Im           {{\rm Im\hskip0.1em}}

\def\cala         {{\cal A}}

\def\calb         {{\cal B}}
\def\calc         {{\cal C}}
\def\cald         {{\cal D}}

\def\calg         {{\cal G}}

\def\calh         {{\cal H}}
\def\cali         {{\cal I}}
\def\calj         {{\cal J}}
\def\calk         {{\cal K}}

\def\calm         {{\cal M}}
\def\caln         {{\cal N}}
\def\calo         {{\cal O}}

\def\calr         {{\cal R}}

\def\calt         {{\cal T}}

\def\calw         {{\cal W}}

\usepackage{xcolor}
\definecolor{colorloc1}{RGB}{0,0,102}  
\definecolor{colorloc2}{RGB}{0,125,253} 
\usepackage[framemethod=default]{mdframed}
\newmdenv[skipabove=10pt,
skipbelow=7pt,
rightline=false,
leftline=true,
topline=false,
bottomline=false,
linecolor=colorloc1,
backgroundcolor=colorloc2!5,
innerleftmargin=4pt,
innerrightmargin=0pt,
innertopmargin=0pt,
leftmargin=2pt,
rightmargin=0pt,
linewidth=2pt,
innerbottommargin=0pt]{lbBox}

\begin{document}

\baselineskip=14pt
\parskip 5pt plus 1pt

\vspace*{-1.5cm}
\begin{flushright}    
  {\small 
 ZMP-HH/22-18}
\end{flushright}

\vspace{2cm}
\begin{center}        

  {\huge  Quantum Gravity Bounds on ${\cal N}=1$  \vspace{2mm} \\ Effective Theories in Four Dimensions   \\
   [.3cm]  }
\end{center}

\vspace{0.5cm}
\begin{center}        
{\large   Luca Martucci$^{1}$, Nicol\`o Risso$^{1}$ and Timo Weigand$^{2,3}$}
\end{center}

\begin{center}  
\noindent 
${}^1$ \emph{Dipartimento di Fisica e Astronomia ``Galileo Galilei",  Universit\`a degli Studi di Padova,} \\
\emph{\& I.N.F.N. Sezione di Padova, Via F. Marzolo 8, 35131 Padova, Italy} 
\smallskip
\\${}^2$ \emph{II.\ Institut f\"ur Theoretische Physik, Universit\"at Hamburg, Luruper Chaussee 149,\\
22607 Hamburg, Germany}\\ 
\smallskip
${}^3$\emph{ Zentrum f\"ur Mathematische Physik, Universit\"at Hamburg,
Bundesstrasse 55, \\ 20146 Hamburg, Germany}

\end{center}

\vspace{2cm}


\begin{abstract}

\noindent We propose quantum gravitational constraints on effective four-dimensional theories with ${\cal N}=1$ supersymmetry. These Swampland constraints arise by demanding consistency of the worldsheet theory of a class of axionic, or EFT, strings whose existence follows from the Completeness Conjecture of quantum gravity. 
Modulo certain assumptions, 
we derive positivity bounds and quantization conditions for the axionic couplings to the gauge and gravitational sector at the two- and four-derivative level, respectively. 
We furthermore obtain general bounds on the rank of the gauge sector in terms of the gravitational couplings to the axions.
We exemplify how these bounds rule out otherwise consistent effective supergravity theories as theories of quantum gravity.
 Our derivations of the quantum gravity bounds are tested and further motivated in concrete string theoretic settings. 
 In particular, this leads to a sharper version of the bound on the gauge group rank in F-theory on elliptic four-folds with a smooth base, which improves the known geometrical Kodaira bounds. We furthermore provide a detailed derivation of the EFT string constraints in heterotic string compactifications including higher derivative corrections to the effective action and apply the bounds to M-theory compactifications on $G_2$ manifolds.

\end{abstract}

\thispagestyle{empty}
\clearpage

\setcounter{page}{1}


\newpage

  \tableofcontents

\newpage


\section{Introduction}

An increasing amount of evidence suggests that to couple a gauge theory to gravity,
severe constraints have to be fulfilled at the quantum level. 
For example, while in quantum field theory, any anomaly free spectrum, no matter how contrived, leads to a consistent quantum gauge theory, it is widely believed that in quantum gravity, the number of degrees of freedom should remain finite. Determining more generally which constraints a quantum gauge theory has to fulfill in presence of gravity is one of the objectives of the Swampland program \cite{Vafa:2005ui} as reviewed, for instance, in \cite{Palti:2017elp,vanBeest:2021lhn,Grana:2021zvf}.

A particularly fruitful idea that has emerged is to derive consistency conditions on an effective field  theory by examining the inclusion of higher dimensional defects as probes in the theory.\footnote{As one of the earlier incarnations of this idea, \cite{Uranga:2000xp} derived for instance the K-theory tadpole cancellation conditions in D-brane models by requiring absence of Witten anomalies on probe branes. See furthermore \cite{Banks:1997zs} and references therein for an analysis of the correspondence between spacetime and worldvolume physics.} 
For instance, if a quantum gravity theory contains higher $p$-form gauge fields, the Completeness Conjecture \cite{Polchinski:2003bq} implies that the theory must contain $p$-dimensional objects charged under this symmetry.
Consistency of the worldvolume theory of these objects poses novel constraints on the  bulk effective field theory which may go far beyond the usual cancellation of gauge and gravitational anomalies. This approach has been successfully applied to minimally supersymmetric gauge-gravity systems in ten \cite{Kim:2019vuc}, eight \cite{Hamada:2021bbz,Bedroya:2021fbu}, six \cite{Kim:2019vuc,Lee:2019skh,Tarazi:2021duw,Angelantonj:2020pyr,Cheng:2021zjh}, and five \cite{Katz:2020ewz} dimensions, as well as to theories with sixteen supercharges in various dimensions \cite{Kim:2019ths}. In this paper we will derive quantum gravity bounds in four-dimensional theories with minimal ${\cal N}=1$ supersymmetry.

An ${\cal N}=1$ supergravity theory in four dimensions typically contains light complex chiral scalar fields parametrizing the field space of the theory. 
In the weak coupling regime, a shift symmetry arises in an emergent axionic sector of the theory, which is only broken by non-perturbative effects. In the limit where the latter are sufficiently suppressed, the axions can be dualized into two-form fields. By the Completeness Conjecture of quantum gravity, these must couple to string-like objects, which in fact can be half-BPS in four dimensions. 
This places the theory precisely into a context, similar to that of \cite{Kim:2019vuc,Hamada:2021bbz,Bedroya:2021fbu,Lee:2019skh,Tarazi:2021duw,Angelantonj:2020pyr,Katz:2020ewz,Kim:2019ths}, in which consistency conditions of these axionic strings can be turned into new constraints on the four-dimensional effective field theory.

As a complication specific to four dimensions, however, string-like objects induce a non-negligible backreaction on the fields in the two directions normal to the string. This is owed to the fact that the string worldsheet is of real codimension two in spacetime \cite{Greene:1989ya}. For general strings, the backreaction questions the validity of applying the probe approximation to the string and of viewing it as weakly coupled.
Interestingly, in
\cite{Lanza:2020qmt,Lanza:2021qsu}
it was understood that for a certain class of half-BPS strings a perturbative treatment of the string worldsheet theory is nonetheless justified.
The strings in question were called  `EFT strings' and have the property that their backreaction on the moduli of the theory is precisely such that close to the string core, the effective theory becomes weakly coupled: More precisely, asymptotically close to the string core, the
axionic shift symmetries which are needed to dualize the axions to two-form fields become exact because the instantons dual to the string become suppressed. As it turns out, the strings with this property are precisely of the form that they cannot decouple from the gravitational sector, making them ideal candidates to probe the quantum gravity nature of the theory. 
The original motivation in studying such strings was the conjecture of \cite{Lanza:2020qmt,Lanza:2021qsu}  that any infinite distance limit in four dimensions can be obtained as the endpoint of an RG flow induced by the backreaction of an EFT string. This idea has been investigated further in \cite{Marchesano:2022avb,Grimm:2022sbl,Cota:2022yjw} from various perspectives.

In the present work, we will explore the role of EFT strings as weakly coupled probes of the four-dimensional ${\cal N}=1$ supersymmetric gauge-gravity sector. In this way, we will be able to constrain the input data of an ${\cal N}=1$ supergravity theory beyond the consistency conditions from anomaly cancellation in the bulk alone. 

The constraints which we will find should be valid for all such theories with a standard coupling to the axionic sector: By this we mean theories in which the gauge field strength and the curvature two-form couple to the axions $a^i$ via terms of the form\footnote{The precise normalisation factors will be given in Section \ref{sec:bulk}.} 
\beq  \label{axcoupIntro}
S_{\rm EFT} \supset   C_i \int a^i \, {\rm tr} F \wedge F  + \tilde C_i \int a^i \,  {\rm tr} R \wedge R \,.
\eeq
As we will see, such couplings induce an anomaly inflow from the four-dimensional bulk to the string worldsheet. As in \cite{Callan:1984sa} and in the higher-dimensional gravitational theories treated in \cite{Kim:2019vuc,Lee:2019skh,Tarazi:2021duw,Katz:2020ewz}, this anomaly inflow must be cancelled by the two-dimensional anomalies on the string worldsheet. Importantly, for EFT strings in the sense of \cite{Lanza:2020qmt,Lanza:2021qsu}, we can reliably compute this worldsheet anomaly in terms of the modes along the string because the theory flows to weak coupling close to the core of the string. This eventually leads to constraints on the four-dimensional bulk theory. A notable difference to the analysis in \cite{Kim:2019vuc,Lee:2019skh,Tarazi:2021duw,Katz:2020ewz} is that we do not assume that the string worldsheet theory flows to a conformal field theory in the infra-red. 
Indeed, in four dimensions this may  a priori not be justified due to the peculiarities of the string, but is also not required as long as the string worldsheet is weakly coupled. The EFT strings are precisely of this type.

Our main results are two types of constraints:
First,  \eqref{tildeCqc} and \eqref{tildehatC0} constrain the quantization and the signs of the
axionic curvature couplings in \eqref{axcoupIntro}.  In particular, \eqref{tildeCqc} fixes the saxionic Gauss-Bonnet term in a large class of gravitational ${\cal N}=1$ supersymmetric effective actions to be positive. Our EFT string analysis therefore complements previous arguments  for the positivity of these higher-derivative terms 
\cite{Kallosh:1995hi,Cheung:2016wjt,GarciaEtxebarria:2020xsr,Aalsma:2022knj,Ong:2022mmm}. 
Second, \eqref{rankbound} bounds the possible ranks of the gauge groups which can be coupled to an axionic sector as in \eqref{axcoupIntro}.

Our derivation of these bounds is subject to certain assumptions on the spectrum of the worldsheet theory of the EFT strings, which we motivate in Section \ref{sec:anomalymatching}. These are manifestly realised in explicit string theoretic settings and moreover appear natural more generally.
Modulo these assumptions, whenever the EFT string derived constraints, in particular the bounds \eqref{rankbound} on the ranks of a gauge sector, are violated in a consistent quantum gravity with minimal supersymmetry,  our analysis implies that the theory cannot exhibit standard axionic couplings of the form \eqref{axcoupIntro}.  This is a rather non-trivial prediction from a purely effective field theoretic point of view. 

Our results can be compared with the concrete constraints imposed on effective field theories in string theory compactifications. 
In the context of F-theory compactifications, we will confront the EFT string bounds \eqref{rankbound} with the geometric bounds from the construction of explicit Weierstrass models. 
The latter have been analysed in a series of works \cite{Kumar:2010ru,Morrison:2011mb,Taylor:2011wt,Grimm:2012yq} in various dimensions to constrain the ranks and matter content of gauge-gravity theories in F-theory.\footnote{More recent examples of the rich corpus of studies investigating the constraints on gauge data imposed by string theory in various dimensions include \cite{Taylor:2019ots,Font:2020rsk,Cvetic:2020kuw,Font:2021uyw,Lee:2021usk,Cvetic:2021sjm}.}
In fact, based on our knowledge of the worldsheet theory of the EFT strings in F-theory following from \cite{Lawrie:2016axq}, we propose \eqref{strictboundF} as a stronger bound on the rank of the gauge group, which should be valid in geometric F-theory compactifications with a smooth base (and minimal supersymmetry).

To test the EFT string bounds in the heterotic context, we will first have to 
extend the analysis in \cite{Lanza:2021qsu} of EFT strings in such setups by including also higher curvature corrections to the effective action. This part of our analysis is in fact interesting by itself and reveals an intriguing modification of the structure of the cone of EFT strings due to said correction terms. We will compare the structure of the heterotic cone of EFT strings with its F-theory dual and discuss
the bounds on the theory.
Furthermore, we give a preliminary analysis of the EFT string derived bounds in M-theory on $G_2$ manifolds, which can serve as a starting point for a more detailed investigation in the future.

The material of this work is organized according to the following structure:
The analysis of Sections \ref{sec:bulk} to \ref{sec:examples} is purely field theoretic in nature and requires no knowledge of string theory.
In Section \ref{sec:bulk} we specify the structure of the four-dimensional ${\cal N}=1$ supersymmetric
gauge-gravity theories for which we will derive quantum gravity constraints. In particular, we will explain the assumption that the effective field theory enjoys axionic couplings of the form \eqref{axcoupIntro}.
In Section \ref{sec:EFTstrings} we first review the main concepts underlying the ideas of EFT strings from \cite{Lanza:2020qmt,Lanza:2021qsu}. 
We then derive the anomaly inflow from the bulk to the string worldsheet induced solely by the axionic couplings \eqref{axcoupIntro}, up to an interesting subtlety discussed in Section \ref{sec:hatC}. Finally, we specify the main ingredients of the weakly coupled non-linear sigma-model governing the ${\cal N} = (0,2)$ supersymmetry worldsheet theory. Of key importance for us is a discussion of the worldsheet modes in Section \ref{sec:anomalymatching}, including our assumptions on their charges.
Section \ref{sec:QGbounds} contains our main results: We derive the EFT string consistency conditions by demanding that the anomaly inflow be cancelled by the  anomalies (reviewed in Appendix \ref{app_Anomalies}) of the weakly coupled worldsheet spectrum. 
In Section \ref{sec:examples} we show how these constraints rule out simple otherwise consistent supergravities as theories with a quantum gravity completion.

In the second part of the article, we test and apply our general results in explicit string theoretic frameworks.
In Section \ref{sec:Ftheory}, we apply our bounds to F-theory compactifications to four dimensions. 
We will illustrate the validity of the assumptions
made in the general setting in Section  \ref{sec:anomalymatching}, and in addition
motivate a sharpened bound on the rank of the gauge group, given by \eqref{strictboundF} in F-theory on smooth three-fold bases.
As another application, we will derive the constraint that in any Type IIB orientifold with O3-planes (and no O7-planes), the number of O3-planes is quantized in units of $16$, a physics post-diction of a mathematical theorem.  Section \ref{sec:heterotic} analyses the intriguing structure of the cone of EFT strings in heterotic string theory including higher derivative corrections. In the interest of readability, we have relegated some of the technicalities to Appendix \ref{app:hetEFT} and \ref{app:M5inst}. 
In Appendix \ref{app:curveNLSM} we corroborate our claims that the EFT string can be described by a weakly coupled non-linear sigma model, by analysing the EFT strings in F-theory and the heterotic theory in more detail.
In Section \ref{sec:Mtheory} we apply our EFT string constraints in the context of M-theory compactifications on $G_2$ manifolds. 
Our conclusions and a list of open questions are presented in Section \ref{sec:concl}.


\section{Perturbative bulk EFT structure}
\label{sec:bulk}

 In the standard Wilsonian interpretation, any effective field theory (EFT) is associated with a given ultraviolet (UV) cut-off energy scale $\Lambda$. We will consider four-dimensional EFTs which preserve minimal $\caln=1$ supersymmetry for a sufficiently high cut-off energy scale $\Lambda$.  
 This minimal supersymmetry may be spontaneously broken at lower energy scales $\Lambda_{\rm SB}\ll \Lambda$, but this will not affect our conclusions as these regard the structure of the EFT defined at the scale $\Lambda$.

 We will be particularly interested in extracting some general constraints on the (massless) EFT gauge sector that is  weakly coupled at the UV cut-off scale $\Lambda$.  The gauge couplings will be regarded as determined by the vacuum expectation values (VEVs) of the scalar fields. This is common in string theory realizations  and also expected from more general quantum gravity principles, which forbid  freely tunable parameters. It is then natural to associate any weakly-coupled gauge sector  with a certain region of the field space which identifies a given perturbative EFT  regime. In the sequel we will confirm this expectation and make it more precise.
 
In order to identify the possible perturbative EFT  regimes, we will  adopt the  general prescription provided in \cite{Lanza:2020qmt,Lanza:2021qsu}, which was proposed to be valid for any four-dimensional $\caln=1$ EFT consistent with quantum gravity and tested  in large classes of string theory models.   Some of its key ingredients will be reviewed below. Combined with additional quantum gravity criteria, this framework will allow us to extract non-trivial  information on the gauge sector and on some higher curvature terms.

 \subsection{Gauge sector}
\label{sec:gauge}
 
Following \cite{Lanza:2020qmt,Lanza:2021qsu,Lanza:2022zyg}, the perturbative regime of an ${\cal N}=1$ supersymmetric EFT in four dimensions is characterized by the presence of a  set of axions $a^i$, with fixed periodicity
\be\label{axionper}
a^i\simeq a^i+1\,,
\ee and a corresponding set of  saxions $s^i$. These  pair up into a set of complex scalar fields 
\be
t^i=a^i+\ii s^i
\ee
forming the bosonic components of ${\cal N}=1$ chiral superfields.  Together with additional chiral fields $\phi^\alpha$, they parametrize the K\"ahler field space $\calm$ of the EFT. 

The saxions in particular determine the exponential suppression factors of the BPS instantons in the theory. 
In a more precise definition, a perturbative EFT  regime is associated with  a set $\calc_{\rm I}$ of non-vanishing BPS instanton charge vectors ${\bf m}=\{m_i\}$, and a   {\em saxionic cone} 
\be\label{scone} 
\Delta=\{{\bm s}\in\mathbb{R}^{\#\text{axions}}| \langle {\bf m},{\bm s}\rangle>0\,, \forall {\bf m} \in \calc_{\rm I}\}\,.
\ee 
Here \be\langle {\bf m},{\bm s}\rangle\equiv m_i s^i\ee is the natural pairing between instanton charges and saxions. 
In terms of this pairing,
 a BPS instanton  with charge vector ${\bf m} \in \calc_{\rm I}$ is suppressed as $|e^{-2\pi\langle {\bf m},{\bm s}\rangle}|$.

We then say that the EFT is in the perturbative regime associated with $\calc_{\rm I}$, or equivalently $\Delta$, if the saxions lie sufficiently deep inside the saxionic cone $\Delta$. In this regime the axionic shift symmetries are broken only by exponentially suppressed non-perturbative corrections dominated by the BPS instantons, which  have the form $\sim |e^{2\pi\ii\langle {\bf m},{\bm t}\rangle}|=|e^{-2\pi\langle {\bf m},{\bm s}\rangle}|\ll 1$ (with $\langle {\bf m},{\bm t}\rangle\equiv m_it^i$). This identifies the perturbative regime with a field space region $\calm^\Delta_{\rm pert}$. Note that for increasing saxionic values inside the saxionic cone, the axionic shift symmetries are better and better preserved. The expected absence of (non-accidental) global symmetries in quantum gravity implies that $\calm^\Delta_{\rm pert}$  can be identified with the neighborhood of a field space boundary component $ \del\calm^{\Delta}_{\infty}\subset \del\calm$ which is at infinite distance. The by now well-tested Swampland Distance Conjecture (SDC) \cite{Ooguri:2006in} implies that one cannot really reach $ \del\calm^{\Delta}_{\infty}$ within the four-dimensional EFT because of the appearance of infinite towers of new microscopic  states which become  light exponentially fast in the field distance. This causes the EFT to break down as soon as the corresponding tower mass scale $m_*$ becomes smaller than the cutoff $\Lambda$. As we will recall in the sequel, \cite{Lanza:2020qmt,Lanza:2021qsu} identify a physically distinguished way to reach the infinite distance points of $ \del\calm^{\Delta}_{\infty}$ and to realize the SDC.

Consider now the gauge theory sector, with gauge group
\be\label{gaugeG}
G=\prod_A U(1)_A\times \prod_I G_I  \,,
\ee 
where $G_I$ denote simple group factors. In the superspace conventions of \cite{Wess:1992cp}, the associated terms in the two-derivative effective action are
\be\label{kingauge0}
\begin{aligned}
&\frac{1}{8\pi\ii}\int\d^4 x\d^2\theta\,2\cE\,  f^{AB}(t,\phi)\cW_A \cW_B+\frac{1}{16\pi\ii}\int \d^4 x\d^2\theta\,2\cE\, f^{I}(t,\phi)\tr(\cW \cW)_I+\text{c.c.}\,,
\end{aligned}
\ee
which includes the bosonic terms
\be\label{kingauge}
\begin{aligned}
-\frac1{4\pi}\int\left( \Im f^{AB} F_A\wedge *F_B+\Re f^{AB}F_A\wedge F_B\right)-\frac1{8\pi}\int \left[\Im f^{I} \tr (F\wedge *F)_I+\Re f^{I} \tr (F\wedge F)_I\right] \,.
\end{aligned}
\ee
Here  $f^{AB}(t,\phi)$ and $f^{I}(t,\phi)$ are holomorphic functions  and we denote chiral superfields and  their bottom scalar components by the same symbols. In \eqref{kingauge}, the trace ${\rm tr}$ on the  algebra $\mathfrak{g}$ of a simple group $G$ is defined by ${\rm tr}\equiv \frac{2}{\ell({\bf r})}{\rm tr}_{\bf r}$,
where ${\rm tr}_{\bf r}$ is the standard trace in   any unitary representation ${\bf r}$, and $\ell({\bf r})$ is the Dynkin index.\footnote{\label{foot:dynkin}We define the Dynkin index  by $\tr_{\bf r} t_a t_b=\frac{\ell({\bf r})}{2h(\mathfrak{g})}\tr_{\bf adj} t_a t_b$,  where $h(\mathfrak{g})$ is the dual Coxeter number of $\mathfrak{g}$. For instance, $\ell({\bf fund})=1$ and $\ell({\bf fund})=2$   if $\mathfrak{g}=\mathfrak{su}(n)/\mathfrak{sp}(n)$    and $\mathfrak{g}=\mathfrak{so}(n)$, respectively, or $\ell({\bf adj})=60$ for $\mathfrak{g}=\mathfrak{e}_8$. In this paper we use {\em hermitian} gauge fields $A=A^at_a$ and field strengths $F=F^at_a$, so that $A^\dagger =A$ and $F^\dagger=F$.
The hermitian generators $t_a$  of the gauge algebra $\mathfrak{g}$ can be normalised so that $\tr t_a t_b=2\delta_{ab}$. The instanton number can be identified with the integer $n=-\frac{1}{16\pi^2}\int \tr(F\wedge F)\in \mathbb{Z}$,
see for instance \cite{Bernard:1977nr}, and  $n=\frac{1}{16\pi^2}\int \tr(F\wedge *F)>0$ for (Euclidean) anti-self-dual ($F=-*F$) instanton  configurations.  Note that in the literature characteristic classes and  anomaly polynomials are often expressed in terms of the  {\em anti}-hermitian field strength $F_{\rm AH}\equiv -\ii F$. } 

In \eqref{kingauge} we assume  that 
\be\label{gaugepert} 
\Im f^{AB}\gg 1\,, \qquad \Im f^I\gg 1\,,
\ee 
so that the gauge sectors can be considered weakly coupled at the cut-off scale $\Lambda$ and the EFT hence admits a sensible perturbative expansion in the gauge couplings. 
Since we would like to focus on the asymptotic field space region $\calm^\Delta_{\rm pert}$ defined above, we furthermore assume that the holomorphic gauge functions $f^{AB},f^{I}$ can be expanded as  
\be\label{gaugef}
\begin{aligned}
f^{AB}(t,\phi)=\langle {\bf C}^{AB},{\bm t}\rangle +\Delta f^{AB}(\phi)+ \ldots\quad,\quad 
f^{I}(t,\phi)= \langle {\bf C}^{I},{\bm t}\rangle +\Delta f^{I}(\phi)+ \ldots \,.
\end{aligned}
\ee
Here we are omitting exponentially suppressed non-perturbative terms $\sim {\cal O}(\left|e^{2\pi\ii \langle {\bf m},{\bm t}\rangle}\right|)$ and we are employing the same index-free notation as in (\ref{scone}) for $t^i = a^i + \ii s^i$, e.g. 
\be 
\langle {\bf C}^{AB},{\bm t}\rangle \equiv {C}^{AB}_i t^i\,.
\ee
 From the expansion \eqref{gaugef} we see that \eqref{kingauge} contains in particular the (s)axionic couplings
\be\label{kingauge2}
\begin{aligned}
-\frac1{4\pi}C^{AB}_i\int  \left(s^i\, F_A\wedge *F_B+ a^i\,F_A\wedge F_B\right)-\frac1{8\pi}C^I_i\int \left[s^i\,\tr (F\wedge *F)_I+a^i\,\tr (F\wedge F)_I\right] \,,
\end{aligned}
\ee
from which one can extract $C^{AB}_i$ and $C^I_i$. Note that the form \eqref{kingauge2} for the (s)axionic couplings is a non-trivial assumption which need not hold for every gauge sector. We will come back to this caveat in the paragraph after \eqref{linmult}.

 Assuming that the gauge instanton configurations are defined on a Euclidean spin manifold, as is natural in supergravity, the compatibility of the gauge instanton corrections with the axion periodicity \eqref{axionper} requires the quantization conditions 
\be\label{CCconst} 
C^{AB}_i,\  C^{I}_i\in \mathbb{Z}\,.
\ee 
In the following we will provide complementary evidence in favor of \eqref{CCconst}, which holds directly in Lorentzian signature. 

For generic values of the fields $\phi^\alpha$, we expect $\Delta f^{AB}(\phi),\Delta f^I(\phi)\sim \calo(1)$ in \eqref{gaugef}. Hence \eqref{gaugepert} suggests that the constants \eqref{CCconst} should obey the constraints
\be\label{sCcond1}
\{\langle {\bf C}^{AB},{\bm s}\rangle\}\geq 0\,, \quad \quad \langle {\bf C}^{I},{\bm s}\rangle\geq  0\quad~~~~~~\forall {\bm s}\in \Delta\,.
\ee
The first condition means that $\langle {\bf C}^{AB},{\bm s}\rangle$ is a positive definite matrix, and we are again using the index-free notation introduced above, e.g.\ $\langle {\bf C}^{AB},{\bm s}\rangle\equiv  C^{AB}_i s^i$ and $t^i = a^i + \ii s^i$. In other words, it is natural to require that  ${\bf C}^{I}\in \calc_{\rm I}$, where we recall from the discussion before \eqref{scone} that $\calc_{\rm I}$ is the cone of BPS instanton charges dual to the saxionic cone $\Delta$, and that $\{{\bf C}^{AB}\}\in \calc_{\rm I}$ in a matrix sense. This is also consistent  with the requirement  that  the perturbative gauge interactions preserve the axionic shift symmetries, while  gauge instantons break them by exponentially suppressed non-perturbative corrections.
To our knowledge the condition  \eqref{sCcond1} is realised in all string theory models and in the following we will assume that it  should hold in any EFT compatible with quantum gravity.


\subsection{Higher curvature terms}
\label{sec:curv}

One can generically write down higher-derivative corrections to the leading two-derivative supergravity. For our purposes it is sufficient to focus on the contribution
\be\label{SGB}
\frac{1}{24\pi\ii}\int\d^4 x \, \d^2\theta\,2\cE\,  \tilde f(t,\phi)Y+\text{c.c.}\,,
\ee
where $\tilde f(t,\phi)$ is another holomorphic function of the chiral multiplets and 
 \begin{equation}\label{YGBdef}
Y\equiv \mathcal{W}^{\alpha \beta \gamma} \mathcal{W}_{\alpha \beta \gamma}-\frac{1}{4}\left(\overline{\mathcal{D}}^{2}-8 \mathcal{R}\right)\left(a \mathcal{R} \overline{\mathcal{R}}+bG^{a} G_{a}\right)\,
\end{equation}
is a composite chiral superfield \cite{Townsend:1979js,Cecotti:1985mf,Cecotti:1987mr} constructed out of the chiral superfields  $\mathcal{W}_{\alpha \beta \gamma}$, $\calr$ and the real superfield $G_a$ of minimal $\caln=1$ supergravity \cite{Wess:1992cp}.  The superspace  contribution  \eqref{SGB} includes in particular the curvature-squared  terms \cite{Cecotti:1985mf,Cecotti:1987mr}
\be\label{PGB}
\begin{aligned}
-\frac1{96\pi}\int \Im\tilde f\, \tr(R\wedge *R)- \frac{1}{96\pi}\int  \Re\tilde f\,{\rm tr}(R\wedge R)+\ldots\,.
\end{aligned}
\ee
Here tr denotes the standard trace on the free indices of the Riemann two-form $R^m{}_n=\frac12R^m{}_{npq}\,\d x^p\wedge \d x^q$ -- see also  \cite{Bonora:2013rta} and \cite{Fotis} for some useful details on the necessary superspace manipulations.\footnote{We thank Fotis Farakos for  useful discussions about the necessary superspace gymnastics.}  In  \eqref{PGB} we have omitted  $R_{mn}R^{mn}$ and $R^2$ terms, since their coefficients depend on the constants $a$ and $b$ and then are not uniquely fixed by the  Pontryagin term containing ${\rm tr}(R\wedge R)$. On the other hand, the arbitrariness is uniquely fixed to $a=2$ and $b=1$ by requiring that the omitted $R_{mn}R^{mn}$ and $R^2$ terms combine with  the first term in \eqref{PGB} to give the Gauss-Bonnet term $\frac{1}{192\pi}\int \Im\tilde f\,  E_{\rm GB}*1$, with 
\be\label{GBcomb} 
E_{\rm GB}\equiv R_{mnpq}R^{mnpq}-4R_{mn}R^{mn}+R^2\,,
\ee 
which is not affected by ghost issues \cite{Zwiebach:1985uq}.

As for the  gauge functions  \eqref{gaugef}, under our general assumptions in the asymptotic field space region $\calm^\Delta_{\rm pert}$ the holomorphic function $\tilde f$ necessarily takes the form 
\be\label{tildef}
\tilde f(t,\phi)= \tilde C_i\, t^i+\Delta \tilde f(\phi)+\ldots \,,
\ee
where again we are omitting exponentially suppressed non-perturbative  terms. In particular, \eqref{PGB} contains the couplings 
\be\label{aRR}
-\frac1{96\pi}\tilde C_i\int\left[ s^i\, \tr(R\wedge *R) +a^i\, \tr(R\wedge R)\right] \,,
\ee
from which one can extract $\tilde C_i$.

According to the normalization  of \eqref{SGB},  the axion periodicity \eqref{axionper} is not broken by possible gravitational instantons  if we require that 
\be\label{tildeCquant}
2\tilde C_i\in \mathbb{Z}\,.
\ee
This can be understood by recalling that  the integral of the first Pontryagin class $p_1(M)=-\frac1{8\pi^2}{\rm tr}(R\wedge R)$ over a Euclidean spin four-manifold $M$ is always a  multiple of 48. In the sequel we will see that quantum gravity constraints require  $\tilde C_i$ to be integral, rather than half-integral as in \eqref{tildeCquant}. 

Unlike for the analogous quantities  discussed in the previous subsection, there is no obvious reason to expect any definite sign of $\Im\tilde f$ and $\langle\tilde {\bf C},{\bm s}\rangle \equiv \tilde C_i s^i$. On the other hand, various results suggest that the positivity of the coefficient of the Gauss-Bonnet term may be a general feature of EFTs consistent with quantum gravity. For instance this is necessary in order to suppress problematic  wormhole effects \cite{Kallosh:1995hi}. More recently, \cite{Cheung:2016wjt}  provides an argument for positivity based on unitarity in pure   in pure $d>4$ gravity, \cite{GarciaEtxebarria:2020xsr} discusses the implications of the sign on the non-perturbative (in)stability of simple dilatonic models, \cite{Aalsma:2022knj} shows how Gauss-Bonnet positivity follows from the WGC for certain black holes and \cite{Ong:2022mmm} analyses constraints based on holography. In our context this would mean that $\Im\tilde f> 0$ and then, as for the  gauge theory sector, it would be natural to require that 
\be\label{tildeCs}
\langle\tilde {\bf C},{\bm s}\rangle>  0\,,
\ee
or equivalently $\tilde {\bf C}\in\calc_{\rm I}$. (Here we are already using the integrality of $\tilde C_i$, anticipated above but not proven yet.)   
We will find that \eqref{tildeCs} follows, under certain additional mild restrictions, from the quantum gravity arguments of the following sections.


\section{EFT strings as quantum gravity probes}
\label{sec:EFTstrings}

The perturbative EFT regimes of Section \ref{sec:bulk} can be characterised in terms of a specific class of BPS axionic strings, called {\em EFT strings} in \cite{Lanza:2020qmt,Lanza:2021qsu} -- see also \cite{Lanza:2022zyg}. 
In this section, after recalling the main properties of such EFT strings, we will describe the anomaly inflow mechanism from the four-dimensional bulk theory to the string worldsheet. We will then argue that the EFT string worldsheet theory can be treated as a weakly coupled non-linear sigma model (NLSM) and characterise the spectrum of its massless fields. This will form the basis for the derivation of quantum gravity constraints for the four-dimensional field theory in Section \ref{sec:QGbounds}.

\subsection{Perturbative EFT regimes and EFT strings}

The presence of the perturbative axionic shift symmetries (\ref{axionper}) naturally leads one to consider axionic strings in four dimensions, around which the axions undergo integral shifts
\be \label{aishift}
a^i\rightarrow a^i+e^i\,.
\ee
In this paper we assume that the  EFT  $U(1)$ gauge fields  do not acquire a St\"uckelberg mass by gauging the axions $a^i$. In this way we exclude  axionic strings  which can break by nucleation of monopole pairs, leaving the investigation of this interesting generalization for future work.

The integers $e^i\in\mathbb{Z}$ can be regarded as the magnetic axionic charges of the string, or as the electric charges under the dual two-form potentials $\calb_{2,i}$ -- see \cite{Lanza:2019xxg,Lanza:2021qsu} for more details on the dualization. In the dual formulation, the strings  contribute to the EFT by a localized term 
\be\label{stringWZ} 
e^i\int_W \calb_{2,i}\,,
\ee
where $W$ denotes the string world-sheet. Furthermore, imposing that these strings are compatible with the bulk supersymmetry  completely fixes \cite{Lanza:2019xxg} the additional contribution $-\int_W \calt_{\bf e}\,\text{vol}_W$ to the effective action: The tension $\calt_{\bf e}$ takes the  form
\be\label{stringNG} 
\calt_{\bf e}\equiv M^2_{\rm P}\,e^i\ell_i\,,
\ee
in terms of the   the {\em dual saxions} $\ell_i$. Together with $\calb_{2,i}$, these form the bosonic components of the linear multiplets dual \cite{Lindstrom:1983rt} to the chiral multiplets $t^i$  and are defined by 
\be \label{linmult}
\ell_i\equiv -\frac12\frac{\del K}{\del s^i}  \,.
\ee 
Here $K$ is the EFT K\"ahler  potential $K$, which is assumed to be invariant under the axionic shift symmetries. Note that the dual formulation in terms of $\calb_{i,e}$ and $\ell_i$, as well as the localised terms \eqref{stringWZ} and \eqref{stringNG}, really make sense only if the axionic shift symmetries are  preserved at the perturbative level, as we are assuming.\footnote{In this dual formulation the non-perturbative corrections can be generated by the mechanism described in \cite{Kallosh:1995hi}.}

At this stage it is important to stress that we are excluding possible  monodromy transformations of the $U(1)$ vectors under the axionic integral shifts \eqref{aishift}. For instance, if we focus on two $U(1)$ field strengths and one chiral field  $t=a+\ii s$, we could impose that the discrete identification $t\simeq t+1$ involves also a shift $F_1\simeq F_1+F_2$. $F_1$ can  then enter the effective Lagrangian only through the monodromy invariant combination $\hat F_1=F_1-a F_2$, which can be completed into the super-field strength $\hat\calw^\alpha_1=\calw^\alpha_1-t\calw_2^\alpha$. These monodromy effects can be immediately generalized to a larger number of  $U(1)$s and axions, and allow for the appearance of quadratic and cubic axion couplings to $F_A\wedge F_B$, rather than the standard linear coupling \eqref{kingauge2}.
As an example of such non-standard couplings to the axions, Kaluza-Klein $U(1)$s have been discussed in \cite{Heidenreich:2021yda}.
Such couplings {\em obstruct} the dualization of the axions to $\calb_{2,i}$ (as well as its supersymmetric completion). Then the EFT contribution of the corresponding axionic strings cannot be described as in \eqref{stringWZ} and \eqref{stringNG}. While these  monodromy effects naturally appear in extended four-dimensional supergravities,  they look more exotic in a minimally supersymmetric context, and we will henceforth only consider effective theories not exhibiting such subtle effects.

Let us come back to the dualization \eqref{linmult} and the fact that this procedure is possible only in presence of a perturbative shift symmetry. This becomes crucial once combined with the observation that the strings coupling to the two-form, being  codimension-two objects,  have a strong backreaction. The backreaction  may force the surrounding bulk scalar fields to flow away from the asymptotic region $\calm^\Delta_{\rm pert}$ associated with the perturbative regime. However, as emphasised in \cite{Lanza:2020qmt,Lanza:2021qsu}, this does not pose any problem at the EFT level: What one really needs for consistency of the dualization and the description of the strings is that the bulk sector is in the weakly coupled region in a small neighborhood of the string, of minimal radius of order  $r_\Lambda\equiv \Lambda^{-1}$. It is  then sufficient to require that the string backreaction is under control in this neighborhood. At such short distances, the string can be well approximated by a straight $\frac12$-BPS string and its backreaction on the saxions is given by
\be\label{sflow}
{\bm s}={\bm s}_0+{\bf e}\,\sigma\,, \quad \quad \sigma\equiv \frac1{2\pi}\log\frac{r_0}{r}\,.
\ee
Here $r$ is the radial distance from the string and ${\bm s}_0=\{s^i_0\}$ represents the initial saxionic values at  $r=r_0$. Note that the additional chiral fields $\phi^\alpha$ do not flow. Hence, if the charge vector ${\bf e}$ belongs to 
\be\label{CSEFT}
\calc_{\rm S}^{\text{\tiny EFT}}\equiv \overline\Delta|_{\mathbb Z}=\{{\bf e}\in \mathbb{Z}^{\text{\#\,axions}}\ |\ \langle {\bf m},{\bf e}\rangle\geq 0\,,\ \forall {\bf m}\in \calc_{\rm I}\}\,,
\ee
a flow \eqref{sflow} starting from any ${\bm s}_0\in \Delta$  never exits the saxionic cone $\Delta$ as one approaches the string. Actually, the scalars are driven more and more inside the asymptotic weakly coupled region $\calm^\Delta_{\rm pert}$ as $\sigma\rightarrow \infty$. This can be taken as the defining property of the EFT strings, which hence admit a controlled  weakly-coupled EFT description. Note that the EFT strings must be considered as fundamental strings\footnote{This notion is not to be confused with that of a {\it critical} string such as the heterotic or Type II string.}, in the sense that they cannot be completed into smooth solitonic objects within a four-dimensional EFT, say by adding a finite number of new degrees of freedom.

In \cite{Lanza:2021qsu} it was shown how the validity  of `weak gravity bounds' \cite{ArkaniHamed:2006dz} for instantons and strings  requires that in $\calm^\Delta_{\rm pert}$  the  K\"ahler potential receives a leading saxionic contribution of the form
\be\label{KlogP}
K=-\log P({\bm s})+\ldots \,,
\ee
where $P({\bm s})$ is a homogeneous function of the saxions. This asymptotic structure of the K\"ahler potential is indeed universally realised in all string theory models, in which the homogeneity degree of $P({\bm s})$ turns out to be integral. In the following, whenever we will need it, we will implicitly assume this asymptotic form of the K\"ahler potential. This in particular implies that  the EFT string flows \eqref{sflow} can be regarded as infinite field distance limits, and the Distant Axionic String Conjecture (DASC) of \cite{Lanza:2021qsu}  proposes that all infinite distance points of $\del\calm^\Delta_\infty$ can actually be reached by an EFT string flow.  

Another important property of the EFT strings is the following. It is not difficult to see that the field-dependent  tension $\calt_{\bf e}$  of an EFT string decreases  along its own saxionic flow \eqref{sflow}, as $\sigma\rightarrow \infty$. On the other hand, since this is an infinite field distance limit the Swampland Distance Conjecture implies that along the flow an infinite tower of microscopic massive modes appears, at a characteristic mass scale $m_*$, which exponentially vanishes with the field distance. By inspecting a large class of string theory models, in \cite{Lanza:2021qsu}  it was found  that along the EFT flows  $m^2_*$ scales to zero as an integral power of $\calt_{\bf e}$, in Planck units. This `experimental' observation was promoted to a possible universal property of EFT strings. In its stronger form proposed in \cite{Lanza:2022zyg}, this is the content of the Integral Weight Conjecture (IWC):
\be\label{IWC}
m^2_*\sim M^2_{\rm P}\left(\frac{\calt_{\bf e}}{M^2_{\rm P}}\right)^w\quad~~~ w\in\{1,2,3\}\,,
\ee
where $w$ is the `scaling weight' associated with the EFT string. Note also that, according to the Emergent String Conjecture (ESC) \cite{Lee:2019wij},  in some duality frame  an EFT string with $w=1$ should coincide with a {\it critical} string, while EFT string flows with $w=2,3$ should correspond to decompactification limits, as further characterised in \cite{Cota:2022yjw}.

In order to motivate our definition of \eqref{CSEFT} we used purely EFT arguments. However, as we already mentioned, EFT strings are `fundamental' and their existence depends on the UV completion of the theory. In the following discussion we will need to assume that the EFT string lattice $\calc_{\rm S}^{\text{\tiny EFT}}$ is actually fully populated. This non-trivial assumption is certainly realized in the large classes of string theory models considered in \cite{Lanza:2021qsu}, and can be more generically motivated by invoking an EFT string version of the Completeness Conjecture \cite{Polchinski:2003bq}, which is one of the better tested quantum gravity criteria.\footnote{
Recently \cite{Montero:2022vva} has found examples of effective field theories in higher dimensions which violate the Completeness Conjecture for {\it BPS} strings; at the same time, there always exists another theory differing only at the massive level which does satisfy the BPS Completeness Conjecture. For our purpose of constraining the massless EFT spectra this would in fact be sufficient.}  

Note that the EFT strings make sense only in a gravitational context.  Indeed, the analysis of \cite{Lanza:2021qsu} implies that the vanishing of an EFT string tension $\calt_{\bf e}=M^2_{\rm P}e^i\ell_i$ identifies a component of the infinite field distance boundary $\del\calm^\Delta_\infty$. Hence there is no (finite distance) point in field space at which  $\calt_{\bf e}/M^2_{\rm P}\rightarrow 0$ and one cannot decouple the string dynamics from the gravitational sector. Furthermore, in all the  string theory realizations considered so far, the EFT strings uplift to brane configurations which can be continuously deformed through the entire internal compactification space. In this sense, the EFT strings can `probe' the entire UV completion of the EFT. We will indeed  see that their quantum consistency provides additional non-trivial constraints on the EFT.

\subsection{World-sheet anomaly from anomaly inflow}

It is well known that  axionic strings can support chiral fermions whose anomaly must be cancelled by a bulk anomaly inflow \cite{Callan:1984sa}. As in this reference, the axion couplings  appearing in \eqref{kingauge} and \eqref{PGB} produce  an anomaly inflow to the  EFT strings. Taking into account \eqref{gaugef} and \eqref{tildef}, the four-dimensional couplings
can be written in the form
\be\label{aIcoupl}
2\pi\int a^i I_{4,i}=-2\pi \int h^i_1\wedge I^{(0)}_{3,i} \,.
\ee
Here we have introduced the one-forms $h^i_1=\d a^i$, which are globally defined even in presence of axionic strings, and 
\be\label{FFterms}
\begin{aligned}
I_{4,i}&\equiv \d I^{(0)}_{3,i}\equiv -\frac{1}{8\pi^2}C_i^{AB} F_A\wedge F_B- \frac{1}{16\pi^2}C^I_{i}\, {\rm tr} (F\wedge F)_I- \frac{1}{192\pi^2}\tilde C_i\,{\rm tr}(R\wedge R)\,.
\end{aligned}
\ee
The anomaly inflow is generated by the term \eqref{aIcoupl} since, in presence of an axionic string of charges $e^i$ and world-sheet $W$, the one-forms $h_1^i$ are not closed, but rather satisfy
\be\label{axisource1}
\d h^i_1=e^i\delta_2(W) \,.
\ee
Under general gauge transformations and local Lorentz transformations we can write $\delta I^{(0)}_{3,i}=\d I^{(1)}_{2,i}$, and then \eqref{aIcoupl} produces the following localized contribution to the corresponding variation of the bulk action: 
\be \label{deltaSbulk}
\delta S_{\rm bulk}=-2\pi e^i \int_W I^{(1)}_{2,i}\,. 
\ee 
The same result can be obtained in the dual formulation, by taking into account that  the couplings \eqref{aIcoupl} modify the dual  field strengths into $\calh_{3,i}=\d\calb_{2,i}+2\pi I^{(0)}_{3,i}$. Since $\calh_{3,i}$ must be gauge invariant, under gauge and local Lorentz transformations $\calb_{2,i}$ must transform non-trivially: 
\be\label{bulkgtr} 
\delta \calb_{2,i}=-2\pi I^{(1)}_{2,i}\,.
\ee 
Hence \eqref{stringWZ} produces precisely the same localized contribution to the gauge variation of the action found above.   

Since a consistent EFT must be anomaly free, the localized contribution \eqref{deltaSbulk} must be cancelled by a string world-sheet anomaly $\delta S^{\rm quantum}_W = + 2\pi e^i \int_W I^{(1)}_{2,i}$. Via the usual descent equation, this worldsheet anomaly corresponds to the anomaly polynomial 
\be\label{polyn2}
\begin{aligned}
e^i I_{4,i}=- \frac{\langle {\bf C}^{AB},{\bf e}\rangle}{8\pi^2} F_A\wedge  F_B -\frac{ \langle {\bf C}^I,{\bf e}\rangle}{16\pi^2} \tr (F\wedge F)_I-\frac{\langle \tilde{\bf C},{\bf e}\rangle}{192\pi^2}\tr (R\wedge R)\,.
\end{aligned}
\ee
The last term is proportional to the bulk first Pontryagin class $p_1(M)=-\frac{1}{8\pi^2}\tr (R\wedge R)$.
Since $TM|_W=TW\oplus N_W$, where $N_W$ is the normal bundle of $W$, $p_1(M)$ can be decomposed into 
\be 
p_1(M)=p_1(W)+c_1(N_W)^2=-\frac{1}{8\pi^2}\tr (R_W\wedge R_W)+\frac{1}{4\pi^2}F_{\rm N}\wedge F_{\rm N}\,,
\ee
where $F_{\rm N}=\d A_{\rm N}$ is the field strength of the $U(1)_{\rm N}$ connection  $A_{\rm N}$ induced on the normal bundle by the bulk Riemannian connection. 
The last term of \eqref{polyn2} then democratically contributes to the gravitational and $U(1)_{\rm N}$ anomaly of the world-sheet theory. 

There can occur an additional, and more subtle, contribution to the $U(1)_{\rm N}$ anomaly inflow. It  is analogous to the additional contribution associated with the type IIA NS5-brane normal bundle,  discussed in  \cite{Witten:1996hc}. Adapted to our context -- see also  \cite{Becker:1999kh} -- the key observation is that the pull-back to the string world-sheet $W$ of the singular right-hand side of \eqref{axisource1} gives a non-vanishing finite term. Indeed, in cohomology
\be\label{deltaN}
\delta_2(W)|_{W}=\chi(N_W)\,,
\ee
where $\chi(N_W)$ is the Euler class of the normal bundle $N_W$. Since we have a distinguished connection $A_{\rm N}$ on $N_W$, it is natural to promote \eqref{deltaN} to an equation for (distributional) differential forms, by identifying
\be
\chi(N_W)=\frac1{2\pi}F_{\rm N}\,.
\ee
The restriction of \eqref{axisource1} to $W$ then contains a finite contribution coming from the delta-source:
\be\label{axisource2}
\d h^i_1|_W=\frac{e^i}{2\pi}F_{\rm N}\,.
\ee
 Topologically, this equation implies that the normal bundle $N_W$ must be trivial.
 Hence, we know that we can globally write $F_{\rm N}=\d A_{\rm N}$.
 Following \cite{Witten:1996hc} we will consider \eqref{axisource2} as  part of the definition of our axionic strings. 

By adopting the `magnetic' axionic formulation, we can now write down the following additional EFT term localised on the string:
\be\label{SN}
S_{\rm N}=-\frac1{24}\hat C_i({\bf e})\int_W h^i_1\wedge A_{\rm N} \,.
\ee 
Here the normalization is chosen for later convenience and we assume that the constants $\hat C_i({\bf e})$ necessarily depend on ${\bf e}$ since otherwise  the term \eqref{SN} would be independent of the EFT string charge, which does not seem  physically reasonable.  By recalling \eqref{axisource2} it is easy to see that \eqref{SN} is not invariant under $U(1)_{\rm N}$ a gauge transformations $\delta A_{\rm N}=\d\lambda_{\rm N}$:  
\be\label{deltaSN}
\delta S_{\rm N}=-\frac1{48\pi}\hat C_i({\bf e})e^i\int_W \lambda_{\rm N}\, F_{\rm N}\,.
\ee

Again, we may obtain this anomalous contribution in the dual description, in which the term \eqref{SN} does not appear and is encoded in a
modification of the $\calh_{3,i}$ Bianchi identities:
\be
\d \calh_{3,i}=2\pi I_{4,i}+\frac1{24}\hat C_i({\bf e})F_{\rm N}\wedge \delta_2(W)\,.
\ee
This can be solved by setting  
\be 
\calh_{3,i}=\d\calb_{2,i}+2\pi I^{(0)}_{3,i}+\frac1{24}\hat C_i({\bf e})A_{\rm N}\wedge \delta_2(W)\,.
\ee
It follows that, in addition to \eqref{bulkgtr},  under local Lorentz transformations the variation of $\calb_{2,i}$  receives also the localised contribution
\be\label{NtrB} 
\delta_{\rm N}\calb_{2,i}=-\frac{1 }{24}\lambda_{\rm N}\,\hat C_i({\bf e})\delta_2(W)\,,
\ee 
which, applied to \eqref{stringWZ}, precisely reproduces \eqref{deltaSN}.

The anomalous contribution \eqref{deltaSN} must be cancelled by a  contribution $2\pi\int_W \hat I^{(1)}_{2,{\rm N}}$ to the world-sheet anomaly, corresponding to the anomaly polynomial
\be\label{Nanomaly0} 
\hat I_{4,{\rm N}}=\frac{\langle \hat {\bf C}({\bf e}),{\bf e}\rangle}{96\pi^2} F_{\rm N}\wedge F_{\rm N} \,,
\ee
where as usual $\langle \hat {\bf C}({\bf e}),{\bf e}\rangle\equiv e^i\hat C_i({\bf e})$. Hence, to sum up, the cancellation of the anomaly inflow requires that the  world-sheet anomaly polynomial must be given by
\be\label{polyn2again}
\begin{aligned}
I^{\rm ws}_{4\,{\bf e}}&= e^i I_{4\,i}+\hat I_{4,{\rm N}}\\
&=- \frac{\langle {\bf C}^{AB},{\bf e}\rangle}{8\pi^2} F_A\wedge  F_B -\frac{ \langle {\bf C}^{I},{\bf e}\rangle}{16\pi^2} \tr (F\wedge F)_I\\
&\quad -\frac{\langle \tilde{\bf C},{\bf e}\rangle}{192\pi^2}\tr (R_W\wedge R_W)+\frac{\langle \tilde{\bf C},{\bf e}\rangle+\langle \hat {\bf C}({\bf e}),{\bf e}\rangle}{96\pi^2} F_{\rm N}\wedge F_{\rm N} \,.
\end{aligned}
\ee
Note that this anomaly polynomial does not include possible mixed $U(1)_{\rm N}$-$U(1)_A$ anomalies. Indeed, they can be  assumed to be cancelled by terms of the form \eqref{SN} involving the $U(1)_A$ gauge fields $A_A$ instead of $A_{\rm N}$ and can then be ignored. 

With the exception of the additional terms in (\ref{Nanomaly0}), the anomalies encoded in
(\ref{polyn2again}) are all proportional to the pairing of the couplings $C^{AB}_i$, $C^{I}_i$ and $\tilde C_i$ with the charges $e^i$ characterising the EFT string.
Recall that the presence of such a string induces a flow of the form (\ref{sflow}) in the saxionic moduli space.
In view of the relations (\ref{gaugef}), we therefore conclude that precisely those gauge 
sectors $U(1)_A$ and $G_I$ that become weakly coupled as a consequence of the backreaction   of an EFT string, as encoded in the flow (\ref{sflow}), can be detected by its world-sheet anomalies. Gauge sectors which stay strongly coupled in presence of an EFT string, on the other hand,
never induce an anomaly on its worldsheet, and more generally the EFT string in question does not detect them via its world-sheet. In particular this applies to gauge sectors with non-standard axionic monodromies of the form mentioned after \eqref{linmult}, which fall outside the class of theories which can be constrained by the analysis of EFT string anomalies as studied in the present work.

The conditions \eqref{sCcond1} can be rewritten in terms of EFT string charges as
\be\label{pos1}
\{\langle  {\bf C}^{AB},{\bf e}\rangle\}\geq 0\,,\quad\langle  {\bf C}^{I},{\bf e}\rangle\geq 0\,, \quad~~~~\forall {\bf e}\in \calc^{\text{\tiny EFT}}_{\rm S}\,,
\ee
where the first inequality means that the matrix $\langle  {\bf C}^{AB},{\bf e}\rangle$ is positive semi-definite. This implies that the coefficients of the gauge world-sheet anomaly polynomial \eqref{polyn2again} have semi-definite sign.

\subsection{UV origin of the additional world-sheet contribution}
\label{sec:hatC}

As will be made more explicit in the next subsection, the constants $\hat C_i({\bf e})$ introduced in \eqref{SN} must satisfy appropriate quantization conditions and should then be associated with some additional discrete structure defining the EFT. In a generic $\caln=1$ four-dimensional EFT there is no natural candidate, but the more constrained higher dimensional supersymmetric theories can indeed be characterized by additional discrete data. This suggests that the presence of a term \eqref{SN} may be interpreted as the manifestation of some hidden higher dimensional structure which can be detected by the EFT string.

More concretely, consider a situation in which the axions $a^i$ come from the dimensional reduction of the $U(1)$ vectors $A^i$ of an $\caln=1$ five-dimensional supergravity. The latter is characterized by a set of quantized constants  \cite{Witten:1996qb} 
\be\label{hatCquant}
\hat C_{ijk}\in\mathbb{Z}\,,
\ee
 which are totally symmetric in the indices and in  particular define the five-dimensional Chern-Simons term
\be\label{5dCS}
S_{\rm 5d}=\frac{1}{6(2\pi)^2}\hat C_{ijk}\int A^i \wedge F^j\wedge F^k\,.
\ee
The four-dimensional axionic strings uplift to five-dimensional monopole strings.
For such monopole strings of charge vector ${\bf e}$, the term  \eqref{5dCS} generates the  inflow contribution  \cite{Freed:1998tg,Becker:1999kh,Boyarsky:2002ck,Katz:2020ewz}
\be\label{CSanom}
-\frac{2\pi}{24}\hat C_{ijk}e^ie^je^k \int_W p_1^{(1)}(\hat N_W)
\ee
 to the $SO(3)$ normal bundle anomaly of the monopole, where $p_1^{(1)}(\hat N_W)$ is the descent two-form associated with the first Pontryagin class of the  normal bundle $\hat N_W$. Under dimensional reduction to four dimensions the normal bundle splits as $\hat N_W=N_W\oplus L$, where $L$ is a trivial real line bundle, and then \eqref{CSanom}  precisely takes the form \eqref{deltaSN} with
\be\label{cubicC}
\hat C_i({\bf e})=\hat C_{ijk}e^je^k \,.
\ee
Hence, with this choice,  \eqref{SN} encodes the information on a microscopic five-dimensional structure that can be detected by the EFT strings. One can also more directly reproduce \eqref{SN} from the dimensional reduction of \eqref{5dCS}, as in \cite{Becker:1999kh}.  Since this effect is local, all that is required is a five-dimensional EFT term of the form \eqref{5dCS}, without the need to assume that the five-dimensional configuration globally preserves eight supercharges.\footnote{Note that a simple circle compactification from five to four dimensions would preserve eight supercharges and contain a KK $U(1)$ field. This KK $U(1)$  would be afflicted by  monodromy effects of the kind discussed in the paragraph after \eqref{linmult} -- see for instance \cite{Heidenreich:2021yda} -- which are not covered by our analysis.} Furthermore, the reduction from five to four dimensions would imply  a leading contribution to the K\"ahler potential of the form \eqref{KlogP} with $P({\bm s})=\hat C_{ijk}s^is^js^k$, inherited by the five-dimensional supersymmetric structure, which would correlate a world-sheet term of the form \eqref{CSanom} to the bulk EFT structure. In the sequel we will more explicitly illustrate these ideas in concrete string theory realizations.

Let us now investigate if the term \eqref{SN} can instead detect a six-dimensional minimally supersymmetric structure. As reviewed for instance in \cite{Kim:2019vuc}, in six dimensions the  $\caln=(1,0)$ tensor multiplet sector is characterized by a symmetric matrix  $\hat C_{ij}$ and the strings act as both electric and magnetic sources for the $\calb_{2,i}$ fields. The Green-Schwarz term produces an anomaly inflow polynomial proportional to $\hat C_{ij} e^ie^j\chi(\hat N_{W})$, where $\chi(\hat N_{W})$ is the Euler class of the string normal bundle $\hat N_W$ which has  $SO(4)\simeq SU(2)_r\times SU(2)_l$ structure group.  It would then be natural to guess that in four dimensions this effect can again be captured by \eqref{SN}, by choosing $\hat C_{\bf e}\propto \hat C_{ij}e^j$. However, this naive guess turns out to be wrong. Indeed, consider a compactification from  six to four  dimensions.  Since the string is point-like in the compact two dimensions, its normal bundle splits  into $\hat N_W=N_W\oplus L\oplus L$ where $L$ is again trivial real line bundle. This implies that $\chi(\hat N_{W})=\chi(N_{W})\chi(L\oplus L)=0$. Hence, even though $\hat C_i({\bf e})\propto \hat C_{ij}e^j$ has the right structure to be associated with a six-dimensional supergravity, apparently no anomaly survives in the  reduction to four dimensions. We are then led to exclude this possibility as being physically irrelevant.

Supergravity theories in $d>6$ dimensions are too rigid to allow for a choice of some discrete data, which could then  enter $\hat C_i({\bf e})$.   For these reasons, in the rest of this paper we will assume \eqref{cubicC} whenever we will need a more explicit form of $\hat C_i({\bf e})$, otherwise keeping it generic.


\subsection{EFT string as weakly coupled NLSM}
\label{sec:anomalymatching}

The world-sheet anomaly polynomial \eqref{polyn2} has been fixed by anomaly inflow arguments, a procedure which basically uses  only the axionic nature of the EFT strings. 
While EFT strings naturally allow for a coupling to a weakly-coupled bulk sector at the EFT cut-off scale $\Lambda$, they generically lead to infra-red divergences due to their large backreaction. In particular, one cannot a priori assume that their world-sheet theory flows to a CFT, as was for instance possible in the higher-dimensional context of \cite{Kim:2019vuc,Katz:2020ewz}.

However, the analysis of explicit UV complete models carried out in Appendix \ref{app:curveNLSM} provides evidence that also the world-sheet sector supported by EFT strings can be considered weakly-coupled at the EFT cut-off scale, and can then be described by a weakly-coupled non-linear sigma model (NLSM). This can be  understood as follows. In string theory models EFT strings are microscopically associated to branes which can freely propagate in the internal compactification space, which then determines the geometry of the effective two-dimensional NLSM. One may be worried that the string backreaction could obstruct  the possibility to make its NLSM weakly coupled. The key point is that for EFT strings this does not occur, and actually the EFT string flow tends to render the NLSM more weakly coupled. 

 Recall the observation made after  \eqref{IWC} and suppose first that $w=2$ or $3$. The Emergent String Conjecture \cite{Lee:2019wij} implies that in some duality frame the EFT flow involves the  decompactification of some internal direction, as has been made more precise for the EFT string limits in \cite{Cota:2022yjw}. The string theory models of Appendix \ref{app:curveNLSM} clearly indicate that this dynamical decompactification makes the effective two-dimensional NLSM more and more weakly coupled.

If instead $w=1$, then according to the Emergent String Conjecture there should exist a duality frame in which the EFT string is a critical string whose flow drives the dilaton to infinity (i.e. weak coupling) \cite{Lanza:2021qsu}. 
The remaining moduli do not flow and we may rescale them in order to make the compactification space arbitrarily large, at least if the dual description admits a geometric phase. 
In this regime the string certainly admits a weakly-coupled NLSM description. In the non-generic cases in which the large volume regime does not exist,  as for instance the rigid Landau-Ginzberg-like theories of the type considered in \cite{Becker:2006ks}, the string is critical in the deep UV and then it should  - and can - instead be directly treated as a CFT.

Motivated by these observations, we propose  that the world-sheet sector supported by EFT strings generically admits a weakly-coupled NLSM description and we will proceed with this working assumption.\footnote{More precisely, some  EFT string may host a `spectator' strongly coupled subsector, which does not participate in the cancellation of the  anomaly inflow and does not interfere with the weakly coupled NLSM dynamics.} Precisely in the non-generic cases in which this is not possible, one can  rephrase the following discussion in CFT terms (along the lines of \cite{Kim:2019ths}) and arrive at the same results. Hence in the following we will focus on the NLSM description.  

The NLSM includes the universal `center of mass' sector, which is described by the Green-Schwarz (GS) formulation of \cite{Lanza:2019xxg}. We can use this formulation to make it  clear why these   strings locally preserve ${\cal N} =(0,2)$ supersymmetry. Besides the world-sheet embedding coordinates, the GS-string \cite{Lanza:2019xxg} supports a GS fermion $\Theta^\alpha$, transforming as a four-dimensional chiral spinor. The preserved supersymmetry is determined by the kappa-symmetry $\delta\Theta_\alpha=\kappa_\alpha$ with
$\kappa_\alpha=\Gamma_\alpha{}^\beta\kappa_\beta$. Here we are using the index notation of \cite{Wess:1992cp} and $\Gamma_\alpha{}^\beta$ is the kappa symmetry operator, which can be identified with a two-dimensional chiral operator.  In a locally adapted reference frame in which the string is locally stretched along the $(x^0,x^3)$-directions, we have that $\Gamma_\alpha{}^\beta=(\sigma_3)_\alpha{}^\beta$. As in \cite{Witten:1993yc}, we can relabel the bulk spinor components as two-dimensional chiral components, e.g.\     $\Theta_\alpha=(\Theta_-,\Theta_+)$ and $\Theta^\alpha=(\Theta^-,\Theta^+)=(\Theta_+,-\Theta_-)$. The projection condition on  $\kappa_\alpha=(\kappa_-,\kappa_+)$ imposes that $\kappa_+=0$ and the kappa-symmetry then implies that $\rho_+\equiv \Theta_+$ is the physical right-moving component  of $\Theta_\alpha$, while $\Theta_-$ is pure gauge and can be set to zero.   
We can also use local static gauge and combine the transversal embedding coordinates into $u=x^1+\ii x^2$. Then $u$ and $\rho_+$ can be reorganized  into the `universal'  ${\cal N} = (0,2)$ two-dimensional chiral superfield  
\be\label{universal}
U=u+\sqrt{2}\theta^+\rho_+-2\ii\theta^+\bar\theta^+\del_{++}u\,.
\ee
We follow the conventions of \cite{Witten:1993yc,Adams:2003zy} -- see also \cite{Melnikov:2019tpl} for an extensive introduction to ${\cal N}= (0,2)$ two-dimensional models.
The components $u$ and $\rho_+$ have charges $1$ and $\frac12$, respectively, under a   $U(1)_{\rm N}$ rotation. 
We can then interpret  $U(1)_{\rm N}$ as an R-symmetry  which acts also  on $\theta^+$ with charge $J_{\rm N}[\theta^+]=\frac12$, so that $J_{\rm N}[U]=1$.

The  ${\cal N}= (0,2)$ NLSM \cite{Hull:1985jv,Dine:1986by} supported by an EFT string  generically includes also an `internal' sector. In particular, it can include $n_{\rm C}$ additional `non-universal' scalar chiral multiplets 
\be\label{chiralexp}
\Phi=\varphi+\sqrt{2}\theta^+\chi_+-2\ii\theta^+\bar\theta^+\del_{++}\varphi \,,
\ee 
where $\varphi$ is a complex scalar and $\chi_+$ is a right-moving fermion. (For simplicity, in \eqref{chiralexp}  we are suppressing indices running from 1 to $n_{\rm C}$, since we will not need them.)
The bosonic components $\varphi$  parametrize the `internal' NLSM target space $\calm_{\text{\tiny NLSM}}$ (while \eqref{universal} has the `external' four-dimensional spacetime as its target space), and the \ fermions $\chi_+$ take values in the corresponding tangent bundle $T\calm_{\text{\tiny NLSM}}$. Representing internal degrees of freedom, the  chiral superfields $\Phi$ are neutral  under the normal  $U(1)_{\rm N}$ rotational symmetry: $J_{\rm N}[\Phi]=0$. Since $J_{\rm N}[\theta^+]=\frac12$, the right-moving fermions $\chi_+$ therefore have  $U(1)_{\rm N}$ charge $J_{\rm N}[\chi_+]=-\frac12$.

Due to the minimal amount of supersymmetry, some of the directions in the NLSM moduli space may be obstructed at higher order.\footnote{In the presence of such obstructions, it is clear that only (some of) the unobstructed directions may admit a gauged axionic shift symmetry, see again Section \ref{sec_Gaugeanomalies} for details.}
The  obstructed directions should  be specified by the vanishing of $n_{\rm N}$ holomorphic superpotentials $J_a(\phi)$ in the ${\cal N}=(0,2)$ theory. In order to implement these constraints, we must include a corresponding number  $n_{\rm N}$ of  neutral Fermi multiplets  $\Lambda^a_-=\lambda^a_-+\ldots$, and add the superpotential term 
 \be\label{Lambdasub} 
\int\d\theta^+ \Lambda^a_- J_a(\Phi)+\text{c.c.}\,.
\ee
Here $\lambda^a_-$, the lowest component of the Fermi superfield $\Lambda^a_-$, is a left-moving chiral fermion.
For such a superpotential to exist, the Fermi multiplets $\Lambda^a_-$ must carry  $U(1)_{\rm N}$ charge $J_{\rm N}[\Lambda^a_-] =\frac12$.  Furthermore we require $\Lambda^a_-$ and $J_a(\Phi)$ to be uncharged under the spacetime gauge group. This requirement is motivated by our experience with explicit string theory models -- see in particular Section \ref{sec:Ftheory} -- and comes from the microscopic Green-Schwarz origin of the fermions $\lambda^a_-$.  Indeed, EFT strings typically uplift to `movable' brane configurations supporting  corresponding Green-Schwarz fermions, which are neutral under the bulk gauge symmetries.\footnote{In the Green-Schwarz formulation, branes are described by their embedding in the bulk superspace, whose odd coordinates correspond to the brane fermions.} 

A priori, the number of $U(1)_{\rm N}$ charged Fermi multiplets $\Lambda^a_-$ and the number of unobstructed moduli of the NLSM might be uncorrelated. However, we propose that the NLSM should obey a certain minimality principle:
The number of Fermi multiplets $\Lambda^a_-$, $n_{\rm N}$, should be given by the minimal number needed in order to account for the potential higher order obstructions in the target space of the NLSM.
In other words,  at the generic point of the moduli space  the number of unobstructed directions  should correspond to  
\be \label{nCeff}
n_{\rm C}^{\rm eff} := n_{\rm C} - n_{\rm N} \,.
\ee
In the sequel, we will assume this principle to hold at least for generators of the cone ${\cal C}_{\rm S}^{\text{\tiny EFT}}$ of EFT strings. The intuition between this non-trivial assumption is that the $U(1)_{\rm N}$ charged Fermi multiplets $\Lambda^a_-$ pair up with  obstructed gauge-neutral chiral multiplets to form the analogue of a `vector-like pair', i.e. an ${\cal N} =(2,2)$ chiral multiplet, leaving behind a collection of $n_{\rm C} - n_{\rm N}$ genuinely chiral ${\cal N} =(0,2)$ supermultiplets associated with the unobstructed directions in the moduli space of the NLSM. The minimality principle then states that there are no extra unpaired $U(1)_{\rm N}$ Fermi multiplets left. 

In particular, the minimality principle implies that $n_{\rm C}-n_{\rm N}\geq 0$, at least for genuinely $(0,2)$ EFT strings.\footnote{\label{foot:bound}Actually, we generically expect that $n_{\rm C}-n_{\rm N}\geq 1$, since generic EFT strings are associated with `movable' internal configurations, whose deformations should be represented by unobstructed chiral multiplets. The extreme case $n_{\rm C}-n_{\rm N}=0$ should instead be associated with the non-generic purely stringy rigid $w=1$ EFT string flows mentioned above.} 
On the other hand, in some cases the EFT string may support an enhanced UV ${\cal N}=(2,2)$ or higher non-chiral supersymmetry. For this to be possible, the Fermi  multiplets $\Lambda^a_-$ necessarily pair with {\it all} the $n_{\rm C}+1$ chiral multiplets, including the universal one \eqref{universal}. As a result, a minimal non-chiral ${\cal N}=(2,2)$ symmetry requires that  $n_{\rm C}-n_{\rm N}=-1$ and $n_{\rm F}=0$. Hence, in general 
\be\label{nCNbound}
 n_{\rm C}-n_{\rm N}\geq - 1 \,,
\ee
where the bound is saturated in presence of enhanced non-chiral world-sheet supersymmetry, while $n_{\rm C}-n_{\rm N}\geq 0$ for strictly chiral world-sheet supersymmetry.

The subset of the bosonic fields $\varphi$ corresponding to the unobstructed moduli of the NLSM
can enjoy an axionic shift symmetry which can be gauged by the bulk gauge fields. We will describe this effect in Section \ref{sec_Gaugeanomalies}.
In principle, some of the chiral multiplets might also carry a linear realization of the bulk gauge algebra.

Finally, the internal world-sheet sector can include also $n_{\rm F}$ chiral  Fermi multiplets $\Psi_-=\psi_-+\ldots $
which are neutral under $U(1)_{\rm N}$,  taking values in some vector bundle defined on the NLSM target space \cite{Hull:1985jv,Dine:1986by,Melnikov:2019tpl}.  
This type of Fermi multiplets can be charged under the four-dimensional gauge symmetry and couple to the corresponding gauge vectors.

To sum up, we will henceforth assume that an EFT string supports a weakly-coupled ${\cal N}=(0,2)$ NLSM, described by $n_{\rm C}+1$ chiral multiplets, $n_{\rm N}$ $U(1)_{\rm N}$ charged Fermi multiplets as well as  $n_{\rm F}$ $U(1)_{\rm N}$ neutral Fermi multiplets.  The relevant possible charges of the corresponding  world-sheet fermions are summarized in the following table:

\begin{equation}
	\label{tab:fcharges}
	\begin{tabular}{|c|c|c|c|c|c|}
		\hline
		 Fermion & \#  & $U(1)_{\rm N}$ charge & $U(1)_A$ charge & $G_I$ repr. & (0,2) multiplet
		\\
		\hline
	 $\rho_+$ & 1 & $\frac12$ & 0 & ${\bf 1}$ & chiral $U$  
		\\
	$\chi_+$ & $n_{\rm  C}$ & $-\frac12\ \ $ & $*$ & $*$ & chiral $\Phi$  
		\\
	$\psi_-$ & $ n_{\rm  F}$ & $0$ & $q_A$ & ${\bf r}_I$ & Fermi  $\Psi_-$
		\\ $\lambda_-$ & $ n_{\rm  N}$ & $\frac12$ & 0 & ${\bf 1}$ & Fermi $\Lambda_-$
		\\
		\hline
	\end{tabular}
\end{equation}

\section{Anomaly matching and quantum gravity constraints}
\label{sec:QGbounds}

We are now in a position to derive the main results of this work, the quantum gravity bounds \eqref{tildehatC0} and \eqref{rankbound}. 
The strategy is to use that the  't Hooft anomaly associated with the world-sheet theory supported by the EFT string must match the expression \eqref{polyn2}, which was obtained by requiring the cancellation of the anomaly inflow contribution from the four-dimensional ${\cal N}=1$ supersymmetric theory.

\subsection{Gravitational/$U(1)_{\rm N}$ anomalies and curvature-squared bounds}

Let us start with the gravitational and $U(1)_{\rm N}$ anomalies. From the fermionic charges \eqref{tab:fcharges} the corresponding anomaly polynomial is  \cite{Alvarez-Gaume:1983ihn}
\be\label{wsAn0}
\begin{aligned}
I^{\rm ws}_{4\,{\bf e}}|_{\text{grav+$U(1)_{\rm N}$}}= -\frac{n_{\rm F}-n_{\rm C}+n_{\rm N}-1}{192\pi^2}\tr (R_W\wedge R_W)+\frac{n_{\rm C}-n_{\rm N}+1}{32\pi^2} F_{\rm N}\wedge F_{\rm N}\,.
\end{aligned}
\ee
Our conventions for the computation of the anomaly polynomial are summarised in Appendix \ref{app_Anomalies}.

 The  't Hooft anomaly \eqref{wsAn0} must match the third line of \eqref{polyn2again}. We then arrive at the identifications
\begin{subequations}\label{CCnn}
\begin{align}
&\langle \tilde {\bf C},{\bf e}\rangle=n_{\rm F}-n_{\rm C}+n_{\rm N}-1\,,\label{CCnn1}\\ &\langle \tilde {\bf C},{\bf e}\rangle+\langle \hat {\bf C}({\bf e}),{\bf e}\rangle=3(n_{\rm C}-n_{\rm N}+1)\,.\label{CCnn2}
\end{align}
\end{subequations}
Note that, according to the above characterization of the world-sheet spectrum, an enhanced non-chiral world-sheet supersymmetry  precisely corresponds to the separate vanishing of   $\langle \tilde {\bf C},{\bf e}\rangle$ and $\langle \hat {\bf C}({\bf e}),{\bf e}\rangle$.

Since by the assumed EFT string completeness  the EFT string charges should generate the entire lattice of string charges, \eqref{CCnn1} implies that 
\be\label{tildeCqc}
\boxedB{\langle\tilde {\bf C},{\bf e}\rangle\in\mathbb{Z}\, \quad \forall {\bf e}\in \calc^{\text{\tiny EFT}}_{\rm S}\,,}
\ee
which refines the EFT quantization condition \eqref{tildeCquant}.  Furthermore, by combining   the matching condition \eqref{CCnn2} with \eqref{nCNbound} we deduce  the bound
\be\label{tildehatC0}
\boxedB{\langle \tilde {\bf C},{\bf e}\rangle+\langle \hat {\bf C}({\bf e}),{\bf e}\rangle\in 3\mathbb{Z}_{\geq 0}\,\quad~~~~\forall {\bf e}\in \calc^{\text{\tiny EFT}}_{\rm S}\,,}
\ee
which should be saturated  for enhanced non-chiral world-sheet supersymmetry. The equations \eqref{CCnn} also imply that
\be\label{hatnF}
4\langle \tilde {\bf C},{\bf e}\rangle+\langle \hat {\bf C}({\bf e}),{\bf e}\rangle =3 n_{\rm F} \,,
\ee
from which, since $ n_{\rm F}\geq 0$, we can then extract another similar bound:
\be\label{tildehatC}
\boxedB{4\langle \tilde {\bf C},{\bf e}\rangle+\langle \hat {\bf C}({\bf e}),{\bf e}\rangle\in 3\mathbb{Z}_{\geq 0}\,\quad~~~~\forall {\bf e}\in \calc^{\text{\tiny EFT}}_{\rm S}\,.}
\ee
We observe  that \eqref{tildehatC} follows from \eqref{tildehatC0} if $\langle \tilde {\bf C},{\bf e}\rangle\geq 0$, and vice versa   if $\langle \tilde {\bf C},{\bf e}\rangle\leq 0$.

If $\hat {\bf C}({\bf e})$ is non-trivial, in principle \eqref{tildehatC0}  and \eqref{tildehatC} allow for negative $\langle \tilde{\bf C},{\bf e}\rangle$, for some ${\bf e}\in\calc_{\rm S}^{\text{\tiny EFT}}$. On the other hand, in the explicit string theory models that we consider in this paper, this does not occur. It is then tempting to promote this observation to a general property of quantum gravity models and, in this sense, we can regard \eqref{tildehatC0} as the strongest bound.
It would certainly be  more satisfactory to derive this additional condition from a self-contained quantum gravity argument, but we leave this interesting question to the future.

\subsection{Gauge anomalies and rank bound} \label{sec_Gaugeanomalies}

We now turn to the world-sheet 't Hooft gauge anomalies, whose polynomial should match the second line of \eqref{polyn2again}.  The positive semi-definite coefficients $\langle  {\bf C}^{AB},{\bf e}\rangle$ and  $\langle  {\bf C}^{I},{\bf e}\rangle$ appearing in \eqref{polyn2again} -- see \eqref{pos1} -- identify  the gauge sector that `interact' with the EFT string, in the sense that the corresponding (s)axionic couplings in \eqref{kingauge2} change along the EFT string flow \eqref{sflow}. In the sequel we will focus on the rank of this gauge sector, which is identified as  
\be\label{rankG}
r({\bf e})\equiv {\rm rank}\{\langle  {\bf C}^{AB},{\bf e}\rangle\}+\sum_{I|\langle  {\bf C}^{I},{\bf e}\rangle>0} {\rm rk}(\mathfrak{g}_I) \,,
\ee
as one can easily realize by going to a basis of $U(1)$ gauge fields in which the symmetric matrix $\langle  {\bf C}^{AB},{\bf e}\rangle$ is diagonalized.\footnote{More explicitly, one can always diagonalize the matrix $\langle  {\bf C}^{AB},{\bf e}\rangle$ by means of an orthogonal matrix $O_{A}{}^B$. Then $\hat A_{A}\equiv O_{A}{}^BA_B$ have diagonal (s)axionic contributions to the kinetic terms couplings, with ${\rm rank}\{\langle  {\bf C}^{AB},{\bf e}\rangle\}$ non-vanishing positive diagonal entries.} 
In particular, we aim at deriving an upper bound on $r({\bf e})$ from the anomaly matching.  

First of all, the Fermi multiplets in  \eqref{tab:fcharges} yield the following  contribution to the anomaly polynomial $I^{\rm ws}_{4\,{\bf e}}|_{\rm gauge}$: 
\be\label{wsAn}
\begin{aligned}
\,-\frac{1}{8\pi^2} k^{AB}_{\rm F}({\bf e}) F_A\wedge F_B-\frac1{16\pi^2} k_{\rm F}^{I}({\bf e}){\rm tr}(F\wedge F)_I\,,
\end{aligned}
\ee
with 
\be\label{kabdef}
 k_{\rm F}^{AB}({\bf e})\equiv \sum_{{\bf q}\in\text{Fermi}}q^Aq^B\quad,\quad  k_{\rm F}^{I}({\bf e}) \equiv \sum_{{\bf r}^I\in\text{Fermi}}\ell({\bf r}^I)  \,.
\ee
The anomaly matching condition implies that  these coefficients provide  positive semi-definite  contributions to the total anomaly coefficients $\langle  {\bf C}^{AB},{\bf e}\rangle$ and $\langle  {\bf C}^{I},{\bf e}\rangle$ appearing in \eqref{polyn2again}. In particular
\be\label{r(e)F}
r_{\rm F}({\bf e})\equiv {\rm rank}\{k_{\rm F}^{AB}({\bf e})\}+\sum_{I|k_{\rm F}^I({\bf e})>0} {\rm rk}(\mathfrak{g}_I) 
\ee
can be regarded as the contribution of the Fermi multiplets to $r({\bf e})$.

 Since we are interested in the rank of the gauge algebra, we can actually focus on the Cartan  sub-algebra $\frak{h}_I\subset \frak{g}_I$ of each simple gauge factor. On each $\frak{h}_I$ we can pick  a basis $H^I_{\alpha}$, where $\alpha=1,\ldots,{\rm rk}(\mathfrak{g}_I)$, normalised so that $\tr(H^I_\alpha H^I_\beta)=2\delta_{\alpha\beta}$. If we turn off all the non-Cartan field strength components in \eqref{wsAn},
the remaining purely $U(1)$  anomaly polynomial is
\be\label{Fabelanom}
-\frac1{8\pi^2}\,k_{\rm F}^{\cala\calb}({\bf e})\,F_{\cala}\wedge F_{\calb} \,,
\ee
where $F_{\cala}=(F_A,F_{I\alpha})$ collectively represent  the remaining $U(1)$  field strengths, and the symmetric matrix $k_{\rm F}^{\cala\calb}$ has block-diagonal entries given by $k_{\rm F}^{AB}$ and $k_{\rm F}^I({\bf e})\delta_{I\alpha,I\beta}$.
The rank of the matrix $k_{\rm F}^{\cala\calb}$ can be identified with $r_{\rm F}({\bf e})$ as defined in \eqref{r(e)F}. On the other hand we can still apply the first formula in \eqref{kabdef} to this extended abelian sector and write
\be\label{kFAB}
k_{\rm F}^{\cala\calb}=\sum_{{\bf q}\in\text{Fermi}}q^\cala q^\calb  \,.
\ee
Hence, the matrix $k_{\rm F}^{\cala\calb}$ is the sum of $n_{\rm F}$ matrices of the form $q^\cala q^\calb$, which have either rank 0 or 1, if either all the $q^{\cala}$ charges are  vanishing or not, respectively. We can now use the general property 
\be\label{rankprop}
{\rm rank}(M_1+M_2)\leq {\rm rank}(M_1)+{\rm rank}(M_2)\,,
\ee
which holds for any pair of matrices $M_1,M_2$. Repeatedly  applied to \eqref{kFAB} and combined with \eqref{hatnF}, it leads to 
\be\label{ranknF}
r_{\rm F}({\bf e})\leq n_{\rm F}=\frac43\langle \tilde {\bf C},{\bf e}\rangle+\frac13\langle \hat {\bf C}({\bf e}),{\bf e}\rangle \,.
\ee

Consider next the anomaly contribution associated with the chiral multiplets $\Phi$. 
If some of these multiplets carry a linear realisation of the gauge group, the charged chiral fermions $\chi_+$ yield a negative definite contribution to the anomaly polynomial. One can convince oneself that this cannot increase the bound on the total rank of the gauge algebra.
We therefore turn to the remaining possibility that some of the chiral fields $\varphi$ corresponding to the unobstructed moduli 
enjoy an axionic shift symmetry which is gauged by the four-dimensional gauge symmetry. 
 According to the minimality principle of the previous section, the number of unobstructed directions is $n_{\rm C}^{\rm eff} := n_{\rm C} - n_{\rm N}$.
One can then introduce a set of `axionic' ${\cal N}= (0,2)$ chiral multiplets $\tau_r\simeq \tau_r+1$, with $r=1,\ldots,n_{\rm A}\leq 
n_{\rm C}^{\rm eff}$, which transform as 
\be\label{phishift}
\tau_r\rightarrow \tau_r+\frac1{2\pi}N_r{}^{\cala}\lambda_{\cala} 
\ee
under the bulk $U(1)$ gauge transformations $A_{\cala}\rightarrow A_{\cala}+\d\lambda_{\cala}$. 
Note that the fermions in the chiral multiplets $\tau_r$  do not transform under \eqref{phishift} and then do not contribute to the quantum anomaly.  On the other hand, following \cite{Blaszczyk:2011ib,Quigley:2011pv} one can write down the following  Green-Schwarz-like contributions to world-sheet effective action, 
\be\label{wsGS}
-M^{\cala r}\int_W \Re\tau_r \, F_\cala-\frac1{8\pi} Q^{AB}\int_W A_\cala\wedge A_\calb \,,
\ee
with
\be
Q^{\cala\calb}=-Q^{\calb\cala}\equiv (MN)^{\cala\calb}-(MN)^{\calb\cala}\,,
\ee 
where $(MN)^{\cala\calb}\equiv M^{\cala r}N_r{}^\calb$.
Note that the term \eqref{wsGS} admits a supersymmetric extension in terms of ${\cal N}=(0,2)$ world-sheet vector multiplets \cite{Blaszczyk:2011ib,Quigley:2011pv}, while it cannot be added to an ${\cal N}=(2,2)$ theory \cite{Adams:2012sh}.

 From \eqref{phishift} we see that the variation of \eqref{wsGS} under a gauge transformation $A_{\cala}\rightarrow A_{\cala}+\d\lambda_{\cala}$ gives  \be 
 -\frac{1}{4\pi}k^{\cala\calb}_{\rm C}({\bf e})\int_W \lambda_\cala\,F_\calb\,,
\ee
where we have introduced the symmetric matrix
\be\label{kCAB}
k^{\cala\calb}_{\rm C}({\bf e})\equiv (MN)^{\cala\calb}+(MN)^{\calb\cala}\,.
\ee
This classical violation of the gauge symmetry has the form of an anomaly associated with the polynomial 
\be\label{Cabelanom}
-\frac1{8\pi^2}\,k_{\rm C}^{\cala\calb}({\bf e})\,F_{\cala}\wedge F_{\calb}\,.
\ee
In particular, we can define a corresponding rank
\be 
r_{\rm C}({\bf e})\equiv {\rm rank}\{k_{\rm C}^{\cala\calb}({\bf e})\}\,.
\ee
Note that the rank  of the matrix $(MN)^{\cala\calb}$, as well as its transposed,  cannot exceed $ n_{\rm A}\leq n^{\rm eff}_{\rm C} =n_{\rm C} - n_{\rm N}$.  Hence applying the property \eqref{rankprop} to \eqref{kCAB} and the identity \eqref{CCnn2}, we obtain the upper bound
\be\label{rCbound}
r_{\rm C}({\bf e})\leq 2n_{\rm C}^{\rm eff} 
 =\frac{2}{3}\langle \tilde {\bf C},{\bf e}\rangle+\frac{2}{3}\langle \hat {\bf C}({\bf e}),{\bf e}\rangle -2\,.
\ee

By combining \eqref{Fabelanom} and \eqref{Cabelanom} we arrive at the  maximally abelian anomaly polynomial
\be\label{Tabelanom}
-\frac1{8\pi^2}\,k^{\cala\calb}({\bf e})\,F_{\cala}\wedge F_{\calb}\,
\ee
with 
\be
k^{\cala\calb}({\bf e})\equiv k^{\cala\calb}_{\rm F}({\bf e})+k^{\cala\calb}_{\rm C}({\bf e})\,.
\ee
By anomaly matching, this matrix must be
positive semi-definite and
its rank provides an upper bound for \eqref{rankG}.\footnote{Possible charged chiral multiplets would provide negative semi-definite contributions, which can only lower this upper bound.}  Hence, by recalling \eqref{ranknF} and \eqref{rCbound} and applying again \eqref{rankprop}, we can conclude that
\be\label{totb0}
r({\bf e})\leq r_{\rm F}({\bf e})+r_{\rm C}({\bf e})\leq n_{\rm F}+2n_{\rm C}^{\rm eff} \,.
\ee
In view of \eqref{ranknF}  and \eqref{rCbound}, we can rewrite this in its final form 
\be\label{rankbound}
\boxedB{r({\bf e})\leq r({\bf e})_{\rm max} \equiv 2 \langle \tilde {\bf C},{\bf e}\rangle+\langle \hat {\bf C}({\bf e}),{\bf e}\rangle-2\,\quad~~~~\forall {\bf e}\in \calc^{\text{\tiny EFT}}_{\rm S}\,.}
\ee

The bounds \eqref{tildehatC0},  \eqref{tildehatC} and \eqref{rankbound} are the main results of this paper. In particular, \eqref{rankbound} encodes a correlation between the ranks of the EFT gauge group  and higher curvature corrections which is completely unexpected from a purely low-energy EFT viewpoint. We will illustrate the power of these bounds in Section \ref{sec:examples}. We stress that the bounds only apply to gauge sectors exhibiting a coupling of the form \eqref{kingauge2} and \eqref{aRR}, from which they were derived.
Furthermore, the bound \eqref{rankbound} (but not \eqref{tildehatC0} and \eqref{tildehatC}) relies on our assumption that the number of unobstructed moduli of the NLSM is given by \eqref{nCeff}.

We expect the bound \eqref{rankbound} to be rather conservative because in many situations only a subset $n_{\rm A}$ of the $n^{\rm eff}_{\rm C}$ unobstructed chiral multiplets enjoy gauged axionic shift symmetries and hence participate in the anomaly matching. 
This will be elucidated further in concrete F-theoretic realisations in Section \ref{sec_MicrRealF}. 
Based on our explicit knowledge of the nature of the chiral multiplets in the NLSM of the EFT strings in F-theory, we will in fact propose the stronger bound \eqref{strictboundF} on the rank of the four-dimensional gauge group, which is expected to hold in F-theory compactifications on smooth geometric backgrounds not admitting extra isometries.
In other cases, however, such as toroidal heterotic orbifolds commented on after \eqref{hetb1},
the bound \eqref{rankbound} can in fact be saturated. Hence, even though in many instances the actual gauge group is of considerably smaller rank than the bound in \eqref{rankbound}, we have to content ourselves with this constraint in a general setting.

\subsection{Some comments}
\label{sec:hatCComments}

So far we have been agnostic about the contribution $\hat C({\bf e})$ to the anomaly inflow and the resulting quantum gravity bounds. In order to extract more precise information from \eqref{tildehatC0}, \eqref{tildehatC} and \eqref{rankbound}, we now analyse the  possibilities identified in Section \ref{sec:hatC} in more detail, that is, either vanishing $\hat C({\bf e})$ or a contribution of the form  \eqref{cubicC}. Let us first assume that $\hat C({\bf e})\equiv 0$. In this case \eqref{tildehatC0} and \eqref{tildehatC} reduce to 
\be\label{tildeCpos}
\langle \tilde {\bf C},{\bf e}\rangle\in 3\mathbb{Z}_{\geq 0}\,\quad~~~~\forall \, {\bf e}\in \calc^{\text{\tiny EFT}}_{\rm S}\,.
\ee
This shows that not only are the constants $\tilde C_i$ integral, rather than half-integral as in \eqref{tildeCquant}, but actually $\langle \tilde {\bf C},{\bf e}\rangle$ should be a non-negative multiple of 3. Hence either $\tilde {\bf C}$ is vanishing or it lies in  the cone $\calc_{\rm I}$. Recalling the definition \eqref{scone} of the saxionic cone,  we therefore see that \eqref{tildeCpos} implies $\langle \tilde {\bf C}, {\bf s}\rangle \geq 0$ (with equality only for $\tilde  {\bf C} =0$), which then follows from the quantum consistency of EFT strings, at least if $\hat C({\bf e})\equiv 0$. Note that the inequality \eqref{tildeCpos} should be saturated precisely by the EFT strings with enhanced non-chiral supersymmetry, while for strict chiral world-sheet  supersymmetry we should actually have  $\langle \tilde {\bf C},{\bf e}\rangle\geq 3$.\footnote{If one considers models admitting a geometric higher dimensional UV-completion, by recalling \eqref{CCnn2} and footnote \ref{foot:bound} this bound  could be further restricted  to $\langle \tilde {\bf C},{\bf e}\rangle\geq 6$.} 

Consider next an anomaly contribution of the form \eqref{cubicC}. This implies that 
\be 
\langle \hat {\bf C}({\bf e}),{\bf e}\rangle=\hat{\bf C}({\bf e},{\bf e},{\bf e})\equiv \hat C_{ijk}\,e^ie^je^k \,,
\ee
where $\hat C_{ijk}\in\mathbb{Z}$ are associated with some underlying five-dimensional $\caln=1$ structure which can be detected by the EFT string. 
Since the EFT string charges form the cone $\calc^{\text{\tiny EFT}}_{\rm S}$, if the condition  \eqref{tildehatC0} holds for a given  string charge vector ${\bf e}\in\calc^{\text{\tiny EFT}}_{\rm S}$, then it should also hold for ${\bf e}'=N{\bf e}$, where $N>0$ is any positive integer. If $\hat{\bf C}({\bf e},{\bf e},{\bf e})\neq 0$, by imposing the condition \eqref{tildehatC0} for ${\bf e}'=N{\bf e}$ with $N\rightarrow \infty$, we obtain $N^3\left[\hat{\bf C}({\bf e},{\bf e},{\bf e})+\calo(1/N^2)\right]\geq 0$ and then 
\be\label{hatCpos}
\hat C({\bf e},{\bf e},{\bf e}) \geq 0\,.
\ee 
We furthermore observe that the condition \eqref{hatCpos}  follows also from the five-dimensional  arguments of \cite{Katz:2020ewz}, and should then hold for our setting as well.

Note that \eqref{cubicC} and \eqref{hatCpos} also imply that $\langle \hat {\bf C}({\bf e}),{\bf e}\rangle\leq \langle \hat {\bf C}(N{\bf e}),N{\bf e}\rangle $ if $N\geq 1$.  
Since the rank of the gauge group interacting with an EFT string does not change if we rescale the string charge vector, that is 
$r(N{\bf e})=r({\bf e})$, we deduce that the strongest  bounds \eqref{rankbound} on the gauge group rank are obtained by using primitive EFT charge vectors. Of course this is also true if  $\hat {\bf C}({\bf e})\equiv 0$. Furthermore, applying \eqref{rankprop} to the matrix $\langle{\bf C}^{AB},{\bf e}_1+{\bf e}_2\rangle$, we obtain
\be
\begin{aligned}
r({\bf e}_1+{\bf e}_2)&\leq r({\bf e}_1)+r({\bf e}_2)\\
&\leq 2 \langle \tilde {\bf C},{\bf e}_1+{\bf e}_2\rangle+\langle \hat {\bf C}({\bf e}_1),{\bf e}_1\rangle+\langle \hat {\bf C}({\bf e}_1),{\bf e}_2\rangle -4\\
& \leq 2   \langle \tilde {\bf C},{\bf e}_1+{\bf e}_2\rangle+\langle \hat {\bf C}({\bf e}_1),{\bf e}_1\rangle+\langle \hat {\bf C}({\bf e}_1),{\bf e}_2\rangle  -2 \,.
\end{aligned}
\ee
Now, a convexity property of the form
\be\label{hatCconv}
\langle \hat {\bf C}({\bf e}_1),{\bf e}_1\rangle+\langle \hat {\bf C}({\bf e}_2),{\bf e}_2\rangle\leq \langle \hat {\bf C}({\bf e}_1+{\bf e}_2),{\bf e}_1+{\bf e}_2\rangle\,
\ee
would guarantee that, if ${\bf e}_1$ and ${\bf e}_2$ satisfy \eqref{rankbound} then   ${\bf e}_1+{\bf e}_2$ satisfies \eqref{rankbound} too. Hence, in this case the strongest constraints would be obtained by considering the generators of the cone $\calc^{\text{\tiny EFT}}_{\rm S}$ of EFT string charges, that is, the {\em elementary} EFT string charges. This property has a clear physical interpretation. One can think of any EFT string as a (possibly threshold) bound state formed by  recombining  elementary EFT strings. According to this property, if these elementary strings are separately consistent, then their bound state should be consistent too. 

The condition \eqref{hatCconv} is trivially satisfied if $\hat {\bf C}({\bf e})\equiv 0$.
If instead  \eqref{cubicC} holds, \eqref{hatCconv} is for instance guaranteed if 
\be\label{postrip}
\hat C({\bf e}_1,{\bf e}_2,{\bf e}_3)\geq 0 \quad~~~~ \forall {\bf e}_1,{\bf e}_2,{\bf e}_3\in \calc^{\text{\tiny EFT}}_{\rm  S}\,.
\ee
In \cite{Katz:2020ewz} an analogous  property has been proposed to hold for general $\caln=1$ five-dimensional supergravities compatible with quantum gravity. This property would descend to  \eqref{postrip} in our four-dimensional EFTs.     
We will see  that \eqref{postrip}, and hence \eqref{hatCconv}, indeed holds in the explicit string models that we will consider.

Note also that these results trivialize in the case  of enhanced non-chiral world-sheet supersymmetry, since in this case $ \langle{\bf C}^{AB},{\bf e}\rangle=\langle  {\bf C}^{I},{\bf e}\rangle=\langle \tilde {\bf C},{\bf e}\rangle=\langle \hat{\bf C}({\bf e}),{\bf e}\rangle=0$.

Finally, according to the general formulation with three-form potentials of \cite{Lanza:2019xxg}, the two-form potentials $\calb_{2,i}$ could be gauged under some two-form gauge transformations. This effect makes some axionic strings `anomalous', forcing them  to be the boundary of membranes. It would be interesting to extend the above arguments to these kinds of strings.


\section{Simple examples}
\label{sec:examples}

In this section we illustrate the implications of the quantum gravity constraints found in Section \ref{sec:QGbounds} for simple EFT models.

Let us start with the most basic possible example:
The gauge group $G$ consists of a simple gauge group factor and the perturbative EFT regime is associated with a single chiral field $t=a+\ii s$, whose saxion $s$ parametrizes the one-dimensional saxionic cone $\Delta=\mathbb{R}_{>0}$, plus possible additional chiral `spectators' $\phi$. According to the general definition \eqref{CSEFT}, $\calc^{\text{\tiny EFT}}_{\rm S}=\mathbb{Z}_{\geq 0}$ then represents a single tower of EFT strings, labeled by the charge $e\in \mathbb{Z}_{\geq 0}$.

By supersymmetry, the relevant structure is specified by the saxionic couplings appearing in \eqref{kingauge} and \eqref{PGB}.
Let us assume that these take the form
\be\label{CsF}
-\frac{1}{8\pi}\int (Cs+\ldots)\,\tr(F\wedge *F)-\frac{1}{192\pi}\int  (\tilde Cs+\ldots) \, E_{\rm GB}*1\,,
\ee
where $E_{\rm GB}$ is the Gauss-Bonnet combination \eqref{GBcomb} and the ellipses represent possible additional constant and  $\phi$-dependent terms. 
As discussed in the previous sections, in general this is a non-trivial assumption about the EFT.

Furthermore we first suppose that the coupling \eqref{SN}   is vanishing. According to the discussion in Section \ref{sec:hatC}, this means that we are considering   models which do not have  a hidden five-dimensional structure. 
At the low-energy EFT level the constants  $C$ and $\tilde C$  should just satisfy the quantization conditions \eqref{CCconst} and \eqref{tildeCquant} and  could otherwise be chosen quite arbitrarily.    
On the other hand, the constraints derived in Section \ref{sec:QGbounds}  imply that this is not true for  EFTs admitting a  quantum gravity completion. First of all, by \eqref{tildeCqc} we must have $\tilde C\in\mathbb{Z}$, rather than just $2\tilde C\in\mathbb{Z}$. Furthermore, since \eqref{SN} is vanishing, \eqref{tildehatC0} and \eqref{tildehatC} reduce to \eqref{tildeCpos} and then
\be
\tilde C=3 k\geq 0 \quad  {\rm with} \quad   k\in\mathbb{Z}_{\geq 0}  \,.
\ee
We could actually be more precise: either $\tilde C=0$, which means that the EFT strings should support an enhanced  non-chiral supersymmetry, or  otherwise $\tilde C= 3\tilde k+3$ with $\tilde k\in\mathbb{Z}_{\geq 0}$.

Consider then the bound \eqref{rankbound}. 
In the present single saxion example, if $C>0$ we can simply set $r({\bf e})={\rm rk}(\mathfrak{g})$ and \eqref{rankbound} reduces to the bound  
\be
{\rm rk}(\mathfrak{g})\leq 2\tilde C-2=6\tilde k+4\,, \qquad \tilde k \in {\mathbb Z}_{\geq 0} 
\ee
on the rank of the Lie algebra associated with the gauge group $G$.  

Let us now allow for a non-trivial coupling \eqref{SN} which, in view of \eqref{cubicC}, takes the form
\be\label{ChA}
-\frac1{24}\hat C e^2\int_\Sigma h_1\wedge A_{\rm N}\,,
\ee
where $\hat C\in\mathbb{Z}$ by  \eqref{hatCquant}.  Now \eqref{tildehatC0} and \eqref{tildehatC} require that 
\be\label{4econ}
\tilde C e+\hat C e^3\in 3\mathbb{Z}_{\geq 0}\,,
\quad 4\tilde C e+\hat C e^3\in 3\mathbb{Z}_{\geq 0}\quad~~~~~~~~ \forall e\in \mathbb{Z}_{\geq 0}\,. \ee
The conditions  can be solved by setting 
\be\label{ccvalues}
\left\{\begin{array}{l}\tilde C=3k -\hat k\\ 
\hat C=\hat k\,,\quad \hat k=0,\ldots, 3k
 \end{array}\right.\,\text{(case I)}\quad~~~\text{or} \quad\left\{\begin{array}{l}\tilde C=-\tilde k\\ 
\hat C=3k+4\tilde k
 \end{array}\right.\,\text{(case II)},
\ee
for all $k,\tilde k\in\mathbb{Z}_{\geq 0}$. The bound \eqref{rankbound} reduces to ${\rm rk}(\mathfrak{g})\leq 2 \tilde C + \hat C -2$ and from \eqref{ccvalues} we then get 
\be
{\rm rk}(\mathfrak{g})\leq  \left\{\begin{array}{ll} 6 k - \hat k  - 2 \quad&\text{(case I)}\\ 
3 k +2\tilde k  - 2 \quad&\text{(case II)}\end{array}\right.\,.
\ee

The above analysis can be immediately  generalized to a non-simple gauge group, with saxionic couplings  
\be\label{CsF1}
-\frac{1}{4\pi}\int (C^{AB}s+\ldots)\,F_A\wedge *F_B-\frac{1}{8\pi}\int (C^Is+\ldots)\,\tr(F\wedge *F)_I\,,
\ee
where by \eqref{sCcond1}  we must impose that $C^I\in\mathbb{Z}_{\geq 0}$ and  $C^{AB}\in \mathbb{Z}$ is a positive semi-definite matrix. Now $r({\bf e})$ is the total rank of the linearly independent gauge fields which interact with the saxion $s$, and it is bounded as in the case of a simple gauge group discussed above.  

Note that the mere presence of some gauge sector coupling to the (s)axion as in \eqref{CsF1} implies that, necessarily, the higher derivative couplings appearing in \eqref{CsF} and \eqref{ChA} cannot be both trivial and the EFT string cannot exhibit enhanced ${\cal N} =(2,2)$ supersymmetry. Suppose for instance that, at the UV cut-off scale $\Lambda$,  our EFT describes an $G=SU(5)$. Then in case I we must necessarily have either $(k,\hat k)=(1,0)$ and then $(\tilde C,\hat C)=(3,0)$, or $k\geq 2$, with any $\hat k=0,\ldots,3k$. In case II we must instead impose $3k+2\tilde k\geq 6$, which for instance implies $\hat C\geq 6$. If we consider instead a $G=SO(10)$ supersymmetric GUT model, then in case I we must necessarily have either $(\tilde C,\hat C)=(6-\hat k,\hat k)$ with $\hat k=0,\ldots,5$ (and $k=2$), or the other values corresponding to $k\geq 3$. In case II one must impose $3k+2\tilde k\geq 7$, which also implies  that $\hat C \geq 7$. According to the proposal of Section \ref{sec:hatC}, the non-vanishing of (\ref{ChA}) should practically mean that the strict weak coupling limit of the gauge theory is an at least partial decompactification limit to five dimensions.

Consider next a model with two saxions $s^1,s^2$ with a general gauge group \eqref{gaugeG}. One can choose a basis in which $(s^1,s^2)$ parametrize the saxionic cone $\Delta=\mathbb{R}^2_{>0}$. In this basis, the EFT strings have charge vectors ${\bf e}=(e^1,e^2)\in\calc^{\text{\tiny{EFT}}}_{\rm S}=\mathbb{Z}^2_{\geq 0}$. By  \eqref{pos1} the components of ${\bf C}^{AB}=(C^{AB}_1,C^{AB}_2)$ are positive semi-definite integral matrices, and ${\bf C}^{I}=(C^{I}_1,C^{I}_2)\in \mathbb{Z}^2_{\geq 0}$. Imposing for simplicity that the coupling \eqref{SN} vanishes, \eqref{tildehatC0} and \eqref{tildehatC} require that $\tilde{\bf C}=(\tilde C_1,\tilde C_2)=(3 k_1,3 k_2)$, with $k_1,k_2\in\mathbb{Z}_{\geq 0}$. Hence the EFT contains the Gauss-Bonnet  term
\be
-\frac{1}{64\pi}\int  (k_1s^1+k_2 s^2+\ldots) \, E_{\rm GB}*1\,.
\ee
The values of $k_1,k_2$ constrain the ranks of the gauge groups coupling to the axions as in \eqref{kingauge2}.
The stricter bounds are obtained by setting ${\bf e}={\bf e}_1\equiv (1,0)$ and ${\bf e}={\bf e}_2\equiv (0,1)$ in \eqref{rankbound}, and read
\be
r({\bf e}_1)\leq 6k_1 -2 \quad,\quad r({\bf e}_2)\leq 6k_2 -2\,.
\ee
Here $r({\bf e}_1)$ and $r({\bf e}_2)$ can be identified with the rank of the gauge sector which interact with the saxions $s^1$ and $s^2$, respectively, through couplings of the form \eqref{CsF1}. If for instance all gauge group factors interact with both saxions, we get a bound on the total rank ${\rm rk}(\mathfrak{g})\leq {\rm min}\{6k_1-2,6k_2 -2 \}$. Otherwise, we can only say that ${\rm rk}(\mathfrak{g})\leq 6(k_1+k_2)-2$, which is what one gets by applying \eqref{rankbound} to ${\bf e}={\bf e}_1+{\bf e}_2$. In any event, a non-trivial gauge group implies that $k_1 + k_2$ cannot vanish, and hence $\tilde C \neq 0$ (where we recall that we have assumed for simplicity that $\hat C =0$).

We could systematically proceed by considering more complicated bottom-up EFTs. Instead, we turn to analyzing how our quantum gravity bounds are realised in large classes of 
$\caln=1$ compactifications of string or M-theory to four dimensions.


\section{Microscopic checks in IIB/F-theory models}
\label{sec:Ftheory}

We begin our string theoretic tests of the quantum gravity bounds with IIB/F-theory models.  Not only will this provide a microscopic  confirmation  of the general bounds derived so far, but we will also find a stronger version of these bounds which is valid for minimally supersymmetric F-theory compactifications on a {\it smooth} three-fold base.

\subsection{F-theory models}
\label{sec:Ftheorymodel}

Let us first analyse the large volume perturbative regime of F-theory compactifications to four dimensions. Our primary interest is in the part of the four-dimensional gauge theory sector that is supported by  7-branes, postponing a discussion of the D3-brane sector to Section \ref{sec:IIBO3}. The 7-brane gauge sector can be weakly coupled, for large volumes of the wrapped four-cycles, even if we relax the ten-dimensional weak string coupling condition. In this way one can also obtain more general gauge groups, including exceptional ones -- see e.g. \cite{Taylor:2011wt,Weigand:2018rez} for reviews.

We then consider a generic F-theory compactification defined by an elliptically fibered  Calabi-Yau four-fold $Y$
with three-fold base $X$. For the time being we assume that the bulk and 7-brane gauge fluxes vanish
and comment on the validity of this assumption in 
 Section \ref{sec_P3ex}.

The type IIB axio-dilaton profile is geometrized into the non-trivial elliptic fibration $\pi: Y\rightarrow X$, described by the Weierstrass model 
\be\label{Weierstrass0}
y^2=x^3+fx z^4 +gz^6 \,,
\ee
where $[x:y:z]\in\mathbb{P}^2_{(2,3,1)}$ and 
\begin{equation}\label{fgdelta}
    f \in \Gamma(\overline K_X^4) \,, \qquad g \in \Gamma( \overline K_X^6)\,, \qquad    \Delta\equiv 4f^3+27 g^2 \in \Gamma(\overline K_X^{12})  \,.
\end{equation}
In our notation a line bundle -- e.g.\ the anti-canonical bundle $\overline K_X$ -- and the corresponding divisor are denoted by the same symbol. The anti-canonical class is required to admit an effective representative in order for the Weierstrass model to exist.
The non-abelian gauge theory sectors  are supported on irreducible  effective components $\cald^I$ of the discriminant locus $\Delta\simeq 12 \overline K_X$. Hence, we can write
\begin{equation}
\label{discrdec}
\Delta= n_I \cald^I +\cald'\simeq 12 \overline K_X \quad~~~~\text{with\quad $n_I\equiv  {\rm ord}(\Delta)|_{\cald^I}$} \,,
\end{equation}
where $\cald'$ is an {\it effective} divisor not supporting any gauge sector.
According to Kodaira's classification \cite{Kodaira2,Kodaira3,Neron} summarized in Table \ref{Kodairatable},  the non-abelian gauge algebra $G_I$ along each component $\cald^I$ is determined by the vanishing orders ${\rm ord}(f,g,\Delta)$ on $\cald^I$.\footnote{The actual gauge algebra in four dimensions may be even smaller due to gauge flux and monodromies that would lead to non-simply laced algebras.}

Abelian gauge algebra factors, on the other hand, are generated by rational sections ${\mathbb S}^A$ of the elliptic fibration independent of the zero-section ${\mathbb S}^0$. As reviewed e.g.\ in \cite{Weigand:2018rez,Cvetic:2018bni}, given a rational section ${\mathbb S}^A$, one defines the Shioda map 
\be \label{Shioda}
\sigma({\mathbb S}^A) = {\mathbb S}^A - {\mathbb S}^0 - \pi^\ast(D({\mathbb S}^A)) + \ell^A_{I_\alpha} E^{I_\alpha} \,,
\ee
where the divisor class $D({\mathbb S}^A)$ on the base $X$ is chosen such that  $\sigma({\mathbb S}^A)$ is orthogonal to the push-forward of all curve classes on $X$ to the total space $Y$.\footnote{Furthermore, the exceptional divisors $E^{I_\alpha}$ are the blowup divisors associated with the resolution of the singularities induced in the Weierstrass model of $Y$ by the appearance of the non-abelian gauge algebra $G_I$.
The coefficients $\ell^A_{I_\alpha} \in \mathbb Q$ are determined by requiring that $\sigma({\mathbb S}^A)$ has vanishing intersection with the generic rational fiber of each $E^{I_\alpha}$.}
The $U(1)_A$ gauge potential is then obtained, in the dual M-theory, by expanding the M-theory three-form $C_3$ in terms of the two-forms dual to the classes $\sigma({\mathbb S}^A)$.

The rank of the non-abelian subsector of the gauge group is constrained geometrically by the so-called {\it Kodaira bound} analysed in \cite{Kumar:2010ru,Morrison:2011mb,Taylor:2011wt,Grimm:2012yq}, which study its consequences for the effective theory that can be obtained from F-theory.
From Table \ref{Kodairatable}, the rank $r_I={\rm rank}(G_I)$ of the non-abelian gauge group $G_I$ on the divisor ${\cal D}^I$ satisfies  the bound 
\begin{equation} \label{Kodairabound}
    r_I   <  n_I\equiv  {\rm ord}(\Delta)|_{\cald^I}  \,.
\end{equation}
This bound is an actual inequality which is never saturated.
It can be translated into a bound on the total rank of the non-abelian part of the gauge group as follows: Consider a curve $\Sigma$ on $X$ with the property that $\Sigma \cdot D_{\rm eff} \geq 1$ for all effective divisors $D_{\rm eff}$. Then 
the rank of the total non-abelian gauge sector is constrained by
\be \label{Kodaira-2}
{ {\rm rk}(G_{\rm non-ab}) \leq} \sum_I  {\rm rk}(G_I) (\Sigma \cdot {\cal D}^I) \leq  \sum_I n_I (\Sigma \cdot {\cal D}^I) + \Sigma \cdot D' =  { \Sigma \cdot \Delta} \,.  
 \ee
We will compare this geometrical Kodaira bound coming from the UV complete description of the model with the EFT string bounds.  
The rank of the abelian sector, on the other hand, is unconstrained by the Kodaira bound  \eqref{Kodaira-2}, while it also enters the EFT string constraints. The latter therefore contain valuable new information which cannot immediately be deduced from the Kodaira bound alone. This was already observed in F-theory compactifications to six dimensions in \cite{Lee:2019skh} and translated into a bound on the rank of the Mordell-Weil group on elliptic Calabi-Yau three-folds.

\begin{table}
\centering
{\footnotesize\begin{tabular}{ | c | c | c | c | c |}
  \hline 
    &${\rm ord}_{\cald}(f)$ & ${\rm ord}_{\cald}(g)$ &
     ${\rm ord}_{\cald}(\Delta)$ & {\rm singularity} \\
     \hline  
     \hline
      I$_0$ & $\geq 0$ & $\geq 0$ & 0 & none  \\
       \hline
      I$_n$, $n\geq 1$  & 0 & 0  & $n$ & $A_{n-1}$  \\
        \hline
      II  & 1 & 1 & $\geq 2$ & none  \\
      \hline
      III &1 & $\geq 2$  & 3  & $A_1$  \\
        \hline
      IV& $\geq 2$ & $2$ & 4  & $A_2$  \\
        \hline
      I$_0^*$ & $\geq 2$ & $\geq 3$&  6 & $D_4$  \\
       \hline
       I$^*_n$, $n\geq 1$  & $2$ & $3$ &$6+n$ & $D_{4+n}$  \\
       \hline
       IV$^*$ &  $\geq 3$ & $4$  &$8$ & $E_6$  \\
       \hline
       III$^*$  & $3$ & $\geq 5$ & $9$ & $E_7$  \\
        \hline
       II$^*$  &  $\geq 4$ & $5$ &$10$  & $E_8$  \\
    \hline  
\end{tabular}
\caption{Kodaira's classification of fiber degenerations.}\label{Kodairatable}}
\end{table}

To complete the description of the EFT data, note that the relevant saxions $s^a$ coupling to these gauge sectors can be defined as 
\be\label{Fsaxions}
s^a=\frac12\int_{D^a}J\wedge J \,,
\ee
where $J$ is the Einstein frame K\"ahler form on $X$ in string units $\ell_{\rm s}=1$ and $D^a$ is a basis of divisors. In Type IIB language, the corresponding axions are then given by $a^a=-\int_{D^a}C^{\text{\tiny RR}}_4$, where $C^{\text{\tiny RR}}_4$ is the R-R four-form potential.  
Let us also introduce a dual basis of two-cycles $\Sigma_a$, such that $D^a\cdot \Sigma_b=\delta_b^a$. In the weak string coupling limit, the coupling to the non-abelian gauge sector appearing in \eqref{kingauge} can be obtained by  expanding the D7-brane CS term, leading to
\be\label{FCalpha}
C^{I}_a=\cald^I\cdot \Sigma_a  \,.
\ee
This result is usually extrapolated to configurations involving regions on $X$ where the string coupling is non-perturbative.

Similarly the couplings $C_a^{AB}$ between the abelian field strengths and the (s)axions can be identified geometrically as the intersection product
\be
{\mathbf C}_a^{AB} = b^{AB} \cdot \Sigma_a \,,
\ee
where 
\be  \label{heightpairingbab}
b^{AB} =  - \pi_\ast(\sigma({\mathbb S}^A) \cdot  \sigma({\mathbb S}^B)) 
\ee
is known as the height-pairing associated with the rational sections underlying the definition of the abelian gauge group factors $U(1)_A$ and $U(1)_B$, see \eqref{Shioda}. In particular, the (effective) divisors $b^{AA}$ can be viewed as the analogue, for abelian gauge groups, of the divisor ${\cal D}^I$ wrapped by a stack of 7-branes supporting the non-abelian gauge group $G_I$.

It will be crucial for us that the anti-canonical divisor determines the  couplings \eqref{aRR} \cite{Grimm:2012yq}.
 By borrowing formula \cite[Eq.~(5.15)]{Lawrie:2016axq}, we obtain 
\be\label{tildeCFtheory}
\tilde C_a=6\int_{\Sigma_a}c_1(B)=6\, \Sigma_a\cdot \overline K_X \,.
\ee
The results of \cite{Grimm:2012yq,Lawrie:2016axq} confirm that this formula \eqref{tildeCFtheory} holds for general F-theory compactifications even if it was identified starting from the weak string coupling limit.

\subsection{EFT strings} \label{sec_EFTstringsF}

The BPS instanton cone $\calc_{\rm I}$ characterizing the above perturbative regime can be identified with the cone of effective divisors  $D_{\bf m}=m_a D^a\in {\rm Eff}^1(X)$. Hence, the associated  saxionic cone $\Delta$ is defined by the conditions  
\be\label{Fscone}
\langle {\bm s},{\bf m}\rangle=\frac12\int_{D_{\bf m}}J\wedge J>0\,.
\ee
The physical interpretation of this condition is clear: Any instanton charge vector  ${\bf m}\in \calc_{\rm I}$ corresponds to a Euclidean D3-brane wrapping an effective divisor $D_{\bf m}$ and \eqref{Fscone} just requires that its  volume is positive. 

As discussed in \cite{Lanza:2021qsu}, the corresponding EFT string charges  ${\bf e}\in\calc^{\text{\tiny EFT}}_{\rm S}$  can be identified with (effective) {\em movable} curves  $\Sigma_{\bf e}=e^i \Sigma_i\in {\rm Mov}_1(X)$. Indeed, 
movable curves are characterized by a non-negative intersection with effective divisors \cite{boucksom2013pseudo} and then $\langle {\bf m},{\bf e}\rangle= D_{\bf m}\cdot \Sigma_{\bf e}\geq 0$. 
The associated EFT strings are obtained from D3-branes wrapping these movable curves. 
A refined classification of the types of movable curves and their associated EFT string limits in F-theory has been obtained in \cite{Cota:2022yjw}.

 By definition, a movable curve can sweep out the entire space $X$. The bosonic fields of the associated  NLSM therefore have a clear geometric interpretation: they parametrize the moduli space $\calm_{\bf e}$ of  $\Sigma_{\bf e}$ inside $X$, including directions which are possibly obstructed at higher order. Its tangent space $T\calm_{\bf e}$ can be identified with the cohomology group $H^0(N\Sigma_{\bf e})$, where  $N\Sigma_{\bf e}$ is the $\Sigma_{\bf e}$ normal bundle. The metric on $T\calm_{\bf e}$ is inherited from the components of the metric on $X$ which are orthogonal to $\Sigma_{\bf e}$. 
These are precisely the directions which are stretched by the flow \eqref{sflow} generated by the EFT string, suggesting that the EFT string NLSM can be assumed to be weakly coupled.

Indeed, the volume of any effective divisor ${\rm Vol}(D_{\bf m})$ changes as follows under $\eqref{sflow}$:
\be
{\rm Vol}(D_{\bf m})=\frac12\int_{D_{\bf m}}J\wedge J=\langle {\bf m},{\bm s}_0\rangle+\langle {\bf m},{\bf e}\rangle\sigma\,.
\ee
Hence we see that it increases precisely if $\langle {\bf m},{\bf e}\rangle=D_{\bf m}\cdot \Sigma_{\bf e}>0$, i.e.\ if the effective divisor $D_{\bf m}$ transversely intersects  the generic curve $\Sigma_{\bf e}$. 
In this case ${\rm Vol}(D_{\bf m})$ provides a measure of the directions transversal to $\Sigma_{\bf e}$, and then of the  metric on the moduli space $\calm_{\bf e}$. Hence these directions are stretched  as we approach the string. From the RG viewpoint, this implies that the EFT string RG-flow scales up the NLSM metric as we move to the UV, in agreement with our general expectation of a weakly-coupled NLSM. In the non generic case in which
$\langle {\bf m},{\bf e}\rangle=D_{\bf m}\cdot \Sigma_{\bf e}=0$, ${\rm Vol}(D_{\bf m})$ remains constant but can anyway be assumed to be large, hence justifying again the weakly-coupled world-sheet description. A more precise justification of these claims is provided in Appendix \ref{app:curveNLSM}. 

This expectation is further supported by the anomaly and central charge  computations of \cite{Lawrie:2016axq}. 
Indeed, by using anomaly inflow arguments, it was found that the world-sheet anomaly and central charges   precisely match what one gets from a massless spectrum obtained by geometrical arguments, combined with the topological duality twist of \cite{Martucci:2014ema}.\footnote{The analysis of \cite{Lawrie:2016axq} assumes that $\Sigma\cdot \overline K_X\geq 0$, which is indeed satisfied by our EFT strings since $\overline K_X$ is effective.} This is  only possible if the world-sheet sector admits a weakly coupled NLSM description. Furthermore, the anomaly matching of \cite{Lawrie:2016axq} uses only bulk terms of the form \eqref{FFterms}, while it does not require any world-sheet term of the form \eqref{SN}. This further confirms that for these EFT strings we can set
$\hat C_i({\bf e})=0$, as suggested by the absence of an apparent five-dimensional  bulk supersymmetric structure.

More precisely, by using the topological duality twist of \cite{Martucci:2014ema}, the spectrum of massless fields of the ${\cal N}=(0,2)$ worldsheet theory was obtained by dimensional reduction of the ${\cal N}=4$ Super-Yang-Mills theory of a single D3-brane wrapping the curve $\Sigma_{\bf e}$.
Apart from one universal chiral multiplet $U$ associated with the center-of-mass motion of the string in four dimensions,
there are two types of extra chiral multiplets, $\Phi^{(1)}$ and $\Phi^{(2)}$, of respective multiplicity
\be
\begin{aligned}
&n^{(1)}_{\rm C} = h^0(\Sigma_{\bf e},N_{\Sigma_{\bf e}/X}) \,, \qquad n^{(2)}_{\rm C} = \overline K_X \cdot \Sigma_{\bf e} + g -1   \,,\\
&n_{\rm C} = n^{(1)}_{\rm C} + n^{(2)}_{\rm C} \,,\\ 
\end{aligned}
\ee
where $g$ denotes the genus of the curve $\Sigma_{\bf e}$.
These are accompanied by two types of Fermi multiplets, $\Lambda^{(1)}_-$ and $\Lambda^{(2)}_-$,
uncharged under the gauge group on the  7-branes, which are likewise obtained from reduction of the fermionic fields on the D3-brane world-volume and which come with multiplicities
\be
\begin{aligned}
&n^{(1)}_{\rm N} = h^1(\Sigma_{\bf e},N_{\Sigma_{\bf e}/X})  = 
h^0(\Sigma_{\bf e},N_{\Sigma_{\bf e}/X})
- \overline K_X \cdot \Sigma_{\bf e} \,,  \qquad  n^{(2)}_{\rm N} = g\,,\\ \,
&n_{\rm N} = n^{(1)}_{\rm N} + n^{(2)}_{\rm N} \,.
\end{aligned}
\ee
The uncharged (with respect to the 7-brane gauge group) Fermi multiplets $\Lambda_-$ are to be contrasted with the 
\bea \label{nFFtheorymicro}
n_{\rm F} = 8 \, \Sigma_{\bf e} \cdot \overline K_X
\eea
charged Fermi multiplets $\Psi_-$ localised at the intersection points of $\Sigma_{\bf e}$ with the divisors wrapped by the 7-branes.
According to the logic of Section \ref{sec_Gaugeanomalies}, only $n_{\rm C} - n_{\rm N}$ many chiral multiplets can potentially contribute to the gauge anomalies on the worldsheet via anomalous couplings of the form (\ref{wsGS}).
From the above multiplicities one finds that
\begin{subequations}
\begin{align}
n_{\rm C}^{(1)} - n_{\rm N}^{(1)} &= \overline K_X \cdot \Sigma_{\bf e}   \,, \label{nCnN1} \\
n_{\rm C}^{(2)} - n_{\rm N}^{(2)} &= \overline K_X \cdot \Sigma_{\bf e} -1\,,  \,  \label{nCnN2}
\end{align}
\end{subequations}
and hence
\bea \label{nCmnN}
n_{\rm C} - n_{\rm N} = 2 \overline K_X \cdot \Sigma_{\bf e} -1 \,.
\eea
Note that $n_{\rm C}^{(1)} - n_{\rm N}^{(1)}$ agrees with the number of unobstructed complex geometric deformation moduli of the curve $\Sigma_{\bf e}$ inside the K\"ahler 3-fold $X$ (see e.g. \cite{Hori:2003ic} for details on how to compute these). In particular, the subtraction of $n_{\rm N}^{(1)}$ accounts for the obstructions of some of the naive $h^0(\Sigma_{\bf e},N_{\Sigma_{\bf e}/X})$ many geometric moduli. 
In this sector, therefore, the minimality principle invoked in Section \ref{sec:anomalymatching} is manifestly realised.

The interpretation of the bosonic components of the chiral multiplets $\Phi^{(2)}$, on the other hand, is rather as a certain type of `twisted' Wilson line moduli of the topologically twisted SYM theory reduced on $\Sigma_{\bf e}$. 
The nature of these fields becomes clearer 
by realising the strings in the dual M-theory compactified on the elliptic Calabi-Yau four-fold $Y$ \cite{Lawrie:2016axq}. In this picture, the strings appear as MSW strings \cite{Maldacena:1997de} obtained by wrapping an M5-brane on the surface $\hat \Sigma_{\bf e}$ which is given by the restriction of the elliptic fibration to $\Sigma_{\bf e}$. 
The expansion of the chiral two-form on the M5-brane in terms of a basis of primitive $(1,1)$  closed forms  leads to left-moving chiral scalar fields on the string worldsheet. Of these, $n_{\rm C}^{(2)}$ many combine with the right-moving scalars obtained from the reduction of the chiral two-form in the $(0,2)$ and non-primitive $(1,1)$ closed forms on the M5-brane into the bosonic components of the chiral superfields $\Phi^{(2)}$, while the remaining chiral scalars can be dualized into the $n_{\rm F}$
Fermi multiplets charged under the bulk gauge group \cite{Lawrie:2016axq}.
This suggests that it is the left-moving scalars of the bosonic components of the fields $\Phi^{(2)}$ which can enjoy an axionic     shift symmetry that can be gauged by the part of the four-dimensional gauge group from the 7-brane sector in F-theory and which can participate in the anomaly cancellation. 
More precisely, taking into account the potential obstructions, the number $n_{\rm A}$ of such axionic moduli is bounded by
\bea
n_{\rm A} \leq n_{\rm C}^{(2)} - n_{\rm N}^{(2)} = \overline K_X \cdot \Sigma_{\bf e} -1 \,
\eea
rather than by $ n_{\rm C} - n_{\rm N} =  2 \overline K_X \cdot \Sigma_{\bf e} -1$. Indeed, the unobstructed complex geometric deformation moduli counted by $n_{\rm C}^{(1)} - n_{\rm N}^{(1)}$ should not exhibit a gauged shift symmetry as long as we are considering F-theory over a smooth three-fold base $X$
with vanishing isometries. 
By contrast, if the three-fold base $X$ does admit isometries,
these can manifest themselves in additional contributions to
the gauge anomaly also from the fields of type
$\Phi^{(1)}$. This, however, either requires that
$X$ is not smooth, for instance of toroidal orbifold type,
or that the theory  exhibits enhanced supersymmetry.
In both scenarios, the additional gauge anomaly contributions from the $\Phi^{(1)}$ sector correspond to Kaluza-Klein gauge symmetry factors in F-theory rather than to the sector from 7-branes.

\subsection{Microscopic realization of EFT predictions} \label{sec_MicrRealF}

We are now ready to test our EFT predictions. 

Observe first that from \eqref{FCalpha} it obviously follows that 
\be \langle {\bf C}^I,{\bf e}\rangle=\cald^I\cdot \Sigma_{\bf e}\geq 0\,,
\ee 
for any ${\bf e}\in\calc_{\rm S}^{\text{\tiny EFT}}$, as in \eqref{pos1}, because $\cald^I$ is effective. 
The second constraint in  \eqref{pos1} implies that 
\be 
 \{\langle {\bf C}^{AB},{\bf e}\rangle\}= \{b^{AB}\cdot \Sigma_{\bf e}\}\geq 0\,,
\ee
in the sense of being a positive semi-definite symmetric matrix. As discussed in \cite{Lee:2018ihr}, it follows from \cite{CoxZucker} that the diagonal components of the height-pairing $b^{AB}$ can in general be written as the sum of effective divisors and hence the diagonal entries of the matrix in question are guaranteed to be positive. 
 More generally, $\langle {\bf C}^{AB},{\bm s}\rangle$ coincides with the kinetic matrix for the abelian gauge sector (see e.g.\cite{Weigand:2018rez}) and must therefore indeed be positive definite, and hence $\langle {\bf C}^{AB},{\bf e}\rangle$ should be positive semi-definite. 

The confirmation of the EFT string bounds on the curvature-squared couplings is immediate. Indeed from \eqref{tildeCFtheory} we obtain 
\be\label{CtildeFtheory} 
\langle \tilde{\bf C},{\bf e}\rangle=6\,\Sigma_{\bf e}\cdot\overline K_X\geq 0\,  \quad~~~~\forall {\bf e}\in\calc_{\rm S}^{\text{\tiny{EFT}}}\,.
\ee 
This shows that \eqref{tildeCpos}, or equivalently \eqref{tildehatC0} or \eqref{tildehatC} with $\hat{\bf C}({\bf e})=0$, is manifestly realized in these F-theory models. Note also that, in fact,  $\langle \tilde{\bf C},{\bf e}\rangle$ is a multiple of 6, which is in agreement with the refined bound which one would get from \eqref{CCnn2} by imposing $n_{\rm C}-n_{\rm N}\geq 1$ rather than \eqref{nCNbound} -- see footnote \ref{foot:bound} -- to strings with genuinely ${\cal N} =(0,2)$ supersymmetry. Indeed, the EFT strings with enhanced non-chiral spectrum  correspond to D3-branes not intersecting the bulk 7-branes, that is, for which $\Sigma_{\bf e}\cdot\overline K_X=0$.

Turning to the bound \eqref{rankbound} on gauge group ranks, we expect it to be related to the Kodaira bound \eqref{Kodairabound}. To address this point, notice first that the rank of the part of the gauge group  which is probed by the EFT string of charge vector ${\bf e}$ takes the form 
\be 
r({\bf e})=\sum_{I| \cald^I\cdot \Sigma_{\bf e}\neq 0} r_I+{\rm rank}(b^{AB}\cdot \Sigma_{\bf e})\,.
\ee 
On the other hand, from  \eqref{CtildeFtheory} and \eqref{discrdec}   we see that \eqref{rankbound} becomes 
\be\label{rboundF}
r({\bf e})\leq r({\bf e})_{\rm max} \equiv 2 \langle \tilde{\bf C},{\bf e}\rangle -2 =12\, \Sigma_{\bf e}\cdot \overline K_X -2 =\Delta\cdot \Sigma_{\bf e} -2  \,.
\ee
Recall from the discussion in Section \ref{sec_Gaugeanomalies} that this bound includes the contribution to the anomaly polynomial both from the charged Fermi multiplets $\Psi_-$ and from the $n_{\rm C}- n_{\rm N}$ unobstructed chiral multiplets, whose bosonic components may in principle transform via a shift under the four-dimensional gauge symmetry, i.e. 
\begin{subequations} \label{rederivationF}
\begin{align}
r({\bf e})_{\rm max} &= r_{\rm F}({\bf e})_{\rm max} + r_{\rm C}({\bf e})_{\rm max}   \,,   \\
r_{\rm F}({\bf e})_{\rm max} &=\frac43\langle \tilde{\bf C},{\bf e}\rangle=8\, \Sigma_{\bf e}\cdot \overline K_X=\frac23\Delta\cdot\Sigma_{\bf e}     \,,    \label{rFFtheory1}\\
r_{\rm C}({\bf e})_{\rm max} &= \frac23\langle \tilde{\bf C},{\bf e}\rangle -2 = 4 \,  \Sigma_{\bf e}\cdot \overline K_X - 2 = \frac13\Delta -2 \label{rCFtheory1} \,.
\end{align}
\end{subequations}
As an important consistency check, note that the value $r_{\rm F}({\bf e})_{\rm max}$ is given by $n_{\rm F} = \frac43\langle \tilde{\bf C},{\bf e}\rangle=8\, \Sigma_{\bf e}\cdot \overline K_X$ identified by \eqref{hatnF}, which perfectly agrees with the microscopic counting \eqref{nFFtheorymicro}   of left-moving Fermi multiplets charged under the gauge group supported by  7-branes, while 
$r_{\rm C}({\bf e})_{\rm max} $ agrees with $2(n_{\rm C} - n_{\rm N})$ as computed in \eqref{nCmnN}.

However, the discussion at the end of Section \ref{sec_EFTstringsF} suggests that the conservative bound \eqref{rboundF} can in fact be sharpened in minimally supersymmetric F-theory models over smooth threefold bases. The point is that at best the unobstructed scalars of type $\Phi^{(2)}$ can contribute to the GS anomaly and hence to the bound on the rank. With this logic, we arrive at a stricter bound in F-theory given by 
\be  \label{strictboundF}
r({\bf e})_{\rm max}^{\rm strict} = r_{\rm F}({\bf e})_{\rm max} + r^{(2)}_{\rm C}({\bf e})_{\rm max}  = \frac{5}{3}   \langle \tilde{\bf C},{\bf e}\rangle -2 = 10 \, \Sigma_{\bf e}\cdot \overline K_X -2 \,, 
\ee
where
\be 
r^{(2)}_{\rm C}({\bf e})_{\rm max} = 2(n_{\rm C}^{(2)} - n_{\rm N}^{(2)}) =
\frac13\langle \tilde{\bf C},{\bf e}\rangle -2 = 2 \,  \Sigma_{\bf e}\cdot \overline K_X
- 2 
\ee
denotes the contribution to the anomaly polynomial from the unobstructed scalars of type $\Phi^{(2)}$, see \eqref{nCnN2}.

Let us stress that the bound \eqref{strictboundF} refers to the gauge subsector which couples to the EFT string of charge ${\bf e}$. 
To bound the rank of the gauge group supported by all 7-branes, we need to take $\Sigma_{\bf e}$ to lie strictly in the interior of the movable cone ${\rm Mov}_1(X)$ so that it has non-zero intersection with all non-abelian divisors ${\cal D}^I$ and height-pairing divisors. The strictest bound is then obtained as the minimal possible value $r({\bf e})_{\rm max}^{\rm strict}$ with $\Sigma_{\bf e}$ in the interior of  ${\rm Mov}_1(X)$.

With this understanding, we propose \eqref{strictboundF}, rather than
\eqref{rboundF}, as a bound on the gauge group rank of minimally supersymmetric F-theory compactifications over smooth threefold base spaces, as these do not admit isometries acting on the internal geometric space.
We now proceed to compare these quantum gravity bounds with the geometric bounds on the gauge group rank in F-theory.

\subsection{Example 1: $X=\mathbb{P}^3$} \label{sec_P3ex}

We begin with the simple choice 
$X = \mathbb P^3$. The one-dimensional group of effective divisors ${\rm Eff}^1(X)$ is spanned by the hyperplane class $H$, in terms of which $\overline K_X = 4 H$. This implies that the sections (\ref{fgdelta}) defining the Weierstrass model correspond to the classes
\be\label{P3Delta} 
\Delta \simeq 48 H \quad,\quad f \simeq 16 H \quad,\quad g \simeq 24  H \,.
\ee
In this case the saxionic cone is one-dimensional and $\calc^{\text{\tiny EFT}}_{\rm S}\simeq {\rm Mov}_1(X)_{\mathbb{Z}}={\rm Eff}_1(X)_{\mathbb{Z}}$ is generated by the curve $ H \cdot H \equiv H^2$. To obtain the strongest quantum gravity bounds from the EFT string, we can take $\Sigma_{\bf e} =H^2$.
Since this curve class intersects all possible divisor classes $\cald^I$, (\ref{rboundF}), and its proposed stronger version \eqref{strictboundF}, constrains
the total rank $r_{\rm tot}$ of  abelian and non-abelian gauge algebra factors supported by all 7-branes,  which can be detected by the EFT string associated with $\Sigma_{\bf e}$:
\begin{equation}\label{P3bound}
 r_{\rm tot} \leq 
\begin{cases}
r({\bf e})_{\rm max}=12\, \Sigma_{\bf e}\cdot \overline K_X -2 =46 \quad~~~~  \text{from \eqref{rboundF}} \,,    \\
r({\bf e})_{\rm max}^{\rm strict}=10\, \Sigma_{\bf e}\cdot \overline K_X -2 =38  \quad~~~ \text{from \eqref{strictboundF}} \,.
\end{cases}
\end{equation}
We can use \eqref{P3bound} to bound the maximal possible $SU(N)$ gauge algebra supported on any irreducible divisor $\cald^{SU(N)}$ appearing in the decomposition \eqref{discrdec}
as 
\bea \label{EFTboundSUN}
    SU(N^{\text{\tiny EFT}}_{\rm max}) = SU(r_{\rm max}+1)=    \begin{cases} SU(47)\quad~~~  \text{from \eqref{rboundF}} \,,  \\
    SU(39) \quad~~~ \text{from \eqref{strictboundF}} \,. \end{cases}  
    \eea
The first line is only slightly stronger than the naive Kodaira bound in its simple form (\ref{Kodairabound}), which together with Table \ref{Kodairatable} would give $SU(N^{\rm Kod}_{\rm max}) = SU(N^{\rm Kod}_{\rm max}) = 49$, as pointed out already in \cite{Morrison:2011mb}.
However, as stressed above, we in fact propose the stricter of the two bounds to hold in F-theory on a smooth base threefold $X$.

To compare this to actual realisations in F-theory, note first that the Kodaira upper bound cannot be saturated because this would require that all branes are mutually local. 
In fact, by an explicit analysis of the Weierstrass model, \cite{Morrison:2011mb} showed that the maximal possible value which can be realised in F-theory on $\mathbb P^3$ corresponds to $N^{\rm F}_{\rm max} = 32$. Consistently, this is
within the stricter of the two bounds in 
(\ref{EFTboundSUN}) based on (\ref{strictboundF}).

The EFT string bounds are potentially very far-reaching when it comes to constraining the maximal rank of abelian gauge group factors \cite{Lee:2019skh}. From (\ref{P3bound}) we immediately constrain the maximal number of abelian gauge algebra factors as
\be \label{U1boundP3}
n^{\text{\tiny EFT}}_{\rm max}(U(1)) = \begin{cases} 46     \qquad \text{from \eqref{rboundF}} \,, \\
38    \qquad \text{from \eqref{strictboundF}} \,.
\end{cases} 
\ee

A comparable bound cannot be deduced in a straightforward manner from the Kodaira bound since the abelian gauge group factors are generated by extra rational sections of the elliptic fibration. While to each such section one can associate
a divisor class on the base via the height-pairing, there does not exist any obvious constraint that forces this divisor class to be bounded by the class of the discriminant. Bounding the number of abelian gauge group factors is equivalent to finding an upper bound for the rank of the Mordell-Weil group of an elliptically fibered Calabi-Yau fourfold, which is an open problem in arithmetic geometry.\footnote{The upper bound  \eqref{U1boundP3} is the analogue, on fourfolds, of the bound $n(U(1)) \leq 32$ for F-theory compactifications to six dimensions on an elliptic three-fold over base $\mathbb P^2$ \cite{Lee:2019skh}. The highest Mordell-Weil rank obtained so far for elliptic threefolds is $r_{\rm max} =10$ \cite{Grassi:2021wii}.}

For the special example of an $SU(N)$ gauge group, an interesting phenomenon occurs:
The geometric bound $N^{\rm F}_{\rm max} = 32$ established in \cite{Morrison:2011mb} is in fact 
only marginally stricter than the  EFT string bound which one would deduce by taking into account, in the computation of the bound \eqref{rederivationF}, only the contribution from the charged localised Fermi multiplets: This would give $r \leq r_{\rm F}({\bf e})_{\rm max} = n_{\rm F} = 8\, \Sigma_{\bf e}\cdot \overline K_X = 32$.
However, it would be incorrect to conclude from this that $r_{\rm F}({\bf e})_{\rm max}$, rather than the weaker bound $r({\bf e})^{\rm strict}_{\rm max}$, represents the upper bound on the rank of the gauge group more generally.

Indeed, it is certainly possible to violate the bound set by $r_{\rm F}({\bf e})_{\rm max}$ in explicit Weierstrass models.\footnote{We thank Wati Taylor for pointing this out to us.}
For example, consider the maximal possible gauge group of type $E_6^{n_1} \times E_7^{n_2}$ by engineering the various gauge group factors on separate divisors in class $H$. 
Compatibility with the (incorrect) bound $r_{\rm F}$ would give that
$6 n_1 + 7 n_2 \leq 32$, with maximal solutions $(n_1,n_2) \in \{(0,4), (1,3), (2,2), (3,2), (4,1), (5,0) \}$.
In actuality, a Weierstrass model allows for the construction of a gauge algebra 
$E_6^{n_1} \times E_7^{n_2}$ with maximal values given by
$(n_1,n_2) \in \{(0,4), (1,4), (2,3), (3,2), (4,1), (5,0)\}$ (again on separate gauge divisors of class $H$). This follows from the vanishing orders of Table \ref{Kodairatable} that need to be engineered in the Weierstrass model. The two configurations $E_6 \times E_7^{4}$ and 
$E_6^2 \times E_7^{3}$ hence overshoot the bound from $r_{\rm F}({\bf e})_{\rm max}$ by $2$ and $1$, respectively, while they respect  the weaker bound from   $r({\bf e})^{\rm strict}_{\rm max}$, which we propose as the correct bound.

Two warnings concerning the validity of these examples are in place: 
First, even if all individual exceptional gauge groups are localised on 7-brane stacks wrapping separate divisors in class $H$, every pair of them intersects in a curve of class $H^2$. Along this curve, non-Kodaira type singularities appear because the vanishing orders of $(f,g)$ are equal to or bigger than $(4,6)$ and hence non-minimal. These non-minimalities in codimension-two arise at finite distance in the complex structure moduli space. 
Extrapolating from the analogous phenomenon in F-theory compactifications on elliptic Calabi-Yau threefolds \cite{Heckman:2018jxk}, one expects a 
genuinely strongly coupled sector localised at the non-minimal loci. To reliably analyse the dynamics one must first remove the non-minimalities by blowing up the intersection loci, thereby separating the exceptional brane stacks. However, this will most likely not affect the original exceptional gauge group factors. 

Another concern is pertinent, in fact, to all examples discussed in this section:
Since the minimal gauge divisor of class $H$ is non-spin, the Freed-Witten quantization condition \cite{Freed:1998tg} requires a half-integer quantized gauge flux in order for the model to be consistent. Depending on the nature of the available flux, the gauge background may reduce the rank of the four-dimensional gauge group or at least break the simple gauge group factors to a group involving abelian factors. Even if the rank were not reduced in this way, the $U(1)$ may become massive via a St\"uckelberg mechanism. If this is the case, the axion obtained by reducing the RR four-form field on $\mathbb P^3$ would likewise acquire a mass - a scenario which we excluded in our derivation of the quantum gravity bounds. 
To explicitly analyse the types of fluxes available and whether or not they necessarily induce a St\"uckelberg mass for the axion, one must resolve the fibral singularities of the Weierstrass model\footnote{In the dual M-theory description, the gauge fluxes are encapsulated in four-form fluxes on the resolved four-fold $\hat Y$ which take values in the vertical component of $H^{2,2}(\hat Y)$ and must be quantized in such a way that $G_4 + \frac{1}{2} c_2(\hat Y) \in H^4(\hat Y, \mathbb Z)$. For details we refer to \cite{Weigand:2018rez} and references therein. Admissible fluxes in F-theory are subject to two transversality conditions and after implementing these it remains to be seen whether fluxes different from the so-called Cartan fluxes are available, which would break the simple gauge group factor. Fluxes which leave the gauge group intact are dual to the matter surfaces \cite{Bies:2017fam}, but their explicit existence can only be determined after performing the resolution steps.} and hence, in view of the non-minimalities in codimension two, first  blow up the base. 
This involved surgery is beyond the scope of this paper, but we stress that from the Weierstrass model alone it not yet clear which part of the gauge group survives in the four-dimensional effective theory after taking the flux background into account.

Let us take the optimistic point of view and assume that none of these technicalities undermines the validity of the models with a rank overshooting the stricter bound from $r_{\rm F}({\bf e})_{\rm max}$. In this case, our analysis shows that 
some of the chiral multiplets, according to our proposal in fact the ones of type $\Phi^{(2)}$, indeed contribute to the 
anomaly matching. We will develop a better intuition behind this in the following section.

\subsection{Example 2: $\mathbb{P}^1\hookrightarrow X\rightarrow B$}
\label{sec:P1P2fibr}

Consider next an  F-theory model in which the base $X$ is a $\mathbb{P}^1$-fibration over a complex surface $B$.
This example will be particularly illuminating for the interpretation of the scalar field contributions to the anomaly matching on the worldsheet and hence to the bound on the gauge group rank.

Quite generally, the twist of a rational fibration $p: X \to B$ is specified by a line bundle ${\cal T}$ on the base $B$,\footnote{The threefold $X$ is constructed as the projectivised bundle $X={\mathbb P}({\cal O} \oplus {\cal T})$.} in terms of which the anti-canonical class of $X$ can be written as \cite{Friedman:1997yq} 
\bea \label{KXgeneral}
\overline K_X = 2 S_- + p^\ast c_1({\cal T}) + p^\ast c_1(B ) \,.
\eea
Here $S_-$ denotes the exceptional section of the rational fibration with the property
\bea \label{Sminusself}
S_- \cdot S_- = - S_- \cdot p^\ast c_1({\cal T}) \,.
\eea
The cone  $\calc_{\rm I} \simeq {\rm Eff}^1(X)$ of effective divisors of $X$ is generated by $S_-$ together with the pullback of the generators of the effective cone of $B $.

Another section, $S_+$, of the fibration is related to $S_-$ as $S_+ = S_- + p^\ast c_1({\cal T})$. It satisfies the relations
$S_+ \cdot S_+ = S_+ \cdot p^\ast c_1({\cal T})$ and $S_- \cdot S_+ =0$.
This section $S_+$ generates the K\"ahler cone of $X$ together with the pullback of the K\"ahler cone generators of $B $ to the rational fibration.

Under F-theory/heterotic duality, this class of theories is dual to the heterotic string compactified on an elliptic fibration over the same base $B $. The gauge groups from 7-branes wrapped along the two sections $S_-$ and $S_+$ map to perturbative heterotic gauge groups in either of the two $E_8$ factors, while gauge groups from 7-branes wrapping divisors pulled back to $X$ from $B $ are of non-perturbative nature in the heterotic frame.

For definiteness, let us specify now  to $B  = \mathbb P^2$
and adopt the notation of \cite[p.\,53]{Lanza:2021qsu},
\bea
D^1 = p^\ast(H)\,, \qquad D^2 = S_+ \,,
\eea
where $H$ is the hyperplane class of $\mathbb P^2$. Parametrizing furthermore the twist of the fibration as $c_1({\cal T}) = n H$, the general formula \eqref{KXgeneral} becomes
\be
\overline K_X=(3-n)D^1+2D^2 \,.
\ee
According to the above discussion, $D^1$ and $D^2$ generate the K\"ahler cone of $X$.
 The dual curves 
 \bea
 \Sigma_1= S_- \cdot p^\ast(H) = (D^2-n D^1) \cdot D^1,\quad \Sigma_2=p^\ast(H) \cdot p^\ast(H) = D^1\cdot D^1
 \eea
 (such that $ D^a\cdot \Sigma_b=\delta_b^a$) are effective and generate the Mori cone. In particular, $\Sigma_1$ can be identified with a  $\mathbb{P}^1$ in the $\mathbb{P}^2$ base, while $\Sigma_2$ can be regarded  as the $\mathbb{P}^1$ fibre of $X$. 

 The cone $\calc_{\rm I} \simeq {\rm Eff}^1(X)$ of effective divisors is generated by $D^1 = p^\ast(H)$ and $S_- = D^2-n D^1$.
The dual cone $\calc_{\rm S}^{\text{\tiny EFT}} \simeq {\rm Mov}_1(X)$ of  movable curves $\Sigma_{\bf e}=e^a\Sigma_a$ is given by
\be \label{EFTcondex}
\calc^{\text{\tiny EFT}}_{\rm S}=\{{\bf e}\in(e^1,e^2)\in\mathbb{Z}_{\geq 0}| e^2\geq ne^1\}
\ee
and is generated by the charges 
\be\label{FelemP1P2}
{\bf e}_1=(1,n)\,,\quad {\bf e}_2=(0,1)\,.
\ee 
Note that $\calc^{\text{\tiny EFT}}_{\rm S}$ is smaller than the cone of possible BPS strings, which can be identified with the Mori cone of effective curves.  In particular, the elementary BPS charge  ${\bf e}=(1,0)$ corresponding to a curve on the base alone does not give rise to an EFT string for $n\geq 1$, but it must be combined with $n$ copies of the charge of an EFT  string associated with the rational fiber.

To evaluate the bound (\ref{rboundF}), we compute the intersection number
\be
\Sigma_{\bf e}\cdot \overline K_X=(3-n)e^1+2e^2.
\ee
The bound \eqref{rboundF} and the proposed stronger form \eqref{strictboundF} then become
\bea\label{exbound}
r({\bf e}) \leq \begin{cases} r({\bf e})_{\rm max}= 12(3-n)e^1+ 24 e^2 - 2   \quad~~~~  \text{from \eqref{rboundF}} \,, \\
r({\bf e})^{\rm strict}_{\rm max}= 10(3-n)e^1+ 20 e^2 - 2 \quad~~~  \text{from \eqref{strictboundF}}\,,
\end{cases}
\eea
which constrains the rank of the gauge group that can be detected by an EFT string with charge ${\bf e}$. 
The strongest constraints are obtained by applying this bound to ${\bf e}_1$ and ${\bf e}_2$. Specifying directly to the proposed stronger bound, one finds
\be\label{exboundagain}
r({\bf e}_1) \leq  28 + 10 n \, ,\quad r({\bf e}_2) \leq  18 \,.
\ee

Suppose first that the semi-simple gauge sector is supported on effective divisors $\cald^I=m^I_1 D^1+m^I_2  (D^2-nD^1)\subset \{\Delta=0\}$ with $m^I_1,m^I_2>0$. In this case both $r({\bf e}_1)$ and $r({\bf e}_2)$ correspond to the rank of the total semi-simple gauge algebra, since it is entirely detected   by both elementary EFT strings. Since $n\geq 0$ we obtain the bound
\be
{\rm rk}(G_{\text{semi-simple}})\leq 18 \,.
\ee

We could instead assume that the discriminant locus splits in disconnected effective components 
\be
\Delta =n_1 \cald^1+n_2\cald^2+\cald' \,,
\ee
where $\cald^1=D^1$ and $\cald^2= D^2-nD^1$ support different semi-simple gauge groups, $G_1$ and $G_2$ respectively. Then $r({\bf e}_1)^{\rm strict}_{\rm max}$ bounds the rank of $G_1$ and  $r({\bf e}_2)^{\rm strict}_{\rm max}$ bounds the rank of $G_2$ as detected by the EFT strings. In this case we obtain
\be\label{r12bounds}
{\rm rk}(G_1)\leq  28 + 10n\, ,\quad {\rm rk}(G_2)\leq 18  \,.
\ee

We furthermore note that as in the case of $\mathbb P^3$, the contribution to the bound from the axionic sector is indeed needed  - taking into account only the charged Fermi multiplets in \eqref{rederivationF} would lead to the (incorrect) bounds
\be\label{r12boundsFermi}
{\rm rk}(G_1)\leq   24 + 8n\, ,\quad {\rm rk}(G_2)\leq 16  \,\qquad \text{based on Fermi multiplets only},
\ee
which are easily violated for instance
in Weierstrass models with exceptional gauge group factors.

Indeed, for simplicity take $n=0$, so that $X= \mathbb P^1 \times \mathbb P^2$.
One can overshoot the first constraint in (\ref{r12boundsFermi}) for instance by realising a configuration with gauge group $G_1 = E_6^3 \times E_7$ and ${\rm rk}(G_1) = 25$, or 
$G_1 = E_6^2 \times E_7^2$ and ${\rm rk}(G_1) = 26$, where the simple gauge factors are supported on four separate (but intersecting) divisors in class $D^1$. 
These examples are analogous to the ones discussed in the previous section for the base $X= \mathbb P^3$, and the same
remarks concerning the viability of these models apply.

More illuminating for us is the second bound in (\ref{r12boundsFermi}) compared to \eqref{r12bounds}. 
It is violated (again for $n=0$) for instance by a configuration of three simple gauge group factors along three separate divisors in class ${D}^2$ such that $G_2 = E_6^3$ or $G_2 = E_7^2 \times SO(8) $ or $G_2 = E_8 \times E_6 \times SO(8)$.
The non-abelian gauge divisors can be separated along their common normal direction on the $\mathbb P^1$-fiber;
nonetheless, there appear non-minimal Kodaira type fibers in codimension two or three, or both.\footnote{Also the second complication encountered already on $\mathbb P^3$ persists, namely that the gauge divisor ${D}^2$ is non-spin and gauge flux must be included to cancel the Freed-Witten anomaly; however we expect that this is possible without affecting the rank of the gauge group.} 
For example, the most general way to engineer the model with gauge group $G_2 = E_6^3$ is to set, in the Weierstrass equation, 
\be \label{WeierstrassP1fibrationE6}
f \equiv 0  \,, \qquad g=p_1^4(u,v) \, q_1^4(u,v) \, r_1^4(u,v) \, s_{18}(u_1,u_2,u_3) \,,
\ee
where $[u_1 :u_2: u_3]$ and $[u:v]$ denote homogeneous coordinates on $\mathbb P^2$ and $\mathbb P^1$, respectively, and $p_1, q_1, r_1, s_{18}$ represent generic homogeneous polynomials of indicated degrees in the coordinates in brackets. Note that  $\Delta = 4 f^3 + 27 g^2 = 27 g^2$ because we were forced to set $f \equiv 0$ to accommodate the $E_6^3$ factor.
As a result, over the divisor $s_{18}=0$, the fiber enhances to Kodaira Type II with ${\rm ord}(f,g,\Delta) = (\infty, 1,2)$. At the intersection locus of $s_{18}=0$ with any of the three $E_6$ divisors $p_1 =0$ or $q_1=0$ or $r_1=0$, the fiber enhances to Kodaira Type II$^{\ast}$, but at the $s_{18} \cdot s_{18} = 18^2$ points on $\mathbb P^2$ given by the self-intersection of $s_{18}=0$, there occurs a non-minimal fiber with vanishing orders ${\rm ord}(f,g,\Delta) = (\infty, 6,12)$. To reliably analyse the model, this non-minimal singularity would first have to be removed via a blowup in the base.

Interestingly, however,
the codimension-three non-minimalities can be avoided by replacing the base $\mathbb P^2$ by a rational elliptic surface dP$_9$, which is an elliptic fibration over $\mathbb P_1$ with generically $12$ singular fibers.
In this case, the section $s_{18}(u_1, u_2, u_3)$ in (\ref{WeierstrassP1fibrationE6}) is replaced by a general section 
$s \in \Gamma( 6 \bar K_{{\rm dP}_9})$.
The anti-canonical class of ${\rm dP}_9$
coincides with the class of the elliptic fiber and hence has vanishing self-intersection, 
$\bar K_{{\rm dP}_9} \cdot \bar K_{{\rm dP}_9}=0$. As a result, there appear no codimension-three non-minimal fibers since $s \cdot s =0$.
For the sake of simplicity we will focus on this geometric example in the sequel.
Clearly, the change of basis from $\mathbb P^2$ to ${\rm dP}_9$ does not affect the second bound in \eqref{r12bounds} or  \eqref{r12boundsFermi}, which only involves the EFT string with charge ${\bf e}_2$ associated with the $\mathbb P^1$-fiber of $X$ (while the first bound in both equations changes).

 Focusing on this second bound, to understand why  \eqref{r12bounds}, rather than \eqref{r12boundsFermi}, gives the correct bound, recall that 
 the EFT string with charge ${\bf e}_2$ is  the critical heterotic string of the dual compactification of the heterotic theory on an elliptic fibration over the base $B$ of $X$.
As discussed at the end of Section \ref{sec_EFTstringsF}, there are two types of chiral scalar fields, $\Phi^{(1)}$  and $\Phi^{(2)}$, in the NLSM of this string.
Since the genus of $\Sigma_{{\bf e}_2} = \mathbb P^1$ vanishes, $n^{(1)}_{\rm N}=0=n^{(2)}_{\rm N}$ and hence both types of scalars correspond to unobstructed moduli directions. 
The $n_{\rm C}^{(1)} - n_{\rm N}^{(1)} = n_{\rm C}^{(1)} = \bar K_{X} \cdot \Sigma_{{\bf e}_2} = 2$ complex moduli of the first type parametrize the motion of the heterotic string along the base of the dual elliptic fibration, while  
$n_{\rm C}^{(2)} - n_{\rm N}^{(2)} = n_{\rm C}^{(2)} = \bar K_{X} \cdot \Sigma_{{\bf e}_2} -1 = 1$ counts the complex modulus of the dual heterotic string along the heterotic torus fiber. 
As long as the base of $X$ admits no isometries - as is the case whenever $X$ is smooth and the theory minimally supersymmetric - the scalars encoded in the chiral fields of type $\Phi^{(1)}$ cannot participate in the gauge anomaly cancellation in F-theory. Indeed, they are geometric moduli in a geometry without extra shift symmetries.
By contrast, the two real scalars encoded in $\Phi^{(2)}$ can enjoy a gauged shift symmetry: They parametrize the motion of the heterotic string along the torus fiber of the dual heterotic geometry, which is responsible for the difference between the two bounds 
 (\ref{r12bounds}) compared to \eqref{r12boundsFermi}.

  The rationale behind this claim is completely analogous to the simpler setup of F-theory compactified on an elliptic K3 surface to eight dimensions.\footnote{Recall that $D^2$ is a copy of $\mathbb P^2$ and intersects the $\mathbb P^1$ fiber of $X$ in a point; similarly, in F-theory on K3, the 7-branes are points on the base of the K3.} 
In such models, a D3-brane wrapped along the $\mathbb P^1$ base of the K3 surface is known to be dual to a heterotic F1 string in 8d.
The zero mode spectrum of the $N=(0,8)$ supersymmetric worldsheet theory can be derived by dimensional reduction, as e.g.\ in \cite{Lawrie:2016axq}. It consists of one ${\cal N}=(0,8)$ hypermultiplet (8 real scalars in the ${\bf 6}$ of $SO(6)_{\rm T}$ and two $SO(6)_{\rm T}$ singlets accompanied by 8 fermions in the ${\bf 4} + {\bf \bar 4}$) as well as 16 left-moving fermions  transforming as $SO(6)_{\rm T}$ singlets and organising into Fermi multiplets. 
The scalars in the ${\bf 6}$ of $SO(6)_{\rm T}$ parametrize the center-of-mass motion of the string in eight dimensions and must therefore be uncharged under the gauge group. 
The only charged fermions are then the $16 = 8 {\rm deg}(\bar K_{\mathbb P^1})$ unpaired left-moving fermions. Taking only these into account would suggest a maximal rank for the gauge group of $r_{\rm max} = 16$, while in fact the maximal rank is known to be $r^{\rm K3}_{\rm max} = 18$. 
This of course matches the counting in the dual heterotic theory compactified on a torus $T^2$, where the gauge group comprises the ten-dimensional rank 16 gauge group together with 2 Kaluza-Klein $U(1)$ factors. 
The Fermi multiplets are charged under the part of the gauge group inherited from the ten-dimensional gauge group (in fact, the left-moving fermions generate the rank 16 current algebra present already in ten dimensions), but do not detect the KK $U(1)$s.
To see these, one must allow for the two left-moving scalars in the ${\cal N}=(0,8)$ hypermultiplet  to shift under the associated gauge transformation and in this way to contribute to the worldsheet anomalies.
At special points in moduli space, the KK $U(1)$s are indistinguishable from the Cartan $U(1)$s inherited from ten dimensions and  combine with part of them 
into higher rank non-abelian group factors.
Examples include the rank 18 exceptional gauge group configurations of the form $E_6^3$, $E_7^2 \times SO(8)$ or $E_8 \times E_6 \times SO(8)$ studied in F-theory on K3 in \cite{Dasgupta:1996ij}. More generally, the classification of maximal non-abelian gauge enhancements for the heterotic string on $T^2$ in \cite{Font:2020rsk} includes many configurations with rank 17 or 18 for the non-abelian sector.

Coming back to the bounds for the four-dimensional F-theory model on $X$,
we see that the contribution to the anomaly from the chiral multiplets of type $\Phi^{(2)}$ is analogous to the way how the heterotic string detects the two KK $U(1)$s in eight dimensions. 
However, in order for such an interpretation to be possible, the dual heterotic compactification space, which is the target space of the NLSM, must be degenerate such as to admit isometries in two directions while at the same time preserving minimal supersymmetry.
In general the dual heterotic string is compactified on an elliptic fibration over the base $B$. 
If this elliptic three-fold were smooth, there would be no room for an interpretation in terms of geometric KK $U(1)$s, not even from the elliptic fiber.\footnote{While the generic elliptic fiber contains two 1-cycles, these do not survive as 1-cycles of the elliptic threefold due to monodromies, as required by the (strict) Calabi-Yau condition.} The way out is that whenever the bound on $r({\bf e}_2) \leq 18$ is saturated, the dual elliptic fibration must degenerate to an orbifold such that the two KK $U(1)$s from the elliptic fiber can survive the projection. In the present example with base $B= {\rm dP}_9$, the dual heterotic elliptic fibration is a Schoen manifold \cite{Schoen}, which can be viewed as a fiber product ${\rm dP}_9 \times_{\mathbb P^1}  {\rm dP}_9$.  Schoen manifolds admit degenerations to toroidal orbifolds \cite{Donagi:2008xy} (see also \cite{GrootNibbelink:2012phj}). We propose such orbifolds as the heterotic duals of the F-theory models with $r({\bf e}_2) =18$. 
The orbifold must act in such a way that only two of the generally possible six KK $U(1)$s survive the projection.\footnote{Heterotic orbifolds with $2+n$ KK $U(1)$s cannot be dual to elliptic fibrations over a smooth base $X$ in F-theory. The  additional $n$ KK $U(1)$s must correspond to KK $U(1)$s, rather than 7-brane $U(1)$s, also on the F-theory side and be encoded in gauged shift symmetries of the scalars of type $\Phi^{(1)}$. Our assumption of a smooth base $B$ of the rational fibration $X$ excludes such situations in the minimally supersymmetric case. If for instance $X=T^4\times \mathbb{P}^1$, the bulk theory preserves sixteen supercharges and the $n^{(1)}_{\rm C}=2$  unobstructed $\Phi^{(1)}$ scalars describing the position of  $\Sigma_{{\bf e}_2}=\mathbb{P}^1$   in the $T^4$ direction enjoy shift symmetries. The corresponding  contribution to the anomaly raises the rank bound from $r({\bf e}_2)^{\rm strict}_{\rm max}=18$ to  $r({\bf e}_2)_{\rm max}=22$, matching the bound found in \cite{Kim:2019ths}.}

It would be desirable to explicitly construct the heterotic duals, also for the blowups of the F-theory models with base $B$ different from ${\rm dP}_9$, whenever the rank bound $r({\bf e}_2) \leq 18$ is saturated.
More ambitiously, we leave it for future work to determine the structure of the NSLM target space for non-heterotic EFT strings in models saturating the bounds \eqref{strictboundF} and to check it for isometries.

\subsection{Perturbative IIB models with O3-planes}
\label{sec:IIBO3}

The EFT strings considered in Section \ref{sec:Ftheorymodel} do not detect possible gauge sectors supported by D3-branes. The corresponding complexified gauge coupling can be identified with the type IIB axio-dilaton. Hence this gauge sector is fully perturbative in the  ten-dimensional  weak coupling limit, in which the compactification space is well described by a Calabi-Yau orientifold. 

For simplicity, we will focus on the simpler compactifications with O3-planes only, but more general configurations with O3/O7 planes can be discussed along the same lines. The relevant chiral field associated with this regime is the type IIB axio-dilaton: $t=a+\ii s\equiv \tau=C^{\rm RR}_0+\ii e^{-\phi}$. The saxionic cone is one-dimensional and ${\cal C}^{\text{\tiny{EFT}}}_{\rm S}=\mathbb{Z}_{\geq 0}$. An EFT string of charge $e\in\mathbb{Z}_{\geq 0}$   corresponds to $e$ D7-branes wrapping the internal space. Note
that the associated EFT string flow  drives the bulk sector to a strongly non-geometric regime in which the string frame volume of the compactification space naively goes to zero  \cite{Lanza:2021qsu}.  On the other hand, according to the estimate of \cite{Lanza:2021qsu}  the scaling weight is $w=1$ and then the Emergent String Conjecture  suggests the existence of a weakly coupled dual description in terms of F1 strings.\footnote{For example,  D7 strings in the toroidal models -- see for instance  \cite{Uranga:2000xp} for a discussion in a similar spirit -- can be T-dualized to a D1-string and S-dualized to an F1-string.} 

Without any additional assumption on the  K\"ahler moduli, in this limit  the only weakly coupled gauge sector  is the one supported by D3-branes, and the total rank of the gauge group is simply given by the number $n_{\rm D3}$ of D3-branes. Taking the gauge group to be completely higgsed to $U(1)^{n_{\rm D3}}$, corresponding to non-coinciding D3-branes, one can expand the standard DBI action and identify the terms
\be
-\frac{1}{4\pi} \sum^{n_{\rm D3}}_{A=1}\int s F_A\wedge * F_A\,.
\ee
By comparison with \eqref{kingauge2}, we see that 
\be\label{D3Cab}
C^{AB}=\delta^{AB}
\ee 
and $C^I=0$. 

The higher curvature terms \eqref{aRR} come from both the D3-brane and O3-plane higher curvature couplings. The D3-brane contribution stems from the action \cite{Green:1996dd,Cheung:1997az,Minasian:1997mm}
\be
-2\pi n_{\rm D3}\int_{\rm D3} C^{\text{\tiny RR}}_0\sqrt{\hat\cala\left(\calr\right)}=\frac{2\pi n_{\rm D3}}{48}\int C^{\text{\tiny RR}}_0\,p_1(M)+\ldots\,,
\ee
where $\hat\cala$ is the $A$-roof genus and we are simplifying the formulas by setting $\ell_{\rm s}\equiv 2\pi\sqrt{\alpha'}=1$.\footnote{Note that the sign is fixed by requiring that the internal cycles wrapped by mutually supersymmetric D-branes are calibrated by $-e^{\ii J}$ so that, for instance, D7-branes would be calibrated by $\frac12J\wedge J$ and then would wrap internal holomorphic cycles. With this choice supersymmetric space-filling D3-branes must be calibrated by $-1$ and may be considered as {\em anti} D3-branes. } The O3-planes instead contribute the terms \cite{Morales:1998ux}
\be
\frac{2\pi n_{\rm O3}}{4}\int C^{\text{\tiny RR}}_0\sqrt{L\left(\frac{1}4\calr\right)}=\frac{2\pi n_{\rm O3}}{4\cdot 96}\int C^{\text{\tiny RR}}_0\,p_1(M)\,,
\ee
where $L$ is the Hirzebruch $L$-polynomial. By using the identification  $C^{\rm RR}_0=a$ and the D3 tadpole cancellation condition
\be\label{D3tadpole0}
n_{\rm O3}=4n_{\rm D3}\,,
\ee
we then obtain the total contribution
\be\label{D3aRR}
\frac{2\pi (8n_{\rm D3}+n_{\rm O3})}{4\cdot 96}\int a\,p_1(M)=- \frac{3}{16}\frac{n_{\rm O3}}{96\pi} \int a\tr (R\wedge R)\,.
\ee
By matching with \eqref{aRR} we conclude that
\be\label{D3tildeC}
\tilde C=\frac3{16} n_{\rm O3}\,.
\ee
Note that these models do not encode any five-dimensional structure and indeed the D7-brane string does not give rise to effective couplings of the form \eqref{SN}. Hence we can set $\hat C=0$.

We can now compare these results with our quantum gravity constraints. Note that the positivity constraints \eqref{pos1} are obviously satisfied. On the other hand, \eqref{tildeCqc} requires that $\tilde C\in\mathbb{Z}$. This is possible only if 
\be\label{no3}
n_{\rm O3} \in 16\mathbb{N}\,.
\ee 
This non-trivial prediction in fact agrees with  Theorem 1.5 of \cite{Favale_2017}, which implies that $n_{\rm O3}=16$ for all (strict) Calabi-Yau covering spaces.  This mathematical result can also be explicitly checked in all the models of the database of \cite{Carta:2020ohw}.\footnote{We thank X.\ Gao and in particular F.\ Carta and R.\ Valandro for discussions on the available examples of Calabi-Yau orientifolds  with just O3 planes. We also thank F.\ Carta for helping us in checking \eqref{no3} by scanning the database \cite{Carta:2020ohw}, and for pointing out to us reference \cite{Favale_2017}.  }

The bound \eqref{rankbound} on the rank of the gauge group detected by the EFT strings becomes
\be
r({\bf e}) \leq \frac{3}{8} n_{\rm O3} -2 \,,
\ee
where we used \eqref{D3tildeC} and $\hat C=0$ and the r.h.s.\ is integral by \eqref{no3}.

From the explicit string theory model, however, we know that the gauge group probed by the EFT strings in question comes from the perturbative D3-brane sector and its rank is therefore bounded by $n_{\rm D3}=\frac14 n_{\rm O3}$. Interestingly, this actual bound coincides with the stronger bound obtained by taking into account only the contribution \eqref{ranknF} from the charged localised Fermi multiplets
in \eqref{totb0}, i.e. 
\be\label{O3strictb}
r_{\rm F}({\bf e}) \leq \frac{4}{3} \tilde C = \frac14 n_{\rm O3}\,.
\ee
This match seems to be characteristic of the perturbative 
(with respect to the $SL(2,\mathbb Z)$ duality group of Type IIB string theory) nature of the D3-brane sector. 

In these models the stricter bound set by \eqref{totb0} is always saturated, see \eqref{D3tadpole0}. However, it is microscopically clear that it remains valid even in presence of internal supersymmetric three-form fluxes, which contribute positively to the D3 tadpole condition and then reduce the number of D3-branes. From the four-dimensional viewpoint, the introduction of these fluxes makes the EFT string `anomalous': The string is forced to be the boundary of a membrane, which microscopically corresponds to a D5-brane ending on the D7 string. We leave for future work  the extension of our four-dimensional quantum gravity arguments to these kinds of anomalous strings.

\subsection{Complex structure EFT strings} \label{subsec_ComplStrEFT}

The third qualitatively different gauge sector in F-theory is associated with the R-R four-form $C^{\rm RR}_4$ expanded in terms of non-trivial elements of $H^3(X,\mathbb{Z})$.
The gauge kinetic function and hence also the characteristic couplings \eqref{kingauge} in this sector depend on the complex structure moduli of the four-fold $Y$ \cite{Grimm:2010ks}. 
Correspondingly, the weak coupling limits are attained in large complex structure limits on $Y$.
An explicit description of the EFT strings associated with this sector is an interesting challenge for future work. 

Suffice it here to mention that under heterotic/F-theory duality the R-R gauge sector is expected to map to the abelian gauge sector from heterotic 5-branes wrapped along curve classes of genus $g \geq 1$ \cite{Witten:1996hc} in the base of the dual heterotic elliptic fibration \cite{Lukas:1998hk,Berglund:1998ej}. 
This implies a duality between the EFT strings associated with this heterotic 5-brane sector and the complex structure EFT strings in F-theory.


\section{Microscopic checks in  heterotic models}
\label{sec:heterotic}

We now turn to compactifications of the heterotic theory on Calabi-Yau three-folds $X$, and their M-theory uplift. If not specified,  we will implicitly be considering the  $E_8\times E_8$ theory, while we will discuss  the $SO(32)$ case only in Section \ref{sec:hetSO(32)}.  

In order to understand how our quantum gravity bounds on the gauge sector and the curvature-squared terms are realized,  we need to revisit and complete the discussion presented in \cite{Lanza:2021qsu}, which neglects corrections  coming from  higher derivative terms in ten/eleven dimensions.  Indeed, we will see that these corrections  modify the structure of the saxionic cone and of the corresponding EFT string spectrum in a non-trivial manner. We will also allow for possible background NS5/M5-branes, not considered in \cite{Lanza:2021qsu}, which significantly broaden this class of vacua and enrich the structure of their EFT in an interesting way.

In the heterotic picture, we  focus on the chiral fields $(\hat t, t^a)$ where $t^a=a^a+\ii s^a$  appear in  the expansion
\be\label{BJexp} 
t^a D_a =a^a D_a+\ii s^a D_a \equiv B_2+\ii J
\ee
and 
\be\label{defhats}
\hat t=\hat a+\ii\hat s\equiv \int_X B_6+\ii e^{-2\phi}V_X\,.
\ee
Here $ D_a $ is Poincar\'e dual to a basis of $H_4(X,\mathbb{Z})$, $B_6$ is the electromagnetic dual of the NS-NS $B_2$ and 
\be
V_X=\frac1{3!}\int_X J\wedge J\wedge J=\frac1{3!}\kappa_{abc}s^as^bs^c
\quad,\quad \kappa_{abc}\equiv D_a\cdot D_b\cdot D_c
\ee
is the string frame volume of $X$ measured in string units $\ell_{\rm s}=2\pi\sqrt{\alpha'}=1$. 

If we neglect higher derivative corrections as in \cite{Lanza:2021qsu}, the saxionic cone described by $(\hat s, s^a)$ can simply be identified with $\mathbb{R}_{>0}\oplus \calk(X)$, where $\calk(X)$ is the K\"ahler cone of $X$. It will be important for us to understand how the higher derivative corrections affect this naive result. To this end we will collect the relevant threshold corrections to the gauge couplings of the four-dimensional EFT, and then argue how these corrections are compatible with a corresponding modification of the relevant sets of BPS instanton and EFT string charges. As we will see, the saxionic cone will also be enlarged into directions corresponding to the moduli of the possible background NS5/M5-branes. We will then check that the  curvature-squared terms obey the expected quantum gravity constraints and microscopically verify the corresponding upper bounds on the ranks of the gauge group.

\subsection{The EFT terms}

As a first step, we need to identify  the couplings \eqref{kingauge2}, including possible contributions coming from heterotic higher derivative corrections. 
All the relevant complexified gauge couplings may be extracted directly from \cite{Blumenhagen:2006ux}, and from various previous partial results cited therein such as \cite{Lukas:1997fg, Lukas:1998ew,Carlevaro:2005bk}. However, in order to render the discussion more self-contained and to carefully pin down the possible convention ambiguities, in Appendix \ref{app:hetEFT} we will briefly go through the main steps of the derivation. Furthermore, we will similarly derive also the curvature-squared terms, which were not explicitly considered in  \cite{Blumenhagen:2006ux}.   

The two heterotic $E_8$ sectors can have a non-trivial bundle structure along the internal space. Let us denote by $\hat F_{1}$ and $\hat F_{2}$ the corresponding fields strength. For simplicity, we assume that $\hat F_{1,2}$ take values in  semi-simple  sub-algebras of $\frak{e}_8$.
This in particular allows us to avoid possible St\"uckelberg masses and non-trivial kinetic mixing between the $U(1)$ factors.\footnote{A (partial)  verification of our quantum gravity constraints in presence of kinetic mixing of $U(1)$ factors can be obtained by duality to other corners of the string landscape considered in this paper.} We also allow for possible NS5-branes wrapping irreducible holomorphic curves $\calc^k$, $k=0,\ldots, N_{\text{\tiny NS5}}$.  The internal gauge bundles and the curves $ \calc^k$ enter the cohomological tadpole condition 
\begin{equation}
\label{Tadpole}
    \lambda(E_1) + \lambda(E_2) + [\calc]= c_2(X) \,,
\end{equation}
where $[\calc]$ is the Poincar\'e dual of the two-cycle $\calc\equiv\sum_k\calc^k$ and
\be\label{lambda}
\lambda(E)\equiv-\frac1{16\pi^2}\tr(\hat F\wedge \hat F)\,, 
\ee
with  the trace defined as in Section \ref{sec:gauge}. 

The background NS5-branes support additional (s)axionic fields. These can be more easily identified by considering the Ho$\check{\rm r}$ava-Witten (HW) M-theory uplift of the heterotic theory \cite{Horava:1995qa,Horava:1996ma}, compactified over $I\times X$, where $I=S^1/\mathbb{Z}_2$ is the HW interval. In the upstairs picture, we parametrize the M-theory circle by  the coordinate  $y\simeq y+2$, on which we impose the $\mathbb{Z}_2$ identification $y\simeq -y$. The $E_8$ sectors are then supported on the walls $\{y=0\}$ and $\{y=1\simeq -1\}$, and we can parametrize $I=S^1/\mathbb{Z}_2$ by $y\in[0,1]$. The background NS5-branes uplift to M5-branes sitting at points $\hat y^k$ along the HW interval $I$. Their worldvolume supports  a self-dual two-form $\tilde\calb^k_2$. This sector hence gives rise to $N_{\text{\tiny NS5}}$ additional axions 
\be \label{def-tildeak}
     \tilde a^k\equiv 
     \int_{{\cal C}^k}\tilde\calb^k_2 \,,
\ee
By supersymmetry the axions pair up with the saxions \cite{Blumenhagen:2006ux}
\be\label{defsk}
\tilde s^k=\left(\hat y^k-\frac12\right)\int_{{\cal C}^k}J=\left(\hat y^k-\frac12\right)m^k_a s^a\,,\quad~~~~~~~~~~~\text{(no sum over $k$)}\,,
\ee
where 
\be\label{defmka}
m^k_a\equiv D_a\cdot \calc^k\,.
\ee

By dimensionally reducing the bulk action, including the Green-Schwarz anomaly cancelling term as well as certain terms supported on the M5-branes, one obtains the following contribution to the four-dimensional axionic couplings:
\be\label{hetF1F2}
\begin{aligned}
&-\frac{1}{8\pi}\int \Big(\hat a+\frac12 p_a a^a - \frac38 q_a a^a - \frac12 \sum_k \tilde a^k \Big)\tr (F_1\wedge F_1)\\  
& -\frac{1}{8\pi}\int \Big(\hat a-\frac12 p_a a^a + \frac18 q_a a^a  + \frac12 \sum_k \tilde a^k \Big)\tr (F_2\wedge F_2) \,.
\end{aligned}
\ee
Here $F_{1,2}$ denote the field strengths of the four-dimensional gauge sectors coming from the two heterotic $E_8$ sectors, respectively, and
\be\label{padef}
p_a\equiv -\int_{D_a}\left[\lambda(E_2)-\frac12 c_2(X)\right]\quad, \quad q_a\equiv D_a\cdot \calc=\sum_k m_a^k\,.
\ee
We furthermore recall the definition of the axions in
\eqref{defhats}, \eqref{BJexp} and \eqref{def-tildeak}. The corresponding saxionic couplings are fixed by supersymmetry.

As discussed in detail in Appendix \ref{app:hetEFT}, one can similarly compute  the relevant curvature-squared terms. By focusing again on the axionic couplings, the final result is 
\be\label{hetaRR1} 
-\frac{1}{96\pi}\int \Big(12 \hat a+n_aa^a - \frac{3}{2} q_a a^a \Big)\tr (R\wedge R) \,,
\ee
where 
\be\label{defna} 
n_a\equiv \frac12\int_{D_a}c_2(X)\,.
\ee
By applying the Hirzebruch-Riemann-Roch theorem to a line bundle $\calo_X(D_a)$ one can easily conclude that $n_a\in \mathbb{Z}$.
Since $\int_{D_a}\lambda(E_2)$ is an instanton number, $p_a\in\mathbb{Z}$ as well. 

Note that the constants appearing in both \eqref{hetF1F2}  and \eqref{hetaRR1} do not satisfy the naively expected quantization conditions \eqref{CCconst} and \eqref{tildeCquant}, respectively. This signals the presence of a rational mixing of the axionic periodicities induced by the higher derivative corrections. In other words, the axions $(\hat a,a^a,\tilde a^k)$ defined in \eqref{defhats}, \eqref{BJexp} and \eqref{def-tildeak} do not satisfy the integral periodicity $a^i\simeq a^i+1$ assumed in this paper. 
This issue can be solved by passing to a better axionic basis, obtained from a rational linear redefinition of the axions. For instance, we can replace the axions $\hat a$ and $\tilde a^k$ with the linear combinations
\be
\begin{aligned}
a^0=\hat a +\frac12 p_a a^a - \frac38 q_a a^a - \frac12 \sum_k \tilde a^k\,,\quad 
a^k=\tilde  a^k+\frac12 m^k_a a^a\,.
\end{aligned}
\ee
In this way the sum of \eqref{hetF1F2} and \eqref{hetaRR1} can be rewritten as 
\be\label{hetF1F2RR}
\begin{aligned}
&-\frac{1}{8\pi}\int  a^0\tr \left(F_1\wedge F_1\right) -\frac{1}{8\pi}\int \Big(a^0- p_a a^a + \sum_k a^k \Big)\tr \left(F_2\wedge F_2\right)\\&-\frac{1}{96\pi}\int \Big(12 a^0- 6 p_a a^a+n_aa^a +6\sum_k a^k \Big)\tr (R\wedge R) \,.
\end{aligned}
\ee
In the axionic basis $a^i=(a^0,a^a,a^k)$, these terms indeed have the form of the axionic terms appearing in \eqref{kingauge2} and \eqref{aRR} and then determine the corresponding constants
\be\begin{aligned}\label{hetM5C}
 {\bf C}^1&=(C^1_0,C^1_a,C^1_k)=(1,\vec 0,\vec 0)\,,\quad \quad {\bf C}^2=(C^2_0,C^2_a,C^2_k)=(1,-p_a,\vec 1) \,, \\
 \tilde{\bf C}&=(\tilde C_0,\tilde C_a,\tilde C_k) = (12,-6p_a+n_a,\vec 6) \,.
\end{aligned}
\ee
The saxionic couplings appearing in \eqref{kingauge2} and \eqref{aRR} are completely fixed by supersymmetry  and involve the saxions $s^i=(s^0,s^a,s^k)$, with 
\begin{subequations}\label{hetfinsaxions}
\begin{align}
&s^0\equiv\hat s +\frac12 p_a s^a - \frac38 q_a s^a - \frac12 \sum_k s^k\,,\label{hetfinsaxionsa}\\
&s^k\equiv\tilde  s^k+\frac12 m^k_a s^a=\hat y^km_a^ks^a\quad \text{(no sum over $k$)}\,.\label{hetfinsaxionsb}
\end{align}
\end{subequations} 
Note that $s^k=0$ if the $k$-th M5-brane sits on the HW wall $y=0$ and $s^k=m^k_a$ if the $k$-th M5-brane sits on the HW wall $y=1$. We also observe that the microscopic symmetry under the exchange of the two HW walls, together with the coordinate change $y\leftrightarrow 1-y$, descends to the invariance of the four-dimensional action under exchange of the two gauge sectors $F_1\leftrightarrow F_2$, together with
\be\label{HWswap}
p_a\leftrightarrow q_a-p_a  \,, \quad \quad s^0\leftrightarrow s^0-p_as^a+\sum_k s^k \,, \quad \quad s^k\leftrightarrow m_a^ks^a-s^k\,,
\ee
plus similar transformations for the axions. This may be regarded as a discrete $\mathbb{Z}_2$ gauge symmetry of the theory, since it describes the same microscopic configuration. 

From \eqref{hetM5C} we note  that the saxions $s^k$ enter the EFT terms \eqref{kingauge2} and \eqref{aRR} in the combination  
\be\label{hetinvcomb} 
p_as^a-\sum_k s^k\,.
\ee 
Interestingly, this combination is continuous under `small instanton transitions' \cite{Ovrut:2000qi} in which some M5-brane is absorbed or emitted by  the HW walls. As an example, take the limit $\hat y^k\rightarrow 1$ in which  all M5-branes are moved on top of the second HW wall, and are then absorbed by a small instanton transition in which $\lambda(E_2)\rightarrow \lambda(E'_2)=\lambda(E_2)+[\calc]$ and then $p_a\rightarrow p_a'=p_a-\sum_km^k_a=p_a-q_a$. Along this process, $p_as^a-\sum_k s^k$ first becomes $p_as^a-\sum_k m^k_as^a $, which is indeed equal to $p'_as^a$. The combination  \eqref{hetinvcomb} is also continuous under a small instanton transition in  which all  M5-branes are absorbed by the first HW wall,  which is simply described by the limit $s^k\rightarrow 0$. Clearly \eqref{hetinvcomb} is also  continuous under   more general small instanton transitions.

Note that the M5-branes can provide an additional gauge sector, coming from the expansion of the M5 self-dual two-form potentials in harmonic one-forms of $\calc^k$, as in \cite{Witten:1997sc}. However, the corresponding gauge couplings  are controlled by the complex structure of the curves $\calc^k$, rather than by their volume. Hence, this sector is generically strongly coupled in the perturbative regime considered here, and one should consider some large complex structure limit in order to identify corresponding axionic strings. The F-theory dual sector was briefly mentioned in Section \ref{subsec_ComplStrEFT}.

\subsection{Saxionic cone}
\label{sec:hetscone}

The saxionic cone is determined by the set of possible BPS instantons, as in \eqref{scone}. In order to clarify its global structure, it is convenient to work in the M-theory frame.
There are three types of BPS instantons to consider for us: Heterotic worldsheet instantons, open membrane instantons ending with one or both ends on an M5-brane, or M5-brane instantons.

Consider first the heterotic world-sheet instantons, which in M-theory are represented by Euclidean open M2-brane wrapping some effective curve $\Sigma\subset X$ and stretching between the two HW walls. Their action is given by $2\pi m_a(\Sigma)s^a$, with $m_a(\Sigma)=D_a\cdot \Sigma$. The condition $m_a(\Sigma) s^a> 0$ defines the standard K\"ahler cone $\calk(X)$.

If in particular we choose $\Sigma=\calc^k$, we obtain the condition $m^k_as^a>0$. Hence, by using the restriction $y\in (0,1)$ in  \eqref{hetfinsaxionsb} we deduce that $s^k$ must satisfy the condition
\be\label{hetskcond}
0 < s^k < m^k_a s^a \,.
\ee
This condition  guarantees the exponential suppression of instanton contributions coming from the second type of instantons, Euclidean open M2-branes  stretched between the background M5-branes and the HW walls \cite{Moore:2000fs,Lima:2001jc,Lima:2001nh}. For instance, consider an  M2-brane  along the curve ${\cal C}^k$ wrapped by the $k$-th M5-brane and connecting the HW wall at $y=0$ with the $k$-th M5-brane. Its action is given precisely by $2\pi s^k$, which is indeed positive by \eqref{hetskcond}. If instead the M2-brane connects the $k$-th M5-brane to the HW wall at $y=1$, then its action is given by $2\pi (1-\hat y^k)m^k_a s^a =2\pi (m^k_a s^a-s^k)$, which is again positive by \eqref{hetskcond}. One can similarly consider an open Euclidean M2-brane stretching between two background M5-branes.

It remains to discuss the more subtle BPS instantons corresponding to Euclidean M5-branes  wrapping the entire Calabi-Yau $X$. A crucial role will be played by the internal M-theory $G_4$ flux, which  is generically non-vanishing.  
In the downstairs picture of the HW orbifold, the internal $G_4$ flux must satisfy specific boundary conditions \cite{Horava:1995qa,Horava:1996ma}.  In our setting, these read 
\begin{subequations}\label{G4boundary}
\begin{align}
\lim_{y\rightarrow 0^+}G_4|_{\{y\}\times X}&=\ell_{\text{\tiny M}}^3\left[ \lambda(E_1) -\frac12 c_2(X)\right]\,,\label{G4boundarya}\\
\lim_{y\rightarrow 1^-}G_4|_{\{y\}\times X}&=-\ell_{\text{\tiny M}}^3\left[ \lambda(E_2) -\frac12 c_2(X)\right]\,,\label{G4boundaryb}
\end{align}
\end{subequations}
where $\ell_\text{\tiny M}$ is the M-theory Planck length, which we choose to coincide with $\ell_\text{s}$ under dimensional reduction.

Inside the HW interval $G_4$ must satisfy the Bianchi identity
\be\label{G4BIhetM}
\d G_4=\ell_{\text{\tiny M}}^3\sum_k\delta(y-\hat y^k)\d y\wedge \delta^{4}_X(\calc^k)\,.
\ee
This implies that the  cohomology of $G_4$ jumps by $\ell_{\text{\tiny M}}^3[\calc^k]$ as one crosses the $k$-th M5-brane.  The combination of \eqref{G4boundary} and \eqref{G4BIhetM} indeed implies the consistency condition \eqref{Tadpole}. 

\begin{figure}[t!]
		\centering
		\includegraphics[width=15.8cm]{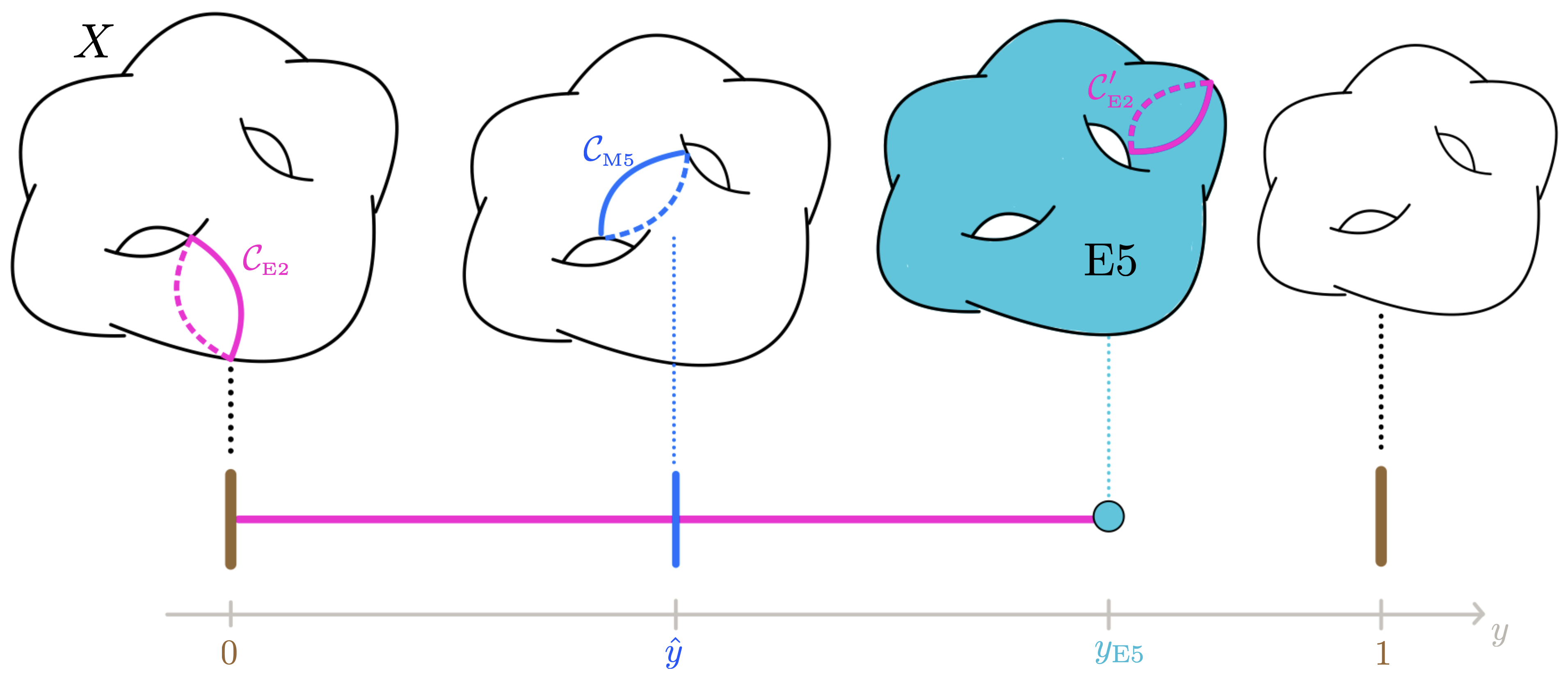}
		\caption{\footnotesize {\bf BPS instantons in} ${\bm E_8}{\bm \times} {\bm E_8}$ {\bf heterotic models}. The internal seven-dimensional space is a fibration of the Calabi-Yau $X$ over the horizontal  M-theory interval $I=\{0\leq y\leq 1\}$. The vertical brown and blue lines denote the HW walls and a background M5-brane wrapping the internal curve $\calc_{\text{\tiny M5}}$, respectively. The picture includes a Euclidean M5/M2-instanton: The light blue Euclidean M5-brane (E5) sits at a given position $y_{\text{\tiny E5}}$ along $I$ and wraps the entire internal Calabi-Yau space.  It is connected to the left  HW wall  by two purple Euclidean M2-branes (E2) wrapping $\calc_{\text{\tiny E2}}$ and $\calc'_{\text{\tiny E2}}$ respectively, such that $\calc'_{\text{\tiny E2}}\simeq \calc_{\text{\tiny E2}}+\calc_{\text{\tiny M5}}$, which reconnect on the bulk M5-brane. See also Appendix \ref{app:M5inst} for further explanations.  \label{fig:hetInst}}   	\end{figure}

As first discussed in \cite{Witten:1996mz}, this non-trivial flux and the HW walls  induce a non-trivial deformation of the internal geometry. Furthermore, the presence of a non-trivial $G_4$  implies that a Euclidean M5 wrapping the entire Calabi-Yau $X$ and sitting at an intermediate position $ 0<y_{\text{\tiny E5}}<1$ is not consistent by itself, because of the world-volume  tadpole condition. Rather,  one must add  Euclidean open  M2-branes ending on the M5-brane and must then consider a composite M5/M2-instanton -- see Figure \ref{fig:hetInst}. All this complicates the direct computation of the instanton Euclidean action and of the corresponding saxionic conditions. However, we can deduce this information by indirect arguments, exploiting the holomorphy of the BPS instanton corrections. 

We devote Appendix \ref{app:M5inst} to a detailed discussion of these arguments and here present only the final result. Namely, the positivity of the action of any  possible BPS M5 instanton (including possible open M2-brane insertions) is guaranteed if we impose the conditions $s^0>0$ and  $s^0-p_as^a+\sum_k s^k>0$. This allows us to complete our identification of the saxionic cone:
\be\label{M5Delta}
\Delta=\Big\{{\bm s}=(s^0,s^a,s^k)\ |\ s^a D_a\in\calk(X)\,, s^0> 0\,, s^0-p_a s^a+\sum_k s^k > 0\,,0< s^k< m^k_a s^a \Big\}\,.
\ee
Note that this saxionic cone is indeed invariant under  \eqref{HWswap}. 

Recalling \eqref{hetM5C}, one immediately checks that the combinations $\langle {\mathbf C}^1,{\bm s}\rangle$ and $\langle {\mathbf C}^2,{\bm s}\rangle$, which define the gauge couplings, are positive within the saxionic cone, as expected. Furthermore, a theorem \cite{miyaoka1985chern} guarantees that
\be\label{nspos}
n_as^a=\frac12\int_X J\wedge c_2(X)>0
\ee
if $J=s^a D_a$ belongs to the K\"ahler cone. Recalling \eqref{hetM5C}, it follows that $\langle \tilde{\bf C},{\bm s}\rangle >0$ within the saxionic cone.

\subsection{EFT strings}
\label{sec:EFTstringsHet}

We now come to specifying the cone of EFT strings. There are three types of BPS strings dual to the three types of BPS instantons discussed in the previous section: The critical heterotic string, i.e.\ an F1 string, corresponding to an open M2-brane stretched between the two HW walls, furthermore the strings obtained by wrapping an M5-brane along an effective divisor on $X$, and finally open M2-branes stretched between two M5-branes or between an M5-brane and an HW wall. 
The cone of EFT string charges, \eqref{CSEFT}, can be identified with the help of the
saxionic cone \eqref{M5Delta} as follows:
\be\label{hetCEFT}
\begin{aligned}
\calc^{\text{\tiny EFT}}_{\rm S}=&\Big\{{\bf e}=(e^0,e^a,e^k)\ |\ D_{\bf e}\equiv e^aD_a\in{\rm Nef}^1(X)\,, \\
&\quad e^0\geq  0\,,\ e^0-p_a e^a+\sum_k e^k \geq 0\,,\ 0\leq e^k\leq m^k_a e^a \Big\}\,.
\end{aligned}
\ee
Recalling \eqref{defmka} and \eqref{padef}, we note that $m^k_a e^a=\calc^k\cdot D_{\bf e}\geq 0$ if $D_{\bf e}$ is nef, and then  $q_a e^a=\sum_k\calc^k\cdot D_{\bf e}\geq 0$ as well. $\calc^{\text{\tiny EFT}}_{\rm S}$ has a richer structure than the cone identified in \cite{Lanza:2021qsu}, not only because of the additional charges $e^k$ associated with the presence of background M5-branes, but also because of the additional condition involving the background constants $p_a$, which come from 10d/11d higher derivative corrections. As a result, the analysis of section 6.1 of \cite{Lanza:2021qsu} must be updated.

A charge vector ${\bf e}=(e^0,\vec 0,\vec 0)$ corresponds to $e^0$ F1 strings and, as in \cite{Lanza:2021qsu},  is associated with an EFT string flow along which the ten-dimensional string coupling vanishes, $e^\phi\rightarrow 0$, while the string frame K\"ahler moduli and the M5 positions $\hat y^k$ remain fixed. As noted already, from the M-theory viewpoint, these EFT strings correspond to M2-branes stretching between the two HW walls.     

A choice  ${\bf e}=(0,\vec 0,e^k)$ corresponds to $e^k$ M2-branes filling two external directions and  stretching between  the $k$-th background  M5 and the second HW wall (at $y=1$). From the ten-dimensional viewpoint, they  appear as `fractional' F1 strings bound to the NS5.   However the condition $0\leq  e^k\leq  m^k_a e^a$ appearing in \eqref{hetCEFT} excludes  a charge vector of the form ${\bf e}=(0,\vec 0,e^k)$ from $\calc^{\text{\tiny EFT}}_{\rm S}$, and hence these strings are {\em not} EFT strings. Note that indeed such strings cannot explore the entire internal space, as would be characteristic for an EFT string. Rather, such BPS strings can become classically tensionless \cite{Mayr:1996sh} at finite distance in the moduli space, where the theory develops a strongly coupled sector in which the open M2-brane instantons discussed in the previous section become unsuppressed.

\begin{figure}[t!]
		\centering
		\includegraphics[width=15.8cm]{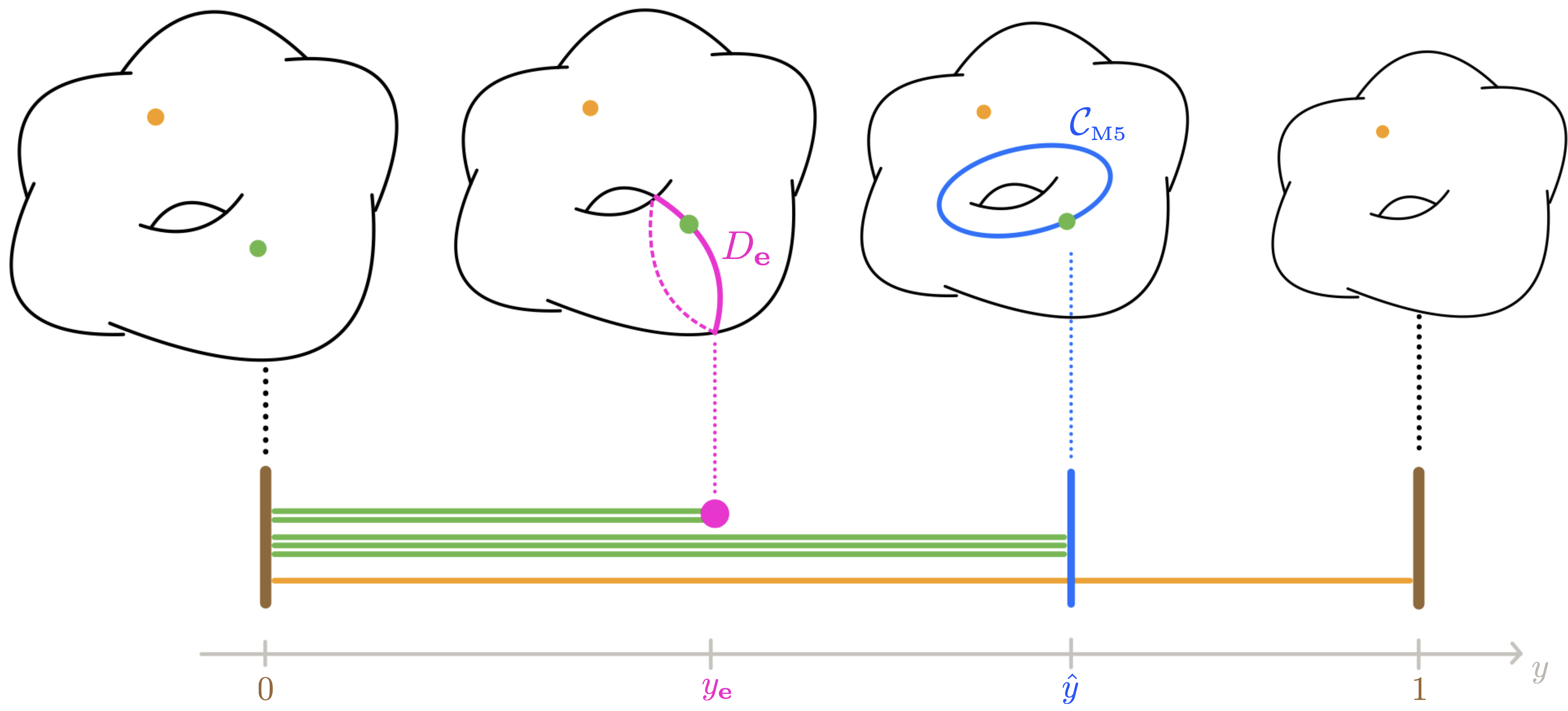}
		\caption{\footnotesize {\bf EFT strings in} ${\bm E_8}{\bm \times} {\bm E_8}$ {\bf heterotic models}. The bulk sector is as in Fig.~\ref{fig:hetInst}, but now the picture includes two  EFT strings: an  orange M2-brane stretching between the HW walls, which descends to the critical heterotic string in ten dimensions; a bound state of $(p_a-m_a)e^a=2$ green open M2-branes and a purple M5-brane wrapping the nef divisor $D_{\bf e}\subset X$ and sitting at $y=y_{\bf e}$. There also appear $m_ae^a=\calc\cdot D_{\bf e}=3$ green open M2-branes, connecting the left HW wall with the background M5-brane, which tie to the purple M5-brane if we move it to $y_{\bf e}>\hat y$.  \label{fig:hetEFTstrings}}   	\end{figure}

In order to interpret the condition $0\leq  e^k\leq  m^k_a e^a$, we must then turn on the charges $e^a$, associated with a string obtained from an M5-brane along a nef divisor $e^a D_a$.  
Assume first that $e^a(p_a-q_a)\geq 0$ (and thus $e^a p_a\geq 0$ as well) and suppose that the M5-string wrapping $D_{\bf e}$ sits on top of  the second HW wall at $y=1$.
From \eqref{padef} and \eqref{G4boundaryb}  we know that $p_a e^a=\frac{1}{\ell_{\text{\tiny M}}^3}\lim_{y_{\bf e}\rightarrow 1}\int_{\{y_{\bf e}\}\times D_{\bf e}}G_4$. In order to move the M5 string away from the HW wall,  there must therefore be $p_a e^a$ M2-branes ending on it from the left (to solve the world-volume tadpole condition along the M5-brane). If $e^k=0$ for any $k$, all these M2-branes must originate on the first HW wall, and hence initially there must be $e^0\geq p_a e^a$ M2-branes connecting the two HW walls, which is precisely the content of \eqref{hetCEFT}.    Now start moving the M5 string to the left, along the $y$ direction. When it crosses the $k$-th  bulk M5, the $G_4$-flux across $D_{\bf e}$ jumps by $-m_a^ke^a$. 
Hence, at a more general point $y_{\bf e}$ the number of M2 strings ending on the M5 string from the left is given by 
\be
\frac{1}{\ell_{\text{\tiny M}}^3}\int_{\{y_{\bf e}\}\times D_{\bf e}}G_4=p_ae^a-\sum_{k|\hat y^k>y_{\bf e}}m_a^ke^a\,,
\ee
while the remaining $m_a^ke^a$  M2 strings stretch between the first HW wall and the $k$-th background M5-brane with $\hat y^k>y_{\bf e}$  -- see Figure \ref{fig:hetEFTstrings} for an example with one background M5-brane. 

More generically, if  $e^k \neq 0$, then initially, when the M5 string is at $y_{\bf e}=1$, there are  $e^k$ open M2-branes connecting the second HW wall to the $k$-th background M5-brane. This implies that in order to allow the M5 string to move away from $y_{\bf e}=1$ it is sufficient to take $e^0\geq p_a e^a-\sum_k e^k$, which is indeed one of the conditions appearing in \eqref{hetCEFT}. 
Moreover,  the last condition  $0\leq e^k\leq m_a^ke^a$ appearing in \eqref{hetCEFT} implies that, after the M5 string crosses a background M5-brane, there remain no open M2-brane ending on it from the right. In the extreme case in which $e^k=m_a^ke^a$ and $e^0=(p_a-q_a) e^a$, when the M5 string arrives at $y_{\bf e}=0$, no M2 strings are left in the bulk.

The case $p_a e^a\leq 0$  can be treated in complete analogy, being related to the case $(p_a -q_a)e^a\geq 0$ by the $\mathbb{Z}_2$ symmetry \eqref{HWswap} which swaps the role of two HW walls.  One may similarly discuss some intermediate case with $p_a e^a\geq 0$ and $(q_a-p_a) e^a\geq 0$. 

As reviewed in Section \ref{sec:EFTstrings}, the EFT strings are associated  to an infinite distance limit. In the present model, the physical properties can be analysed by slightly adapting the discussion of \cite{Lanza:2021qsu}, taking into account two important differences. First, as highlighted by the above discussion, the flows corresponding to NS5/M5 strings can generically involve also the dilaton $s^0$ and the 5-brane moduli $s^k$, in addition to the K\"ahler moduli $s^a$. Furthermore, the identifications \eqref{hetfinsaxions} complicate the microscopic interpretation of the various EFT string flows and in general affect also the value of the corresponding scaling weight. However, a conclusion of \cite{Lanza:2021qsu} still holds: the EFT string flows such that
\be
\kappa({\bf e},{\bf e},{\bf e})\equiv \kappa_{abc}e^ae^be^c=D_{\bf e}^3>0
\ee
 lead to a rapid growth of the HW interval, so that we have a dynamically generated sharp hierarchy between its length and the length scale of the internal Calabi-Yau. Hence, close to the string, one internal direction opens up, and the 4d string uplifts to a string/membrane bound state in an HW-like 5d theory on $M_4\times I$, where $I=S^1/\mathbb{Z}_2$. In particular, the string probes a local $\caln=1$ 5d supergravity, whose 8 supercharges are spontaneously broken to 4 by the presence of the HW walls. These types of EFT have been studied for instance in \cite{Lukas:1998tt} but the relevant term can be quite easily obtained by reducing on the Calabi-Yau $X$ the M-theory CS term  (in upstairs picture). By using the decomposition 
\be
C_3=\frac{\ell_{\text{\tiny M}}^3}{2\pi}A^a\wedge D_a
\ee
one obtains the five-dimensional CS term
\be\label{5dCSagain}
S^{\rm 5d}_{\rm CS}=\frac{\pi}{6(2\pi)^3}\kappa_{abc}\int_{M_4\times S^1}A^a\wedge F^b\wedge F^c\quad~~~~\text{with}\quad \kappa_{abc}\equiv D_a\cdot D_b\cdot D_c\,,
\ee
in the upstairs picture, in which $y\in [-1,1]$.
Note that $C_3$ is odd under parity. Hence the five-dimensional $U(1)$ gauge fields $A^a$ are odd under the $\mathbb{Z}_2$ parity $\iota: y\mapsto -y$. The $S^1/\mathbb{Z}_2$ orbifold projection then imposes that $\iota^*A^a=-A^a$. If we restrict to zero modes along $S^1$, we are then forced to set $A^a=2\pi a^a\d y$, where $a^a$ are our 4d axions. This in particular implies that, from the 4d viewpoint, the 5d gauge fields $A^a$ contain a finite number of massless axions plus an infinite tower of massive vectors and pseudoscalars. Furthermore, the restriction of   \eqref{5dCSagain} to the zero modes identically vanishes. 

On the other hand, as in Section \ref{sec:hatC} this additional CS term produces an extra world-sheet term of the from \eqref{SN}, with 
\be\label{hethatC}
\hat C_i({\bf e})=\delta_i^a\kappa_{abc}e^be^c\,.
\ee
This provides an explicit microscopic realization of effect described in Section \ref{sec:hatC}, with 
\be
\hat C_{ijk}=\left\{\begin{array}{ll}
\kappa_{abd} & \text{if $(i,j,k)=(a,b,c)$}\\
0 & \text{otherwise}
\end{array}\right.\,.
\ee

\subsection{Microscopic check of quantum gravity bounds}\label{nmultiple3}

We are now ready to test our EFT quantum gravity constraints. From  \eqref{hetM5C} we obtain the relations
\be\label{hetCe}
\begin{aligned}
&\langle {\bf C}^1,{\bf e}\rangle=e^0\quad,\quad \langle {\bf C}^2,{\bf e}\rangle=e^0-p_ae^a+\sum_k e^k\,,\\
&\langle \tilde{\bf C},{\bf e}\rangle=6\langle {\bf C}^1,{\bf e}\rangle+ 6\langle {\bf C}^2,{\bf e}\rangle+n_a e^a\,.
\end{aligned}
\ee

 First of all, the positivity bounds \eqref{pos1} are satisfied by definition of \eqref{hetCEFT}. It follows that not only is \eqref{tildeCqc} obeyed, but actually $\langle \tilde{\bf C},{\bf e}\rangle\in\mathbb{Z}_{\geq 0}$,  since \eqref{nspos} implies that 
 \be
 n_ae^a=\frac12\int_{D_{\bf e}}c_2(X)\geq 0
 \ee
for any nef divisor $D_{\bf e}$. In turn, this guarantees that  the combinations appearing in \eqref{tildehatC0} and \eqref{tildehatC} are non-negative,
since 
\be \label{Chatiskappa}
\langle \hat C({\bf e}), {\bf e}\rangle =\hat C({\bf e},{\bf e},{\bf e})= \kappa({\bf e},{\bf e},{\bf e}) =D_{\bf e}^3
\ee
and the triple self-intersection of a nef divisor is non-negative.

 On the other hand  \eqref{tildehatC0}  makes the stronger prediction that  $\langle \tilde{\bf C},{\bf e}\rangle+\langle \hat{\bf C}({\bf e}),{\bf e}\rangle$ should actually be a (positive) integral multiple of 3. The contributions to $\langle \tilde{\bf C},{\bf e}\rangle+\langle \hat{\bf C}({\bf e}),{\bf e}\rangle$ coming from the first two   terms appearing in the second line of \eqref{hetCe} clearly satisfy this property, being positive multiples of $6$. It then remains to check that 
 \be\label{2cD3} 
 n_ae^a+\langle \hat{\bf C}({\bf e}),{\bf e}\rangle = \frac12\int_{ D_{\bf e}}c_2(X)+D_{\bf e}^3\in 3\mathbb{Z}_{\geq 0}\,.
 \ee  In order to prove it,  it is sufficient to prove that    $D_{\bf e}^3-\int_{ D_{\bf e}}c_2(X)$ is a multiple of 3. The latter statement follows from the index theorem applied to the signature complex twisted by the line bundle $\calo_X(D_{\bf e})$ -- see for instance \cite{Eguchi:1980jx} -- which implies that\footnote{Here $L(X)$ is the Hirzebruch $L$-polynomial and $\widetilde{\rm ch}(V)$ is the Chern character in which one replaces the curvature $F$ of $V$  by $2F$ in all expressions.}
\be
\begin{aligned}
&\int_XL(X)\wedge \widetilde{
\rm ch}(\calo_X(D_{\bf e}))=\frac23\int_{D_{\bf e}}p_1(X)+\frac43D_{\bf e}^3=\frac43\left(D_{\bf e}^3- \int_{D_{\bf e}}c_2(X)\right)\in\mathbb{Z}\,.
\end{aligned}
\ee
Since $D_{\bf e}^3- \int_{D_{\bf e}} c_2(X)$ is guaranteed to be integral, this result
confirms
the quantization condition \eqref{tildehatC0} (and then \eqref{tildehatC}).

Let us now test \eqref{rankbound}, which bounds the rank of the gauge group detected by an EFT string of charge ${\bf e}$ and presently takes 
the form
\be\label{rankbound2}
r({\bf e})\leq 2 \langle \tilde {\bf C},{\bf e}\rangle+ D_{\bf e}^3 -2 \,.
\ee
As explained in Section \ref{sec:QGbounds}, the strongest bounds are obtained by picking the generators of $\calc^{\text{\tiny EFT}}_{\rm S}$. First of all, we can consider
\be
{\bf e}=(1,\vec 0,\vec 0) \,,
\ee
which corresponds to the saxionic direction $s^0$ and `detects'  {all} perturbative gauge groups. 
With such a choice  $D_{\bf e}^3=0$  and $\langle \tilde {\bf C},{\bf e}\rangle=12$ and then the bound \eqref{rankbound2} gives 
\be\label{hetb1}
r({\bf e})\leq 22 \,.
\ee
Here $r({\bf e})$ includes the rank of the perturbative $E_8\times E_8$ gauge group present already in ten dimensions plus a maximal extra contribution of six to the total rank as encoded in the chiral superfields. In the following we will refer to this gauge sector as the `perturbative' one. As we will see more explicitly at the end of Section \ref{sec:elliphet}, compactifications on singular Calabi-Yau three-folds can host also `non-perturbative' gauge sectors, which are not accounted for by \eqref{hetb1}.  
The bound \eqref{hetb1} agrees with expectations from heterotic compactifications for instance on toroidal orbifolds, which can admit a maximum of six additional $U(1)$ group factors associated with the six KK $U(1)$ gauge fields \cite{Ibanez:1987pj}.\footnote{In such situations, all of the chiral superfields can contribute in the anomaly cancellation, showing that the bound \eqref{rankbound} can indeed be saturated.}

On the other hand,  the bound \eqref{hetb1} does not include possible $U(1)$s coming from the M5-branes, whose gauge coupling is of order one for generic complex structure, or more generally any other  gauge factors which avoid a coupling of the form (\ref{kingauge2}) to the axion associated with the fundamental heterotic string.

Assume next that there exists a nef divisor $D_{\bf e}=e^aD_a$ such that  $p_ae^a=0$ 
 and $q_ae^a=0$, so that ${\bf e}=(0,e^a,0)$ belongs to the cone $\calc_{\rm S}^{\text{\tiny EFT}}$. In this case $\langle {\bf C}^1,{\bf e}\rangle=\langle {\bf C}^2,{\bf e}\rangle=0$.\footnote{These strings are related  to the $(0,4)$ supergravity strings in five dimensions discussed in \cite{Katz:2020ewz} by dimensional reduction on the HW interval. If in addition $D^3_{\bf e}=0$ and $n_ae^a=0$, then they support an enhanced $(4,4)$ or $(8,8)$ non-chiral supersymmetric spectrum.}  The corresponding EFT string does not interact with the heterotic gauge bundles and does not provide any bound on their rank.\footnote{By contrast, based on duality with F-theory, we will argue at the end of this section that EFT strings with $e^a \neq 0$ detect a certain non-perturbative gauge sector of the heterotic compactification.}

More generally, let us assume that $(p_a-q_a)e^a> 0$. Then we can pick
\be\label{hete}
{\bf e}=((p_a-q_a)e^a,e^a,m^k_ae^a)\,.
\ee
In this case $\langle {\bf C}^1,{\bf e}\rangle=(p_a-q_a)e^a$ and $\langle {\bf C}^2,{\bf e}\rangle=0$, where we recall \eqref{hetCe} and \eqref{padef}. Hence, as far as the perturbative gauge group is concerned, $r({\bf e})$ is sensitive to the rank of the gauge sector coming only from the first HW wall. Since
$\langle \tilde {\bf C},{\bf e}\rangle=6(p_a-q_a)e^a+n_ae^a$, with $n_ae^a = \frac{1}{2} \int_{D_{\bf e}} c_2(X)$, \eqref{rankbound2} yields the bound
\be\label{hetb2}
r({\bf e})\leq 12(p_a-q_a)e^a+ \int_{D_{\bf e}}c_2(X) + D_{\bf e}^3 -2 \,.
\ee
This bound is clearly satisfied by the pertubative gauge sector supported on the first HW wall, since we know from the microscopic uplift that its rank is at most $8$, and can provide a non-trivial bound on the non-perturbative gauge sector.  By applying the $\mathbb{Z}_2$ symmetry \eqref{HWswap}, we obtain an analogous bound for the second gauge sector if $p_ae^a<0$.

If we specialise to  models with standard embedding $\lambda(E_1)=\lambda(X)$, from \eqref{padef} and \eqref{defna} we read off that $p_a=n_a$ and $q_a=0$ (i.e.\ there are  no background M5s).\footnote{The case $\lambda(E_2)=\lambda(X)$ can again be obtained by  applying \eqref{HWswap}.} Noting that in this case $p_ae^a\geq 0$,  we could pick ${\bf e}$ as in \eqref{hete} and the bound \eqref{hetb2} becomes 
\be\label{hetb3}
r({\bf e})\leq  7 \int_{D_{\bf e}}c_2(X) + D_{\bf e}^3 - 2\,.
\ee

We now turn to discuss some concrete models. As we will see, at least in these models, the bounds on the perturbative gauge sector coming from M5 strings are always weaker than \eqref{hetb1}.

\subsubsection{Example 1: The quintic}

The quintic three-fold $X$ is defined as the vanishing locus of a section of $\calo_{\mathbb{P}^4}(5)$ inside $\mathbb{P}^4$. In this case the effective  divisor $D_{\bf e}$ generating the (1-dimensional) nef cone is obtained by restricting the hyperplane  $H\subset\mathbb{P}^4$ to the hypersurface. Hence $D_{\bf e}^3=5H^4=5$, and by using the adjunction formula one obtains $c_2(X)=10[H^2]|_X$ and then  $\int_D c_2(X)=50$. Hence, if we consider a standard embedding, the bound \eqref{hetb2} becomes $r({\bf e})\leq 353$ which, if applied to the perturbative gauge sector, is clearly much weaker than \eqref{hetb1}. 

\subsubsection{Example 2: Elliptically fibered CYs and duality with F-theory}
\label{sec:elliphet}

Consider a smooth CY three-fold $X$ which is given by an elliptic fibration $\pi: X\rightarrow B$ over some weak-Fano two-fold $B$ and which is described by a smooth Weierstrass model. The weak Fano condition implies that $c_1(B)=c_1(\overline K_B)$ is Poincar\'e dual to an effective nef divisor.

One can  compute the second Chern class of $X$ as 
\be
c_2(X)=\pi^*c_2(B)+12 S\cdot \pi^*\overline K_B+11\pi^*\overline K_B^2 \,,
\ee
where Poincar\'e duality is implicit,  $\overline K_B$ is identified with its divisor and $S$ denotes the divisor associated with the global section of the Weierstrass model. 
The effective curves are generated by the elliptic $T^2$ fibre  and the push forward $\sigma_*(c)$ of the base effective curves $c\subset B$.  
The nef cone is generated by the vertical divisors $V=\pi^* j$ which project to a nef divisor $j$ of the base and the `horizontal' divisor $H=S+\pi^*\overline K_B$. 

Notice that, if we pick an EFT charge vector ${\bf e}=(e^0,e^a,e^k)$ with  $e^a$ such that $D_{\bf e}=e^aD_a=V$, then $\hat{\bf C}({\bf e})=D_{\bf e}^3=0$ and -- see \eqref{hetCe} --
\be \label{tildeCe-het}
\langle\tilde{\bf C},{\bf e}\rangle=6\langle{\bf C}^1,{\bf e}\rangle+6\langle{\bf C}^2,{\bf e}\rangle+6\, j\cdot \overline K_B  \,.
\ee

Heterotic compactifications on such threefolds are dual to F-theory compactified on an elliptic four-fold whose base  is a $\mathbb P^1$-fibration over the same weak-Fano two-fold $B$, blown up in the fiber over curves on $B$ wrapped by heterotic 5-branes.
To avoid confusion we call this F-theory threefold base $X_F$ in this section. In the notation of Section \ref{sec:P1P2fibr}, the anti-canonical class of $X_F$ is\footnote{For simplicity we are considering single blowups over separate curves.}
\bea
\overline K_{X_F} = 2 S_- + p^\ast c_1({\cal T}) + p^\ast c_1(B) - \sum_k E_k \,.
\eea
Here the twist bundle ${\cal T}$ characterising the rational fibration is related to the heterotic invariants $p_a$ as
\bea
p_a = \int_{d_a} c_1({\cal T}) \,,
\eea
with $d_a$ a basis of divisors on $B$.
In particular, for a positive bundle, i.e.\ $c_1({\cal T})$ effective, the gauge sector on the second HW wall  maps to the gauge theory on a stack of 7-branes wrapping the exceptional section $S_-$ in the F-theory base $X_F$.\footnote{Consistently, for
bigger and bigger effective twist class $c_1({\cal T})$, or larger positive values of $p_a$, the gauge flux on the second HW wall becomes smaller and smaller, as follows from the definition \eqref{padef}. This eventually results in a non-Higgsable remnant gauge group, which in F-theory must be localised on the rigid section $S_-$, with negative self-intersection \eqref{Sminusself}, rather than on $S_+$.}
Furthermore $E_k$ denotes a blowup divisor on $X_F$ over the curve $\calc^k$ wrapped by the $k$-th M5-brane.
Under the duality, an EFT string with charges  ${\bf e}=(e^0,e^a,e^k)$ in the heterotic theory maps to an EFT string obtained by wrapping a D3-brane along the curve
\be
\Sigma_{\bf e} = e^0 F_0 + e^a S_- \cdot p^\ast(d_a) + e^k F_k \,,
\ee
where $F_0$ denotes the generic rational fiber of $X_F$, $d_a$ continues to denote a basis of divisors on $B$ and $F_k$ is the rational fiber of the exceptional divisor $E_k$ with $E_i \cdot F_j = - \delta_{ij}$. The inequalities \eqref{hetCEFT} translate into the condition for the curve $\Sigma_{\bf e}$ to lie inside the movable cone. 
For instance, in the simple example discussed in Section \ref{sec:P1P2fibr} (with all $e^k=0$), the EFT string condition \eqref{EFTcondex} maps to the condition \eqref{hetCEFT} once we identify $p_a$ with the positive integer $n$ defining the twist. With these identifications, one can convince oneself that \eqref{tildeCe-het}
agrees with the dual expression \eqref{tildeCFtheory}, more precisely with $6 \bar K_{X_F} \cdot \Sigma_{\bf e}$, in F-theory.

Let us now turn to the bounds:
By  assuming for instance a standard embedding,   the bound \eqref{hetb3} resulting from a charge \eqref{hete} with $q_a=0$ and $p_a = n_a$ and with $D_{\bf e} = V = \pi^\ast(j)$ becomes
\be
r({\bf e})\leq  7 c_2(X)\cdot D_{\bf e} -2 =84 j\cdot \overline K_B -2 \,.
\ee
This bound does not contain any new information. If  $j\cdot \overline K_B=0$ then $\langle {\bf C}^1,{\bf e}\rangle=\langle {\bf C}^2,{\bf e}\rangle=0$, because in the charge vector \eqref{hete} the F1 component vanishes for the standard embedding: $p_a  e^a = n_a e^a = \frac{1}{2} \int_{\pi^\ast(j)} c_2(X) =0$; in this case we already know that 
 $r({\bf e})=0$. If  $j\cdot \overline K_B\geq 1$ and we focus just on  the perturbative gauge sector, we arrive at a bound much weaker than \eqref{hetb1}. 
As another example, pick a charge of the form \eqref{hete} but with $D_{\rm \bf e}=H = S + \pi^\ast(\bar K_B)$, still assuming a standard embedding. Then  the bound \eqref{hetb2} becomes
\be
r({\bf e})\leq  7 \chi(B)+ 78\overline K_B\cdot \overline K_B -2 \,.
\ee
For $B$ a weak-Fano space,  $\overline K_B\cdot \overline K_B\geq 1$, and then again this bound does not improve the bound \eqref{hetb1} for the perturbative gauge sector.

We have emphasized that these bounds test the perturbative part of the heterotic gauge group probed by the EFT string with charge ${\bf e}$ as given.  In F-theory this gauge sector corresponds to gauge symmetry from 7-branes on the two sections $S_-$ and $S_+$ of $X_F$.
On the other hand, on the F-theory side there can also be non-abelian gauge groups supported on a vertical divisor $D_V = p^\ast{d}$ with $d$ an effective divisor on the base $B$ of $X_F$. Note that such gauge sectors are independent of the existence of blowup divisors in F-theory, which map to the background M5-branes in heterotic M-theory whose gauge coupling is controlled by the complex structure of the heterotic theory.
An example of such a vertical gauge sector
 was discussed in Section \ref{sec:P1P2fibr} by taking the 7-branes in class $D^1$.
The vertical gauge sector in F-theory must be dual to a non-perturbative gauge sector in the heterotic theory which arises when the elliptic fibration becomes singular over the divisor $d\subset B$.
The EFT strings detecting the vertical part of the gauge group in F-theory map to heterotic EFT strings with non-vanishing charges $e^a$. By duality, therefore, such heterotic EFT strings must contain information about the non-perturbative sectors in question. It would be very interesting to study this effect further from the heterotic point of view.


\subsection{$SO(32)$ heterotic/Type I models}
\label{sec:hetSO(32)}

The $SO(32)$ heterotic models can be discussed in a similar manner, but are somewhat simpler and so we will be brief. We first assume that there are no background NS5-branes. The relevant EFT terms can be derived as in the $E_8\times E_8$ case.  The details are provided in Appendix \ref{app_SO32}. Assume an internal gauge bundle and suppose for simplicity that it takes values in a semi-simple sub-algebra $\frak{g}\subset \frak{so}(32)$ with vanishing forth order Casimir.
Then the four-dimensional EFT contains the terms
\be\label{so(32)hetEFT} 
-\frac{1}{8\pi}\int s^0 \tr (F\wedge *F)-\frac{1}{96\pi}\int (12 s^0+3n_a s^a)\tr (R\wedge *R)\,,
\ee
where $F$ takes values in the commutant of $\frak{g}\subset \frak{so}(32)$,  $n_a$ is defined as in \eqref{defna}, and \be\label{so(32)hets} 
s^0=\hat s-\frac16n_as^a\,,
\ee
with $\hat s$ as in \eqref{defhats}. Hence
\be
{\bf C}=(C^0,C^a)=(1,0)\quad,\quad \tilde{\bf C}=(\tilde C^0,\tilde C^a)=(12,3n_a)\,.
\ee

One can also discuss the instanton corrections and the saxionic cone as in Section \ref{sec:hetscone}. In this case the ten-dimensional curvature corrections do not affect the form of the saxionic cone, but only the microscopic definition \eqref{so(32)hets} of $s^0$. The saxionic cone is simply given by $\mathbb{R}_{>0}\oplus \calk(X)$, where $\mathbb{R}_{>0}$ is parametrized by $s^0$. Correspondingly the EFT string charges are given by $\calc^{\text{\tiny EFT}}_{\rm S}=\{{\bf e}=(e^0,e^a)|e^0\geq 0\,, D_{\bf e}\equiv e^a[D_a]\in \text{Nef}^1(X)\}$. It is then easy to see that \eqref{pos1} and our quantum gravity constraints  \eqref{tildehatC0} and \eqref{tildehatC}  are satisfied. 

In particular, by picking 
 ${\bf e}_{\text{\tiny F1}}=(1,\vec 0)\in \calc^{\text{\tiny EFT}}_{\rm S}$ in  \eqref{rankbound} we obtain $r({\bf e}_{\text{\tiny F1}})\leq 22$, correctly reproducing the rank of the `perturbative' gauge sector detected by the perturbative heterotic string: the $\frak{so}(32)$ sector explicitly appearing in \eqref{so(32)hetEFT}, plus the up to six possible additional KK $U(1)$s, depending on the type of background. (As before, potential gauge group factors whose axionic couplings are not of the form \eqref{kingauge2} cannot be detected in this manner.)

One may equivalently start from the S-dual Type I description of these backgrounds. In particular, the EFT string charge ${\bf e}_{\text{\tiny F1}}=(1,\vec 0)$ corresponds to a D1-brane. However, as discussed in \cite{Lanza:2021qsu}, the D1-brane string flow drives the Type I dilaton  to $+\infty$, and then the heterotic formulation is better suited for describing the UV completion of the corresponding perturbative regime. 
 
We next consider  EFT strings of charges ${\bf e}_{\text{\tiny NS5}}=(0,e^a)$, which correspond to NS5-branes wrapping internal nef divisors $ D_{\bf e}\equiv e^a[D_a]\in \text{Nef}^1(X)$. These EFT strings can detect the `non-perturbative' gauge sector supported by bulk NS5-branes, which is not included in \eqref{so(32)hetEFT}. These NS5-branes can appear through small instanton transitions \cite{Witten:1995gx} and correspond to D5-branes in the dual Type I description. In fact, as in the $E_8\times E_8$ case, the heterotic dilaton  $e^{2\phi_{\rm het}}=6(\kappa_{abc}s^as^bs^c)/\hat{s}$ generically diverges along the flows of these  EFT strings. Hence the type I description is better suited.\footnote{The microscopic description of these infinite distance limits can be done as in \cite{Lanza:2021qsu}, but  should  be  revised by taking into account the curvature correction entering \eqref{so(32)hets}.} 

The bound \eqref{rankbound} now takes the form
\be\label{NS5bound}
r({\bf e}_{\text{\tiny NS5}})\leq 3\int_{D_{\bf e}}c_2(X)-2
\ee
In order to show that this bound is indeed satisfied, recall the $SO(32)$ counterpart of the tadpole condition \eqref{Tadpole}:
\be\label{TadpoleSO}
\lambda(E)+[\calc]=c_2(X)\,.
\ee
Here $\lambda(E)$ is defined as in \eqref{lambda} and 
\be
\calc=N_A\calc^A\,,
\ee
where $N_A>0$ counts the NS5-branes wrapping the irreducible curve $\calc^A$. The supersymmetry condition on the internal bundle implies that $\int_{D_{\bf e}}\lambda(E)\geq 0$. Hence from \eqref{TadpoleSO} we get
\be\label{NS5tadbound}
\calc\cdot D_{\bf e}\leq \int_{D_{\bf e}}c_2(X)\,.
\ee

We can now use the dual type I description to compute the rank of $r({\bf e}_{\text{\tiny NS5}})$, which just counts the bulk D5-branes which intersect the nef divisor $D_{\bf e}$:
\be
r({\bf e}_{\text{\tiny NS5}})=\sum_{A|\calc^A\cdot D_{\bf e}\neq 0}N_A\,.
\ee
Since $D_{\bf e}$ is nef, we know that $\calc^A\cdot D_{\bf e}\geq 0$ and then
\be\label{reNS5b}
r({\bf e}_{\text{\tiny NS5}})\leq \sum_{A}N_A(\calc^A\cdot D_{\bf e})=\calc\cdot D_{\bf e}\leq \int_{D_{\bf e}}c_2(X)\,,
\ee
where in the last step we have used \eqref{NS5tadbound}. The microscopic bound \eqref{reNS5b}  implies our quantum gravity bound \eqref{NS5bound} (in the non-trivial case $\langle \tilde{\bf C},{\bf e}_{\text{\tiny NS5}}\rangle =\frac32\int_{D_{\bf e}}c_2(X)>0$), which is then always satisfied.


\section{Microscopic checks in $G_2$  M-theory models}
\label{sec:Mtheory}

As a last class of models, we consider M-theory compactified on  a $G_2$-manifold $X$. In this case, the 4d $U(1)$  gauge sector comes from the expansion of the M-theory three-form $C_3$ into a basis of integral harmonic two-forms $[\Gamma^A]_{\rm harm}\in H^2(X,\mathbb{Z})$, where $\Gamma^A\in H_5(X,\mathbb{Z})$ denotes the Poincar\'e dual basis:
\be\label{MC3exp}
C_3=\frac{\ell^3_{\text{\tiny M}}}{2\pi}A_A\wedge [\Gamma^A]_{\rm harm}\,.
\ee
One can also have  non-abelian gauge sectors localised at singularities, whose $U(1)$ Cartan sector can be identified by resolving the singularity. Our EFT constraints should also hold in the singular case, but in order to check them  we will assume that all these singularities have been resolved.  

The (s)axions are obtained from the expansion
\be
C_3+\ii \Phi=\ell^3_{\text{\tiny M}}(a^i+\ii s^i)[\Pi_i]_{\rm harm} \,,
\ee
where $\Pi_i\in H_4(X,\mathbb{Z})$ is a basis of 4-cycles, $[\Pi_i]_{\rm harm}\in H^3(X,\mathbb{Z})$ is the Poincar\'e dual basis of harmonic representatives, and $\Phi$ is the associative three-form. As a key property of $G_2$ manifolds, any harmonic two-form $\omega$ satisfies the relation \cite{10.2307/1971360}
\be\label{staromega}
*_X\omega=-\Phi\wedge \omega \,,
\ee
where $*_X$ is the Hodge star associated with the internal $G_2$ metric. By expanding the eleven-dimensional terms
\be
-\frac{\pi}{\ell^9_{\text{\tiny M}}}\int G_4\wedge * G_4+\frac{2\pi}{\ell^9_{\text{\tiny M}}}\int C_3\wedge G_4\wedge G_4
\ee
one obtains the four-dimensional terms of the form \eqref{kingauge}, with $C^{AB}_i=-\Pi_i\cdot\Gamma^A\cdot \Gamma^B$ 
and $C^I_i=0$. Note that \eqref{staromega} implies that the matrix 
\be\label{MthposC} 
\langle {\bf C}^{AB},{\bm s}\rangle=-\int_X \Phi\wedge [\Gamma^A]\wedge [\Gamma^B]
\ee is positive definite.  

Consider now the eleven-dimensional term 
\be\label{I8term}
\frac{1}{192(2\pi)^3\ell_{\text{\tiny M}}^3}\int C_3\wedge \left[\tr{\bf R}^4-\frac14(\tr{\bf R}^2)^2\right]\,,
\ee
where ${\bf R}$ is the eleven-dimensional curvature two-form. By splitting ${\bf R}=R+\hat R$ according to the $4+7$ split,  one finds an axionic coupling of the form appearing in \eqref{aRR}, with 
\be
\tilde C_i=-\frac{1}{4}\int_{\Pi_i} p_1(X)\,.
\ee

A four-dimensional BPS axionic string of charges $e^i$  is obtained by wrapping an M5-brane along a  coassociative 4-cycle $\Pi=e^i\Pi_i$. Imposing that this string is EFT,  ${\bf e}\in\calc^{\text{\tiny EFT}}_{\rm S}$, corresponds to requiring that the Poincar\'e dual three-cocycle $[\Pi]$ can be represented by  an associative three-form $\Phi_{\bf e}$, or a  limit thereof reached by approaching the boundary of the corresponding saxionic cone \cite{Lanza:2021qsu}. From the positive-definiteness of \eqref{MthposC}, we conclude that $\langle {\bf C}^{AB},{\bf e}\rangle$
is positive semi-definite, hence realizing \eqref{pos1}. Furthermore an argument of footnote 2 of \cite{Harvey:1999as} and Lemma 1.1.2 of \cite{joyce1996compact} implies  that 
\be\label{quantcoass}
\langle \tilde{\bf C},{\bf e}\rangle=-\frac{1}{4}\int_X\Phi_{\bf e}\wedge p_1(X)\in \mathbb{Z}_{\geq 0}\,.
\ee

By splitting $TX|_\Pi=T\Pi\oplus N\Pi$ and using the identifications $N\Pi\simeq \Lambda^2_+\Pi$ (the bundle of self-dual two-forms) and $p_1(\Lambda^2_+\Pi)=p_1(\Pi)+2e(\Pi)$ \cite{mclean1998deformations}\footnote{Our self-dual forms corresponds to the  anti-self-dual ones of \cite{mclean1998deformations}, and viceversa.}, we can rewrite $\langle \tilde{\bf C},{\bf e}\rangle=-\frac{1}{4}\int_\Pi p_1(X)$ in the form
\be\label{quantcoass1}
\begin{aligned}
\langle \tilde{\bf C},{\bf e}\rangle= b_2^-(\Pi)-2b_2^+(\Pi)+b_1(\Pi)-1\,.
\end{aligned}
\ee
By the anomaly matching argument, this formula should agree with \eqref{CCnn1}.  We can indeed check  this result microscopically, following \cite{Gukov:2002jv}.
More precisely, the numbers of massless fields on the string are counted by
\bea \label{modes-M5}
n_{\rm C} = b_2^+(\Pi) \,, \quad 
n_{\rm F} = b_2^-(\Pi) - b_2^+(\Pi) \,, \quad n_{\rm N} = b_1(\Pi)  \,.
\eea
To see this, note first that the world-sheet  theory contains $b_2^+(\Pi)$ massless real scalars, which parametrize the geometric deformations of $\Pi$ \cite{mclean1998deformations}. Furthermore, by dimensionally reducing the self-dual M5 two-form on $\Pi$ one obtains $b_2^+(\Pi)$ right-moving plus $b_2^-(\Pi)$  left-moving real chiral scalars. Supersymmetry then fixes the orientation of $\Pi$ so that $b_2^-(\Pi)-b_2^+(\Pi)\geq 0$ and we can combine the above modes into the $b_2^+(\Pi)$ complex scalars. These form the scalar components of  $n_{\rm C}=b_2^+(\Pi)$  chiral multiplets, whose $b_2^+(\Pi)$ right-moving fermions should come from the reduction of the M5 fermions. The remaining $b_2^-(\Pi)-b_2^+(\Pi)\geq 0$ left moving real chiral bosons can be fermionized and, completed by $b_2^-(\Pi)-b_2^+(\Pi)$ left-moving fermions coming from the M5-brane fermions, form $n_{\rm F}= b_2^-(\Pi)-b_2^+(\Pi)\geq 0$  Fermi multiplets. Furthermore, reducing the M5 self-dual two-form on the harmonic one-forms yields $b_1(\Pi)$ $U(1)$ vectors, which are completed into vector multiplets by a corresponding number of  left-moving fermions $\lambda_-$ with charges as in Table \eqref{tab:fcharges}. The corresponding $n_{\rm N}=b_1(\Pi)$ Fermi superfields $\Lambda_-$ can be identified with the super field strengths of the vector multiplets, which contribute to our anomaly matching in the same way as fundamental Fermi multiplets.\footnote{ 
In two dimensions a $U(1)$ gauge field does not carry propagating degrees of freedom and, in absence of charged matter, its field strength can be traded for the auxiliary field of the Fermi multiplet. This is the two-dimensional counterpart of the relation between three-form multiplets and chiral multiplets in four-dimensions discussed in \cite{Farakos:2017jme,Lanza:2019xxg}.}
Altogether, we have therefore derived the values of $n_{\rm C}$, $n_{\rm F}$ and $n_{\rm N}$ in \eqref{modes-M5}, which together with \eqref{CCnn1} reproduce the geometric prediction \eqref{quantcoass1}.

From \eqref{totb0} and with the help of \eqref{modes-M5}, the bound on the rank of the gauge group detected by the EFT string is given by\footnote{Recall that the subtraction of the $n_{\rm N}$ Fermi multiplets accounts for potential obstructions of the scalar moduli. In the present situation, these would have to be due to non-perturbative effects from M2-brane instantons ending on the M5-brane, since the $b_2^+$ massless modes describing the geometric deformations of a coassociative cycle are classically unobstructed \cite{mclean1998deformations}.
}
\bea \label{general-Mbound}
r({\bf e})\leq n_{\rm F} + 2 (n_{\rm C} - n_{\rm N}) = b_2(\Pi) - 2 b_1(\Pi).
\eea
On the other hand, \eqref{rankbound} and \eqref{quantcoass1}
give
\bea \label{ranktopM}
r({\bf e})\leq    2 \langle \tilde{\bf C},{\bf e}\rangle  +  \langle \hat{\bf C}({\bf e}),{\bf e}\rangle  -2 = 2 b_2(\Pi) - 6 b_2^+(\Pi) + 2 b_1(\Pi)   -4 +\langle \hat{\bf C}({\bf e}),{\bf e}\rangle \,.
\eea
These two bounds must coincide and their difference  can therefore be read as an equation for $\langle \hat{\bf C}({\bf e}),{\bf e}\rangle$.

As an example,  we  may assume that $\Pi$ has trivial normal bundle $N\Pi\simeq \Lambda^2_+\Pi$ as in \cite{bryant2000calibrated}.  In this case    $p_1(N\Pi)=0$ and then $\Pi$ is spin \cite{bryant2000calibrated}. Hence by Rochlin’s theorem $\int_\Pi p_1(\Pi)\in 48\mathbb{Z}$ and $\langle\tilde {\bf C},{\bf e}\rangle=-\frac{1}{4}\int_\Pi p_1(X)\in 12\mathbb{Z}$. For instance, one can consider a so called `Twisted Connected Sum' $G_2$ manifold \cite{+2003+125+160}, which can be viewed as a coassociative K3-fibration  over an $S^3$.  By fiber-wise duality it should be dual to a heterotic compactification on a Calabi-Yau three-fold -- see for instance \cite{Braun:2017uku}. More precisely, the M5-brane wrapping the K3 fiber is the dual heterotic fundamental string and our quantum gravity bounds derived from this string must therefore coincide with the bounds obtained in the heterotic case. Indeed, $\int_\Pi p_1(X)=-48$ and then $\langle\tilde {\bf C},{\bf e}\rangle=12$, as found for the heterotic fundamental string  in Section \ref{nmultiple3}. Furthermore, by comparing the bounds \eqref{general-Mbound} and \eqref{ranktopM} and using the K3  Betti numbers
$b_2^+(\Pi)=3$, $b_2^-(\Pi)=19$ and $b_1(\Pi)=0$, one gets  $\langle \hat{\bf C}({\bf e}),{\bf e}\rangle =0$ as expected.
The bound \eqref{ranktopM} -- or equivalently \eqref{general-Mbound} -- then gives
 \be
r({\bf e})\leq  2 \langle \tilde{\bf C},{\bf e}\rangle    -2 = 22,
\ee
which reproduces the general bound \eqref{hetb1} on the rank of a heterotic perturbative gauge group in four dimensions.

In the large volume heterotic regime, we have encountered EFT strings corresponding to NS5-branes wrapping nef divisors. These should also correspond to M5 EFT strings in the M-theory picture. This in particular implies that there should exist EFT strings with non-vanishing $\hat C_i({\bf e})=\hat C_{ijk}e^je^k$. At a first sight, this may appear in contradiction with our general expectation that  such a $\hat C_i({\bf e})$  should be associated with some hidden five-dimensional structure, since apparently no such structure is present. However, the strings with non-trivial $\hat C_i({\bf e})$ may dynamically generate a preferred fifth direction along their flow. 

In order to support this proposal, let us consider an M-theory compactification admitting a weakly-coupled type IIA limit, in which the $G_2$ manifold $X$ becomes the orbifold
\cite{Kachru:2001je}
\be\label{XYS} 
X_\circ=(Y\times S^1)/(\sigma,-1)\,.
\ee
Here $Y$ represents a Calabi-Yau three-fold admitting an O6-plane involution $\sigma: Y\rightarrow Y$. In this limit, each irreducible component of the fixed locus of $\sigma$ is occupied by one O6-plane  and four D6-planes. Furthermore the associative three-form becomes $\Phi=\Re\Omega+\d y\wedge J$, where $J$ and $\Omega$ are respectively the K\"ahler form and (appropriately normalized) holomorphic $(3,0)$ form of $Y$. Then the coassociative four-cycles are calibrated by $*_{X_\circ}\Phi=\frac12 J\wedge J+ \d y\wedge \Im\Omega$. Note that $(J,\Omega)$ must satisfy the orientifold projection $\sigma^*J=-J$ and $\sigma^*\Re\Omega=\Re\Omega$.

Let us then focus on those coassociative four-cycles $\Pi$ in $X_\circ$ which  are calibrated by $\frac12 J\wedge J$. These can be regarded as orientifold-even effective divisors $D$ in $Y$. In this case, an  M5-brane wrapping $\Pi$ reduces to an NS5-brane wrapping $D$ and represents an EFT string if $D$ is a nef divisor. The analysis of these EFT strings is completely analogous, {\it mutatis mutandis}, to the analysis carried for the $E_8\times E_8$ models  in Section \ref{sec:heterotic} (and in \cite{Lanza:2021qsu}), without the complications due to the higher derivative corrections discussed therein. In particular, the EFT strings should support a term of the form \eqref{SN} with $\hat C_i({\bf e})$ as in \eqref{hethatC}, where  $\kappa_{abc}$ now denotes the triple intersection number defined on $H^2_-(Y,\mathbb{Z})$.   

Now the key point is that, along the string flows associated with the EFT strings  with non-vanishing $\hat C_i({\bf e})$, the M-theory circle decompactifies much faster than the Calabi-Yau $Y$. So, even if one starts from a more generic field space point, in which the factorized geometry of the form \eqref{XYS} is not manifest, a preferred fifth direction would dynamically emerge. 

In type IIA language, up to  
$U(1)$ mixing effects \cite{Camara:2011jg},  these EFT strings more naturally detect  the R-R $U(1)$ gauge sectors, while the EFT strings corresponding to D4-branes on appropriate special Langrangian three-cycles detect the D6 gauge sectors. The M-theory analysis nicely unifies these sectors. The corresponding EFT constraints may be tested as already done above in other models but, since ${X_\circ}$ is not complex, in this case it is harder to both extract further general  results or perform a case by case analysis. We leave this interesting task for future explorations.

\section{Conclusions}
\label{sec:concl}

In this article we have derived a number of quantum gravity constraints on the effective action of an ${\cal N}=1$ supergravity theory in four dimensions.
The constraints arise, modulo certain assumptions, by demanding the consistent cancellation of the gauge and gravitational anomalies induced by inflow from the bulk on the worldsheet of certain axionic, or EFT strings, introduced in \cite{Lanza:2020qmt,Lanza:2021qsu}. 
This logic is generally in the spirit of the analysis of
\cite{Kim:2019vuc,Hamada:2021bbz,Bedroya:2021fbu,Lee:2019skh,Tarazi:2021duw,Angelantonj:2020pyr,Katz:2020ewz} treating effective theories with more supersymmetry or in a larger number of dimensions.

Our derivation of the anomaly inflow and hence the quantum gravity bounds is valid for theories in which  the gauge and gravity sector couple in a standard way to the axionic sector of the supergravity as in \eqref{kingauge2} and  \eqref{aRR}. For theories with this property, 
we  have derived the positivity bounds and quantization conditions \eqref{tildeCqc}, \eqref{tildehatC0} and \eqref{tildehatC} on the axionic couplings and furthermore argued for the bound \eqref{rankbound} on the rank of the gauge sector coupling to the EFT strings. 
These bounds indeed rule out many supergravity theories as quantum gravity theories which would otherwise seem perfectly consistent, as we have demonstrated in simple settings in Section \ref{sec:examples}. 

At a technical level the derivation of the bounds rests on the fact that the worldsheet theory of the EFT strings can be assumed to be weakly coupled. This is a consequence of the backreaction of the strings on the supergravity background in four dimensions \cite{Lanza:2020qmt,Lanza:2021qsu} and distinguishes EFT strings from intrinsically strongly coupled strings. At the same time, the class of EFT strings cannot be decoupled from the gravitational sector, and in this sense are similar to the supergravity strings in higher  dimensions \cite{Kim:2019vuc,Katz:2020ewz}. This makes them ideal candidates to constrain the quantum gravity effective action. 

Our derivations furthermore rest on the assumption, specified around \eqref{nCeff}, that the spectrum of the weakly coupled NLSM obeys a certain minimality principle: The $n_{\rm N}$ Fermi multiplets charged under the group of transverse rotations, $U(1)_{\rm N}$, are all assumed to participate in the lifting of a subset of the $n_{\rm C}$ chiral multiplets in such a way that at best the difference $n_{\rm C} - n_{\rm N}$ of remaining unobstructed chiral multiplets can contribute to the worldsheet anomalies via a Green-Schwarz type coupling. 
This assumption is natural in concrete string theory realizations of the effective theory.
As one of the classes studied in more detail in this paper, we have analyzed EFT strings in F-theory compactifications by wrapping D3-branes on movable curves on the base of the elliptic Calabi-Yau four-fold.
In this context, we have seen that the difference $n_{\rm C} - n_{\rm N}$ is a topological index; in the geometric subsector it corresponds precisely to the number of unobstructed geometric moduli which enter the NLSM. In fact, the minimality principle is manifestly satisfied for EFT strings obtained from D3-branes wrapping movable curves of genus $g=0$, as in this case  $n_{\rm N} =0$.
If the cone of movable curves is generated by rational curves, 
our assumptions are thus indeed realised. Incidentally, for Fano 3-folds this is the case (see e.g. \cite{BarkowskiThesis} and references therein), and it is tempting to speculate if this pattern generalises to non-Fano base spaces of Calabi-Yau fourfolds with minimal holonomy. 
More generally, it would be important to better understand the validity of our working assumption \eqref{nCeff} independently of concrete realizations in string theory.

In full generality, the quantum gravity bounds depend not only on the axionic couplings $C$ and $\tilde C$ appearing in the four-dimensional effective field theory, \eqref{kingauge2}  and  \eqref{aRR}, but also on a contribution to the worldsheet anomalies encoded in the quantity $\hat C$ as defined in \eqref{SN}. We propose that such contributions should be present only for EFT strings whose backreaction probes a five-dimensional substructure of the theory. Examples of such EFT strings can arise in the heterotic theory, for strings obtained by dimensional reduction of heterotic 5-branes along nef divisors with non-vanishing triple self-intersection on the Calabi-Yau threefold. In a general setting, the appearance of the coupling $\hat C$ could be used to deduce an underlying five-dimensional structure of the four-dimensional effective theory whenever a comparison of our bounds with the four-dimensional effective action implies that $\hat C \neq 0$. We believe that this interesting effect deserves further investigation.

The bounds on the rank of the gauge algebra in their general form \eqref{rankbound} are oftentimes rather conservative. 
In the context of minimally supersymmetric F-theory on elliptic fibrations with a {\it smooth} base, we have proposed a sharper bound, \eqref{strictboundF}, based on a detailed understanding of the NLSM massless fields in this case. 
On the other hand,
the more general bound \eqref{rankbound} can in fact be saturated for instance in heterotic orbifolds (and therefore also their F-theory duals, which necessarily involve a non-smooth base) and is hence to be regarded as the more general bound. 
Irrespective of this, it would be very important to verify if the sharper bound \eqref{strictboundF} indeed withstands scrutiny in F-theory on smooth three-fold bases, as our preliminary analysis in a handful of examples is currently suggesting. 
We believe that the analysis of EFT strings adds a fruitful new line of investigation to the active program \cite{Kumar:2010ru,Morrison:2011mb,Taylor:2011wt,Grimm:2012yq,Taylor:2019ots,Font:2020rsk,Cvetic:2020kuw,Font:2021uyw,Lee:2021usk,Cvetic:2021sjm} of constraining the possible effective field theories which can be obtained in F-theory. 
An important achievement would be to establish a universal bound on the rank of the gauge group in this class of minimally supersymmetric compactifications, as was possible for the abelian subsector in six dimensions in \cite{Lee:2019skh}.
In view of \eqref{strictboundF}, this challenge can be re-phrased as the task of
finding a universal bound for the intersection product between a curve in the interior of the movable cone ${\rm Mov}_1(X)$ and the anti-canonical class $\bar K_X$ on all possible three-fold base spaces $X$ over which an elliptic Calabi-Yau can be constructed.

Finally our focus in concrete string realizations has been on strings probing the K\"ahler sector e.g. of F-theory or heterotic compactifications. By contrast, the sector of EFT strings probing the complex structure sector of these theories is far less understood. We leave an investigation of this sector of EFT strings as a challenge for future work.


\vspace{1cm}

\centerline{\large\em Acknowledgments}

\vspace{0.5cm}

\noindent We thank B.\ Acharya, R.\ Alvarez-Garcia, B.\ Bellazzini, F.\ Carta, D.\ Cassani, C.\ Cota,  F.\ Farakos, A.\ Font, X.\ Gao, A.\ Grassi,  G.\ Inverso, S.-J. Lee,  F.\ Marchesano, J.\ McNamara, A.\ Mininno, M.\ Montero, H.-P.\ Nilles, A.\ Ruip\'erez, G.\ Shiu, H.-C.\ Tarazi, W.\ Taylor, C.\ Vafa, R.\ Valandro, I.\ Valenzuela, P.\ Vaudrevange and M.\ Wiesner for useful discussions and correspondence. 
We are particularly grateful to W.\ Taylor for important discussions on the validity of the rank bounds in F-theory.
LM and NR are supported in part by MIUR-PRIN contract 2017CC72MK003.
TW is supported in part by Deutsche Forschungsgemeinschaft under Germany's Excellence Strategy EXC 2121  Quantum Universe 390833306 and by Deutsche Forschungsgemeinschaft through a German-Israeli Project Cooperation (DIP) grant ``Holography and the Swampland”.
TW furthermore thanks LPENS, Paris, for hospitality during the final completion of this work.

\vspace{2cm}

\newpage

\centerline{\LARGE \bf Appendix}
\vspace{0.5cm}

\appendix

\section{Conventions for 2d anomalies} \label{app_Anomalies}

In this appendix we summarise our conventions for the computation of the two-dimensional gauge and gravitational anomalies -- see for instance  \cite{Bilal} for a review and references to the original literature.

The notion of positive and negative chirality fermions along the two-dimensional worldsheet of the string is induced from the four-dimensional bulk theory by identifying the two-dimensional chirality matrix $\gamma_*$ with  $ -\Gamma_\alpha{}^\beta=-(\sigma_3)_{\alpha}^{\, \, \beta}$, using the notation introduced in Section \ref{sec:anomalymatching}. In matrix notation, 
\be 
\gamma_\ast= \begin{pmatrix}
   -1 & 0 \\
   0 & 1
   \end{pmatrix}
   \ee
and  a two-dimensional Dirac spinor decomposes as 
   \be 
   \psi =\begin{pmatrix}
   \psi_- \\
   \psi_+
   \end{pmatrix}\,.
   \ee
In order to be compatible with the terminology adopted in several related works, we will refer to the negative (positive) chirality spinors $\psi_-$ ($\psi_+$) as left-moving (right-moving). (Admittedly, this may be confusing, since on-shell $\psi_\mp$  depends only on $y^{\mp\mp}\equiv y^0\mp y^1$.)    
  
The variation of the quantum effective action can be written as
\begin{equation}
   \delta \Gamma\equiv  (\delta_{\rm gauge} + \delta_{\rm Lorentz}) \Gamma = 2\pi \int I^{(1)}_2 \,,
\end{equation}
where the polynomial $I^{(1)}_2$ obeys the descent relations
\begin{equation}
  I_4 = \d I^{(0)}_3 \,, \qquad \delta  I^{(0)}_3 = \d I^{(1)}_2 
     \,.
\end{equation}
In $2n$ dimensions, the anomaly polynomial $I_{2n+2}$ associated with a complex positive-chirality Weyl fermion  transforming in a representation ${\bf r}$ of the gauge group is given by the following   general formula:
\begin{equation}
    I_{2n+2} = [\mathcal{A}(M)\operatorname{ch}_{\bf r}(-{F})]_{2n+2} \,,
\end{equation}
where $\mathcal{A}(M)$ is the Dirac genus of the manifold $M$ and $\operatorname{ch}(-F)$ the Chern character:
\begin{equation}\begin{aligned}
    \mathcal{A}(M) &= 1+ \frac{1}{12(4\pi)^2} \operatorname{tr}R^2 + \frac{1}{(4\pi)^4}\left[\frac{1}{360}\operatorname{tr}R^4+\frac{1}{288}(\operatorname{tr}R^2)^2\right] + \ldots \\
    \operatorname{ch}_{\bf r}(-{F}) &=  \operatorname{tr}_{\bf r} \operatorname{exp}\left(-\frac{1}{2\pi} F\right) =  \dim {\bf r}  -\frac{1}{2\pi} \operatorname{tr}_{\bf r} F + \frac{1}{2(2\pi)^2}\operatorname{tr}_{\bf r} F^2 + \ldots \,.
\end{aligned}\end{equation}
In two dimensions, the gauge and gravitational anomaly generated by a single complex Weyl fermion with positive chirality (i.e.\ a right-moving fermion) is therefore encoded in
\begin{equation}
\begin{aligned}
    I_4 = \frac{1}{8\pi^2}\operatorname{tr}_{\bf r} F^2 + \frac{\dim{\bf r} }{192\pi^2} \operatorname{tr}R^2\,.
\end{aligned}
\end{equation}

Notice that in this form the possibility of mixed abelian anomalies is automatically taken into consideration. In fact, if the gauge group $G$ can be written as a direct product as in \eqref{gaugeG}, the field strength has the form  $F=\sum_A F_A + \sum_I F_I$ and we obtain
\begin{equation}
\begin{aligned}
    I_4 &= \frac{1}{8\pi^2}\operatorname{tr}_{\bf r} F^2 + \frac{\dim{\bf r}}{192\pi^2} \operatorname{tr}R^2   \\
    &= \frac{1}{8\pi^2}\operatorname{tr}_{\bf r} (\sum_A F_A + \sum_I F_I)(\sum_B F_B + \sum_J F_J) + \frac{\dim{\bf r}}{192\pi^2} \operatorname{tr}R^2  
    \\
    &= \frac{1}{8\pi^2}  \sum_{i}q_i^A q_i^B F_A \wedge F_B + \sum_I\sum_k\frac{\ell ({\bf r}^I_k)}{16\pi^2}\operatorname{tr} F_I^2 + \frac{\dim{\bf r}}{192\pi^2} \operatorname{tr}R^2\,,
\end{aligned}
\end{equation}
where $q^A_i$ denotes the $U(1)_A$ charge of the $i$-th component of the representation ${\bf r}$ and we have decomposed ${\bf r}=\bigoplus_k {\bf r}^I_k$ of ${\bf r}$ into $G_I$ representations ${\bf r}^I_k$. In the second line we have used the  trace $\tr\equiv \frac{2}{\ell({\bf r})}\tr_{\bf r} $ introduced in section \ref{sec:gauge}.  A negative chirality (i.e.\ left-moving) complex fermion contributes with the opposite sign. Taking this into account and summing over all the possible contributions of (0,2) scalar and Fermi chiral multiplets, one gets \eqref{wsAn0} and \eqref{wsAn}.


\section{Weakly coupled NLSMs on EFT strings}
\label{app:curveNLSM}

In this appendix we provide some non-trivial evidence that the NLSM supported on EFT strings can be treated as weakly coupled. 

\subsection{EFT strings in F-theory models}
\label{app:curveNLSM1}

Consider an F-theory model of the type described in Section \ref{sec:Ftheory} and focus on the EFT strings corresponding to D3-branes wrapping movable curves. Note that by the EFT Completeness Conjecture the  movable curves should admit an effective representative, which is furthermore expected to be  generically smooth. The geometric moduli space  $\calm_{\rm geom}$ of such a movable curve $\Sigma\subset X$ should provide part of the target space $\calm_{\text{\tiny NLSM}}$ of the NLSM model supported by the EFT string. Other possible directions along $\calm_{\text{\tiny NLSM}}$ are represented by $SL(2,\mathbb{Z})$-twisted `Wilson lines', which are part of the chiral fields $\Phi^{(2)}$ in the language of Section \ref{sec_EFTstringsF}.
By appropriately tuning the initial value of the moduli, the NLSM should remain weakly coupled along the EFT string flow. In this appendix we more explicitly check these expectations.

If we  introduce some complex coordinates $Z^i$ on $X$, $\varphi^I$ on $\calm_{\rm geom}$ and  $\zeta$ along $\Sigma$, then the moduli space corresponds to a family of local embeddings $\zeta\mapsto Z^i(\zeta;\varphi)$. A first-order infinitesimal deformation of $\Sigma$   corresponds to an element of $T\calm_{\rm geom}|_\Sigma\simeq H^0(N\Sigma)$, where $N\Sigma=TX|_\Sigma/T\Sigma$.  
By introducing a basis $\omega_I$ of $H^0(N\Sigma)$ we can regard these elements as vectors $\omega_I=\omega^i_I\del_i$ describing the deformations of the local embedding: 
\be\label{deltaZ}
\delta Z^i=\omega^i_I(\zeta;\varphi)\delta\varphi^I\,.
\ee

The IIB Einstein frame spacetime metric takes the form
\be
\d s^2=e^{2A}\d s^2_{4}+\ell^2_{\rm s}\d \hat s^2_{X}\,,
\ee
with $e^{2A}=\frac{M^2_{\rm P} \ell^2_{\rm s}}{4\pi V_X}$, where the dimensionless quantity $V_X$ is the volume of $X$ in string units (i.e.\ measured by $\d \hat s^2_{X}$, which we also take to be dimensionless).

Along the D3-brane world-volume $W\times \Sigma$, where $W$ is a two-dimensional world-sheet in the external directions, we can use coordinates $\xi^A=(\sigma^\alpha,u^a)$, where $u^a=u^a(\zeta,\bar\zeta)$ are real coordinates  along $\Sigma$. The complete embedding is then defined by $X^\mu(\sigma), Z^i(\zeta;\varphi(\sigma))$, and we can consider $\varphi^I(\sigma)$ as NLSM fields.  The metric $\tilde h_{AB}$ induced  on $W\times \Sigma$ splits as follows:
\be
\begin{aligned}
\d s^2|_{W\times \Sigma}=&\, \tilde h_{AB}\d\xi^A\d\xi^B=\tilde h_{\alpha\beta}\d\sigma^\alpha\d\sigma^\beta+2\tilde h_{\alpha a}\d\sigma^\alpha\d u^b  +\ell^2_{\rm s}\hat h_{ab}\d u^a\d u^b\\
=&\, \left[e^{2 A}h_{\alpha\beta}+ 2\ell^2_{\rm s}\hat g_{i\bar\jmath}(Z,\bar Z)\omega^i_I\overline\omega^{\bar\jmath}_{\bar J}\del_\alpha\varphi^I\del_\beta\bar\varphi^{\bar J}\right]\d\sigma^\alpha\d\sigma^\beta\\
&+2\ell^2_{\rm s}\left[\hat g_{i\bar\jmath}(Z,\bar Z)\omega^i_I\del_\alpha\varphi^I\del_b\bar Z^{\bar\jmath}+\text{c.c.}\right]\d\sigma^\alpha\d u^b+ \ell^2_{\rm s}\,\hat h_{ab}\d u^a\d u^b\,,
\end{aligned}
\ee
where $h_{\alpha\beta}\equiv g_{\mu\nu}(X)\del_\alpha X^\mu\del_\beta X^\nu$ is the pull-back to the EFT string world-sheet of the four-dimensional Einstein frame   metric, and $\hat h_{ab}\equiv \hat g_{i\bar\jmath}(Z,\bar Z)\del_a Z^i\del_b\bar Z^{\bar\jmath}+\text{c.c.}$\,. 
Then
\be
\begin{aligned}
\det(\tilde h_{AB})=\ell^4_{\rm s}\det\left[e^{2 A}h_{\alpha\beta}+\ell^2_{\rm s}\left(\hat g^\perp_{i\bar\jmath}(Z,\bar Z)\,\omega^i_I\overline\omega^{\bar\jmath}_{\bar J}\del_\alpha\varphi^I\del_\beta\bar\varphi^{\bar J}+\text{c.c.} \right)\right]\det(\hat h_{ab}) \,,
\end{aligned}
\ee
where 
\beq
\hat g^\perp_{i\bar\jmath}\equiv \hat g_{i\bar\jmath}\,-\hat g_{i\bar k}\,\hat g_{l\bar\jmath}\,\hat h^{\bar\zeta \zeta}\delbar_{\bar\zeta}\bar Z^{\bar k}\del_\zeta Z^l
\eeq
is the projection of the metric in the orthogonal directions, in the sense that $\hat g^\perp_{i\bar\jmath} V^i=\hat g^\perp_{i\bar\jmath} \bar V^{\bar\jmath}=0$ if $V^i\del_i\in TD$ and $\hat g^\perp_{i\bar\jmath} V^i \bar V^{\bar\jmath}=\hat g_{i\bar\jmath} V^i \bar V^{\bar\jmath}$ if $V^i\del_i\in T^\perp_D$. 
Neglecting the flux contributions, the D3-action becomes
\be\label{D3NG}
\begin{aligned}
S_{\rm D3}&=-\frac{2\pi}{\ell_{\rm s}^4}\int_{W\times \Sigma}\d^4\xi \sqrt{-\det \tilde h_{AB}}+\ldots\\
&=- \frac{\calt_\Sigma}{V_\Sigma}\int_W \d^2\sigma\d^2 u\sqrt{-\det\left[h_{\alpha\beta} +\frac{2\pi V_\Sigma}{\calt_\Sigma}\left(\hat g^\perp_{i\bar\jmath}(Z,\bar Z)\,\omega^i_I\overline\omega^{\bar\jmath}_{\bar J}\del_\alpha\varphi^I\del_\beta\bar\varphi^{\bar J}+\text{c.c.} \right)\right]\det(\hat h_{ab})}+\ldots  \,.
\end{aligned}
\ee
Here we have introduced  the volume $V_\Sigma$ of $\Sigma$ in string units, and the string tension
\be
\calt_\Sigma=\frac{2\pi e^{2 A} V_\Sigma}{\ell_{\text{s}}^2}\,.
\ee

We could now tune the bulk moduli so that, at energies $E\leq \Lambda \ll m_*$, we can  expand the square root appearing in the second line of \eqref{D3NG}. We now argue that the validity of this property is preserved, or even improved, along  the flow \eqref{sflow} generated by the EFT string. Indeed, we can make the identifications
\be 
m_*^2=\frac{e^{2A}}{\ell^2_{\rm s}L^2_\perp}\quad~~~~~~\text{and}\quad~~~~~~ \hat g^\perp_{i\bar\jmath}= L^2_\perp\, \hat g^\perp_{ 0 i\bar\jmath}\,,
\ee
where $L_\perp$ is a length scale (in string units) associated with the directions orthogonal to $\Sigma$. For instance, we may identify $L^4_\perp\sim V_D$,
where $D$ is some effective divisor with $D\cdot \Sigma\geq 1$. This implies that  $L^2_\perp$ scales as $\sqrt{\sigma}$ along the EFT string flow \eqref{sflow} and $m_*$ can be identified with the lightest KK scale. Hence
\be
\frac{2\pi V_{\Sigma}}{\calt_\Sigma}\hat g^\perp_{i\bar\jmath}=\frac{1}{m_*^2}\,\hat g^\perp_{ 0 i\bar\jmath} \,.
\ee
If we pick local inertial coordinates $\sigma^\alpha$ in which $h_{\alpha\beta}\simeq \eta_{\alpha\beta}$, at energy scales of order $E$ the second term under the square root in the second line of \eqref{D3NG} is of order $E^2/m_*^2\ll 1$.

We can then expand the square root, getting 
\be
S_{\rm D3}\simeq -\calt_\Sigma\int_W \d^2\sigma\sqrt{-\det h_{\alpha\beta}}- \int_W \d^2\sigma\sqrt{-\det h_{\alpha\beta}}\,\calg_{I\bar J}(\varphi,\bar\varphi)h^{\alpha\beta}\del_\alpha\varphi^I\del_\beta\bar\varphi^{\bar J}+\ldots
\ee

The first term is the standard contribution of the universal sector. The leading correction to  takes the form of an NLSM, with  
target space metric 
\be\label{D3NLSMm}
\calg_{I\bar J}\equiv 2\pi\int_\Sigma\d^2 u \sqrt{\det\hat h_\Sigma}\,g^\perp_{i\bar\jmath}(Z,\bar Z)\,\omega^i_I\overline\omega^{\bar\jmath}_{\bar J}\,.
\ee

It remains to investigate 
the scaling of this target space metric along the EFT flow, which depends also on the scaling of $\hat h_\Sigma$. This can be understood by using the classification of possible EFT string flows provided in \cite{Cota:2022yjw}. 
By adopting the terminology of this reference, one can distinguish three main cases labelled by an integer $q=0,1,2$. These are associated with so-called quasi-primitive EFT strings, which form the building blocks of EFT string limits; generalisations beyond these quasi-primitive limits are then possible along the lines of \cite{Cota:2022yjw}.

In limits of type $q=0$, $X$ can be regarded as a $\mathbb{P}^1$ fibration  over a base two-fold $B$, and   $\Sigma$ can be identified with the $\mathbb{P}^1$ fiber. This EFT string is dual to a heterotic fundamental string moving in a Calabi-Yau which is elliptically fibered over $B$. In the dual heterotic description it is clear that an EFT string flow is compatible with a weakly coupled regime -- see section \ref{sec:anomalymatching}. This can also be understood from the F-theory viewpoint, at least if we focus on the geometric moduli space $\calm_{\rm geom}$.
Indeed, any vertical divisor in $X$ has vanishing intersection with $\Sigma$ and then, asymptotically, does not scale with $\sigma$. On the other hand, the base volume scales like $\sigma$.
This implies that $\hat h_\Sigma$ scales like  $\frac1{\sqrt{\sigma}}$, and then that $\calg_{I\bar J}$ does not scale. This is consistent with the identification of $\calg_{I\bar J}$ with  the string frame metric of base $B$ in the dual heterotic description.

In the case $q=1$, $X$ can be identified with the fibration of a surface $S$ over a $\mathbb{P}^1$ base, and $\Sigma$ is a movable curve inside the $S$ fibre with $(\Sigma\cdot\Sigma)_S\geq 1$. 
In this case $S\cdot \Sigma=0$ and then the volume of $S$ does not scale.  
Consider now the effective divisor $D_\Sigma$ of $X$ obtained by fibering $\Sigma$  - or in fact any curve in the fiber $S$ with non-vanishing intersection with $\Sigma$ - over the base $\mathbb P^1$. Hence $\Sigma\cdot D_\Sigma=(\Sigma\cdot\Sigma)_S\geq 1$ and the volume of $D_\Sigma$ scales as $\sigma$ along the EFT string flow. The analysis of \cite{Cota:2022yjw} shows that $V_\Sigma$ is constant and the base $\mathbb{P}^1$ volume scales as $\sigma$. From \eqref{D3NLSMm}  we see that in this case $\calg_{I\bar J}$ scales as $\sigma$. Note that this case corresponds to $w=2$. 

Finally, consider the case $q=2$. This case 
describes a homogeneous decompactification.
Hence, $V_\Sigma\sim \sqrt{\sigma}$ and $g^\perp_{i\bar\jmath}\sim \sqrt{\sigma}$ and then
 $\calg_{I\bar J}$ scales as $\sigma$.  Also in this case the scaling weight is $w=2$. 

In summary, we have found that 
\be
\calg_{I\bar J}\sim  \sigma^{w-1}\calg^0_{I\bar J} \,.
\ee
Hence, if $w=1$ we can either choose  the NLSM  to be weakly coupled or even exactly quantize it, as in perturbative superstring theory, since $g_{\rm s}=e^\phi$ is automatically driven to zero. If $w>1$, the EFT string does not uplift to a critical string but the corresponding flow automatically drives the bulk moduli to a large distance limit in which the $\calm_{\rm geom}$ sector of the  NLSM is weakly coupled. By applying similar scaling arguments  to the complete DBI-action, one can extend these conclusions to the twisted Wilson-line sector as well.

\subsection{EFT strings in heterotic models}
\label{app:curveNLSM2}

\label{sec}

We now consider the EFT strings of the heterotic models discussed in Section \ref{sec:heterotic}.

Take first an EFT string that uplifts to a heterotic F1 string. Its flow is compatible with a weak coupling regime, since along it  the (string frame) K\"ahler moduli do not change, while the ten-dimensional string coupling goes to zero \cite{Lanza:2021qsu}. Hence one may even exactly quantize it in superstring perturbation theory. 

Let us then turn to an EFT string corresponding to an NS5-brane wrapping a nef divisor $D\subset X$. In this case $g_{\rm s}$ can increase along the EFT string flow -- see \cite{Lanza:2021qsu} 
 and below -- and then it is convenient to uplift the model to an HW M-theory compactification on $I\times X$, with $I=S^1/\mathbb{Z}_2$. The NS5-brane uplifts to an M5-brane  wrapping $\{\hat y\}\times D\subset I\times X$ and, as discussed in section \ref{sec:EFTstringsHet},  may require the insertion of open M2-branes. The M5's geometric moduli space can be identified with $I\times \calm_D$, where $\calm_D$ is the moduli space of $D\subset X$.
Any nef divisor $D$ is expected to be automatically effective, as conjectured in  \cite{Katz:2020ewz}. One can more precisely describe $\calm_D$ as in \cite{Maldacena:1997de,Katz:2020ewz}, but we will not need such a description.   

We will focus on the NLSM description of the M5's geometric moduli space $I\times \calm_D$, proceeding as for the D3-branes discussed in Appendix \ref{app:curveNLSM1}. We introduce some complex coordinates $z^i$ on $X$, $\varphi^I$ on $\calm_D$ and  $\zeta^\cali$ along $D$. $\calm_D$ describes a family of local embeddings $\zeta^\cali\mapsto Z^i(\zeta;\varphi)$ and a first-order infinitesimal deformation of $D$ corresponds to an element of $T\calm|_D\simeq H^0(ND)$, where $ND=TX|_D/TD\simeq \calo_X(D)|_D$.
The elements of a basis $\omega_I=\omega^i_I\del_i$ of $H^0(ND)$ describe the deformations of the local embedding as in \eqref{deltaZ}. 

As in section \ref{sec:heterotic}, we set the M-theory Planck length $\ell_{\text{\tiny M}}\equiv \ell_{\rm s} $ and parametrize $I\equiv S^1/\mathbb{Z}_2$ by a coordinate $y\simeq y+2$, with $y\simeq - y$. We can then restrict to $y\in[0,1]$ with $y=0, 1$ hosting the ten-dimensional $E_8$  gauge sectors.  The  eleven dimensional M-theory metric takes the form   
\be
\begin{aligned}
\d s^2_{11}=e^{2 A}\d s^2_{4}+\ell^2_{\text{\tiny M}}\left(\d \hat s^2_X+e^{\frac{4\phi}{3}}\,\d y^2\right)\,
\end{aligned}
\ee
with
\be\label{defwarphetM}
e^{2 A}=\frac{\ell^2_{\text{\tiny M}}M^2_{\rm P}e^{-\frac{2\phi}3}}{4\pi \hat V_X}\,,
\ee
where $\hat V_X$ is the volume of $X$ in eleven-dimensional Planck units.
We can identify $e^\phi$ with the heterotic string coupling and $e^{\frac{2\phi}{3}}$
with the length of the HW interval. Here we are neglecting subleading backreaction effects due to non-trivial gauge bundles or $G_4$ field strength, and to bulk M5-branes, since they will not be relevant in the following.

An M5-brane has world-volume $\Gamma=W\times \{y=\hat Y\}\times D$, on which we introduce adapted real coordinates $\xi^A=(\sigma^\alpha,u^a)$ (where $u^a=u^a(\zeta,\bar\zeta)$). The embedding is defined by $X^\mu(\sigma), \hat Y(\sigma)$, $Z^i(\zeta;\varphi(\sigma))$, and we can consider $X^\mu(\sigma),\hat Y(\sigma),\varphi^I(\sigma)$ as NLSM fields. By repeating the same steps followed in Appendix \ref{app:curveNLSM1}, the geometrical sector supported by the M5-brane is described by the effective action
\be \label{M5exp}
\begin{aligned}
S_{\rm M5}&=-\frac{2\pi}{\ell_{\text{\tiny M}}^6}\int_\Gamma\d^6\xi \sqrt{-\det g|_\Sigma}+\text{(flux terms)}\\
&=- \frac{\calt_D}{\hat V_D}\int_\Gamma \d^6\xi\sqrt{-\det\left\{h_{\alpha\beta}+\frac{2\pi 
 \hat V_D}{\calt_D}\left[e^{\frac{4\phi}{3}}\del_\alpha\hat Y\del_\beta\hat Y+\left(\hat g^\perp_{i\bar\jmath}(Z,\bar Z)\,\omega^i_I\overline\omega^{\bar\jmath}_{\bar J}\del_\gamma\varphi^I\del_\beta\bar\varphi^{\bar J}+\text{c.c.}\right)\right]\right\}}\\
 &\quad~~~~~~~~~~~~~~~~~~~~~~\cdot \sqrt{\det\hat h_{ab}}+\text{(flux terms)}\,.
\end{aligned}
\ee
Here we have introduced the projection of the Calabi-Yau metric to the orthogonal directions
\beq
\hat g^\perp_{i\bar\jmath}\equiv \hat g_{i\bar\jmath}\,-\hat g_{i\bar k}\,\hat g_{l\bar\jmath}\,\hat h^{\bar\cali\calj}\delbar_{\bar\cali}\bar Z^{\bar k}\del_\calj Z^l\,,
\eeq
 the world-sheet metric $h_{\alpha\beta}=g_{\mu\nu}(X)\del_\alpha X^\mu\del_\beta X^\nu$, the induced metric $\hat h_{ab}\equiv \hat g_{i\bar\jmath}(Z,\bar Z)\del_a Z^i\del_b\bar Z^{\bar\jmath}+\text{c.c.}$ along $D$, the volume $\hat V_D$ of $D$ in eleven-dimensional Planck units, and the EFT string tension:
\be
\calt_D=\frac{2\pi e^{2 A} \hat V_D}{\ell_{\text{\tiny M}}^2}\,.
\ee
We now note that the combination appearing in front of $(\del Y)^2$ in \eqref{M5exp} is given by
\be\label{defm*het}
\frac{2\pi 
 \hat V_D e^{\frac{4\phi}{3}}}{\calt_D}=\ell_{\text{\tiny M}}^2e^{\frac{4\phi}{3}}e^{-2A}\equiv \frac{1}{m_*^2}\,,
\ee
where $m_*$ is the KK mass along $I$. As in the D3 case of Appendix \ref{app:curveNLSM1}, we can write 
\be 
\hat g^\perp_{i\bar\jmath}= L^2_\perp\, \hat g^\perp_{ 0 i\bar\jmath}\,,
\ee 
where $L_\perp$ is identified with the length scale (in eleven dimensional Planck units) of the directions in $X$ transversal to $D$, and $g^\perp_{0i\bar\jmath}$ is at most of order $\calo(1)$.
By introducing the corresponding KK scale $m_{\text{\tiny KK}}=e^{A}/(\ell_{\text{\tiny M}} L_\perp)$, we obtain the relation 
\be
\frac{2\pi 
 \hat V_D }{\calt_D}\hat g^\perp_{i\bar\jmath}=\frac{1}{m_{\text{\tiny KK}}^2}\hat g^\perp_{0i\bar\jmath}\,.
\ee
We can then expand the square root
appearing in \eqref{M5exp}, since the second term appearing therein is at most of order $\calo(\frac{E^2}{m_*^2},\frac{E^2}{m_{\text{\tiny KK}}^2})$, which is small in the EFT regime  $E\leq \Lambda\ll \text{min}\{m_*,m_{\text{\tiny KK}}\}$.

Hence, the leading contribution to the world-sheet action splits into a standard universal term and a NLSM term:
\be\label{M5NLSM}
\begin{aligned}
&-\calt_D\int\d^2\sigma \sqrt{-\det\,h_{\alpha\beta}}-\pi e^{\frac{4\phi}{3}}\int\d^2\sigma \sqrt{-\det\,h_{\alpha\beta}}\,h^{\alpha\beta}\del_\alpha \hat Y\del_\beta\hat Y\\
&\quad~~~~~~~~ - \int\d^2\sigma \sqrt{-\det\,h_{\alpha\beta}}\,h^{\alpha\beta}\calg_{I\bar J}(\varphi,\bar\varphi)\del_\alpha\varphi^I\del_\beta\bar\varphi^{\bar J}+\ldots
\end{aligned}
\ee
with
\be\label{NS5defmetric}
\calg_{I\bar J}\equiv 2\pi\int_D\d^4 u \sqrt{\det\hat h_D}\,\hat g^\perp_{i\bar\jmath}(Z,\bar Z)\,\omega^i_I\overline\omega^{\bar\jmath}_{\bar J}\,.
\ee
 Note also that $\calg_{I\bar J}$ approximately scales like $\hat V_X$ under a deformation of the internal space.   

We can now discuss the scaling behaviour of the NLSM terms along the possible EFT string flows, following the classification of \cite{Lanza:2021qsu}. In order to lighten the discussion, we will assume that there are no background M5-branes. Their inclusion can be treated similarly.

As a preliminary step, we observe that combining \eqref{defhats} and \eqref{hetfinsaxionsa} (plus the absence of background M5s) we get
\be
s^0=\frac1{3!}e^{-2\phi}\kappa({\bm s},{\bm s},{\bm s})+\frac12 p_as^a\,,
\ee
and then
\be\label{phihetinv}
e^{2\phi}=\frac{\kappa({\bm s},{\bm s},{\bm s})}{6(s^0-\frac12p_as^a)}\quad~~~\text{and}\quad~~~ \hat V_X= s^0-\frac12 p_as^a\,.
\ee
Furthermore, recalling \eqref{defwarphetM} we can rewrite the microscopic mass scales $m_*$  and $m_{\text{\tiny KK}}$ as follows:
\be\label{defm*mKK}
m_*^2= \frac{M^2_{\rm P}e^{-2\phi}}{4\pi \hat V_X}=\frac{3M^2_{\rm P}}{2\pi \kappa({\bm s},{\bm s},{\bm s})}\quad,\quad m^2_{\text{\tiny KK}}=\frac{M^2_{\rm P}e^{-\frac23\phi}}{4\pi L^2_\perp\hat V_X}=\frac{6^{\frac13}\,M^2_{\rm P}}{ 4\pi L^2_\perp[\kappa({\bm s},{\bm s},{\bm s})]^{1/3}\hat V_X^{2/3}} \,.
\ee

Consider now an EFT string with charge vector ${\bf e}=(e^0,e^a)$. Without loss of generality, we can assume that $p_ae^a\geq 0$, since the case $p_ae^a\leq 0$ can be recovered by applying the symmetry \eqref{HWswap}. We will actually restrict to the case $p_ae^a> 0$, in which the EFT string detects the perturbative gauge sector. 

From \eqref{hetCEFT} we see that $D_{\bf e}\equiv e^aD_a$ must be nef and furthermore we must impose that $e^0\geq p_ae^a$. We will then make the minimal choice $e^0= p_ae^a$, since the extension of the following discussion to the case $e^0> p_ae^a$ is immediate. Note in particular that the condition $e^0= p_ae^a$ forces the internal volume $\hat V_X$ to asymptotically scale like 
\be
\hat V_X\simeq \frac12 p_ae^a \sigma\,,
\ee
under any EFT string flow of this class. This implies that  $\calg_{I\bar J}$ asymptotically scales like $\sigma$ too, guaranteeing the weak-coupling description of the corresponding NLSM sector. These effects are induced by the inclusion of the higher derivative corrections and of the corresponding deformation \eqref{hetCEFT} of the saxionic cone identified in \cite{Lanza:2021qsu}, which was not taken into account in that paper. Hence, the discussion of \cite{Lanza:2021qsu}  actually holds only if $p_ae^a=0$, and we now revisit the three cases considered therein in the more general case $p_ae^a>0$.

\bigskip

{\em{\large Case 1: $\kappa({\bf e},{\bf e},{\bf e})>0$}}

\smallskip

\noindent The first of \eqref{phihetinv} implies that  $e^{2\phi}\simeq  \frac{\kappa({\bf e},{\bf e},{\bf e})}{3p_ae^a}\sigma^2$ asymptotically along the EFT string flow \eqref{sflow}.   We then see that the two NLSM terms appearing in  \eqref{M5NLSM}  scale like $\sigma^{4/3}$ and $\sigma$ respectively, hence naturally guaranteeing the weakly-coupled description. Furthermore, note that the length of the HW interval scales like $\sigma^{\frac23}$, while the Calabi-Yau characteristic length scales like $\sigma^{\frac16}$. Hence the EFT string backreaction generates a hierarchy and induces an intermediate decompactification to five dimensions. Furthermore   $L^2_\perp\sim \sigma^{1/3}$ and from \eqref{defm*mKK} we get the asymptotic scalings   $m_*^2\sim M^2_{\rm P}\sigma^{-3}$ and $m^2_{\text{\tiny KK}}\sim M^2_{\rm P}\sigma^{-2}$, hence confirming the scaling weight $w=3$ found in \cite{Lanza:2021qsu}, but microscopically realizing it in a different way.

\bigskip

{\em\large Case 2: $\kappa({\bf e},{\bf e},{\bf e})=0$ but $\kappa({\bf e},{\bf e},{\bf e}')> 0$ for some ${\bf e}'\in{\rm Nef}_{\mathbb{Z}}(X)$}

\smallskip

\noindent  In this case $X$ can be realised as the $T^2$ fibration over a two-fold $B$ and the divisor $D_{\bf e}$ is `vertical'.
Along the flow both  $e^{2\phi}$ and $\calg_{I\bar J}$ scale like $\sigma$. This implies that the two NLSM terms appearing in \eqref{M5NLSM} scale like $\sigma^{2/3}$ and $\sigma$ respectively, again justifying the weak-coupling assumption. 

Note also that the effective curve  $\calc=D^2_{\bf e}$ is a multiple of the $T^2$ fibre and has  constant string frame volume ${\rm vol}(\calc)=\kappa({\bf e},{\bf e},{\bm s})=\kappa({\bf e},{\bf e},{\bm s}_0)$. In the M-theory frame,  $\widehat{\rm vol}(\calc)=e^{-\frac{2\phi}{3}}\kappa({\bf e},{\bf e},{\bm s}_0)$
and then $\calc$ shrinks to zero-size as  $e^{-2\phi/3}\sim\sigma^{-1/3}$.  This implies that the M-theory base volume $\widehat{\rm vol}(B)$ should diverge like $\sigma^{4/3}$. Hence, the corresponding length goes like $\sigma^{1/3}$, as the length of the HW interval. This means that the EFT string flow induces again  a hierarchy of compactification scales, so that  the configuration could be described by first reducing  M-theory theory to nine dimensions along a $T^2$ fibre, and then by further compactifying along $B\times I$. Furthermore  $L^2_\perp\sim \sigma^{2/3}$ and \eqref{defm*mKK} gives the asymptotic scalings   $m_*^2\sim M^2_{\rm P}\sigma^{-2}$ and $m^2_{\text{\tiny KK}}\sim M^2_{\rm P}\sigma^{-2}$, confirming the scaling weight $w=2$ found in \cite{Lanza:2021qsu}, but again microscopically realizing it in a different way.

\bigskip

{\em\large Case 3: $\kappa({\bf e},{\bf e},{\bf e}')=0$ for any ${\bf e}'\in{\rm Nef}_{\mathbb{Z}}(X)$}

\smallskip

\noindent In this case $e^{2\phi}$ does not scale asymptotically along the EFT string flow, while as in the above cases $\calg_{I\bar J}$ scales  like $\sigma$, again allowing for  a  weakly-coupled NLSM description. The internal space $X$ can be regarded as a $T^4$ or K3-fibration over $\mathbb{P}^1$. The divisor $D_{\bf e}$  is a multiple of the fibre and has constant string-frame volume $\frac12\kappa({\bf e},{\bm s}_0,{\bm s}_0)$. Hence $\hat V_{D_{\bf e}}=\frac12e^{-\frac{4\phi}{3}}\kappa({\bf e},{\bm s}_0,{\bm s}_0)$ is also constant. Since $\hat V_X$ diverges as $\sigma$, the volume of the $\mathbb{P}^1$ base  should also diverge like $\sigma$, so that $L^2_\perp\sim \sigma$. We then see that the EFT string flow induces a decompactification along the $\mathbb{P}^1$ base, while the other directions do not decompactify. In this case we get a scaling weight which is different from the one obtained in \cite{Lanza:2021qsu}. Indeed \eqref{defm*mKK}  now gives $m_*^2\sim M^2_{\rm P}\sigma^{-1}$ and $m^2_{\text{\tiny KK}}\sim M^2_{\rm P}\sigma^{-2}$, corresponding to a scaling weight $w=2$, which is different from the value $w=1$ found in \cite{Lanza:2021qsu}.

In conclusion, in all cases a weakly-coupled NLSM description is allowed, if not even  favored, by the EFT string flow.


\section{Derivation of  heterotic (s)axionic couplings}
\label{app:hetEFT}
In this section, we recall the derivation of the threshold corrections to the (s)axionic couplings in both the $E_8 \times E_8$ and the $SO(32)$ heterotic string models in order to obtain the expression that we used in this paper to check the validity of the EFT string constraints.

\subsection{$E_8\times E_8$ models}

We begin by reviewing the derivation of the threshold corrections to the (s)axionic couplings in heterotic $E_8 \times E_8$ string models, referring to \cite{Blumenhagen:2006ux} and references therein for further details. Let us denote the two heterotic $E_8$ field strengths by ${\bm F}_{1}$ and ${\bm F}_{2}$ respectively.
Our conventions for the traces are the same ones as  described after \eqref{kingauge} and in Footnote \ref{foot:dynkin}. Similarly, we denote the ten-dimensional curvature two-form by ${\bm R}$. 

These objects enter the Bianchi identity
\be\label{hetBI}
\d H_3=-\frac{\ell_{\rm s}^2}{16\pi^2}\left[\tr ({\bm F}_1\wedge {\bm F}_1)+\tr ({\bm F}_2\wedge {\bm F}_2)+\tr ({\bm R}\wedge {\bm R})\right]+\ell^2_{\rm s}\delta_4(\Gamma) \,,
\ee
where $\Gamma=\mathbb{R}^4\times {\cal C}$ with ${\cal C}\equiv \bigcup_k {\cal C}^k$ denotes the overall world-volume of a set of NS5-branes, labelled by $k$, wrapped on  irreducible  internal curves ${\cal C}^k\subset X$.  

We can now split ${\bm F}_{1,2}$ and $\bm R$ into external and internal contributions: ${\bm F}_{1,2}=F_{1,2}+\hat F_{1,2}$ and $\bm R=R+\hat R$.  
For simplicity, we assume that the possible non-trivial internal gauge bundles associated with $\hat F_{1,2}$ correspond to semi-simple sub-algebras of $\frak{e}_8$.
This leads to the following cohomological condition for the tadpole of the internal contribution,
\be 
    \lambda(E_1) + \lambda(E_2) = c_2(X) - [{\cal C}],
\ee
where we have defined
\be
\lambda(E)\equiv-\frac1{16\pi^2}\tr(\hat{F}\wedge \hat{F})\,. 
\ee
We are interested in the higher curvature terms originating from the 10-dimensional Green-Schwarz counterterm \cite{Green:1984sg,MR952374}
\be\label{hetGS}
\Delta S_{\rm GS}=\frac{1}{\ell_{\rm s}^2} \int B_2\wedge I_8 \,,
\ee
written in terms of the anomaly polynomial 
\be
I_8=\frac{1}{192(2\pi)^3} \Big[2(\operatorname{tr} {\bm F}_1^2)^2 + 2(\operatorname{tr} {\bm F}_2^2)^2 -  2\operatorname{tr} {\bm F}_1^2 \operatorname{tr} {\bm F}_2^2 + (\operatorname{tr} {\bf F}_1^2 + \operatorname{tr} {\bm F}_2^2) \operatorname{tr} {\bm R}^2+ \operatorname{tr} {\bm R}^4 + \frac{1}{4} (\operatorname{tr} {\bm R}^2)^2 \Big]\,,
\ee
and from the following two terms 
associated with the background NS5-branes,
\begin{equation}\begin{aligned}
    S^{(1)} &= \frac{1}{192 \pi }  q_a\int_{ M_4} a^a  \left ( \operatorname{tr} F_1^2 + \operatorname{tr} F_2^2 + \operatorname{tr} R^2 \right) \,,\\
    S^{(2)} &= \frac{1}{16\pi}\sum_k
    \int_{M_4} \tilde a^k  \left ( \operatorname{tr} F_1^2 - \operatorname{tr} F_2^2 \right) \,.
\end{aligned}\end{equation}
Here we are using a compact notation in which the wedge product is implicit, which we will often adopt also in the following. Furthermore we have introduced the intersection numbers
\be \label{chargeC}
q_a \equiv \int_{\cal C} \omega_a = D_a \cdot {\cal C} \,,
\ee
and the NS5 axions defined as
\be
     \tilde a^k\equiv \int_{{\cal C}^k}\tilde{\calb}^k_2  \,,
\ee
where  $\tilde{\calb}^k_2$ is the self-dual gauge two-form living on the $k$-th NS5-brane wrapping the curve ${\cal C}^k$. The derivation of the corrections to the heterotic action from the 5-branes is particularly elegant in the framework of heterotic M-theory, where they can be attributed to the M5-brane sector. For further details on the derivation, we refer to \cite{Blumenhagen:2006ux}.

Under our assumptions on the internal bundles, we can split the field strengths as
\be
\begin{aligned}\label{TRFF}
&\tr {\bm F}_1^2= \tr F_1^2+\tr \hat{F}_1^2\,,\quad \tr {\bm F}_2^2=\tr F_2^2+\tr \hat{F}^2_2\,,\\
&\tr {\bm R}^2=\tr R^2+\tr\hat R^2\,,\quad \tr {\bm R}^4=0 \,.
\end{aligned}
\ee
Taking this into account and expanding the perturbative heterotic $B$-field as $B_2= \ell^2_{\rm s} a^a \omega_a$, we arrive at the total threshold corrections for the gauge sector,
\begin{equation}\begin{aligned}\label{KineticCorrection}
    &-\frac{1}{8\pi}\int \left(\hat a+\frac12 p_a a^a - \frac38 q_a a^a  - \frac12 \sum_k \tilde a^k \right)\tr (F_1\wedge F_1) \\
    &-\frac{1}{8\pi}\int \left(\hat a-\frac12 p_a a^a + \frac18 q_a a^a  + \frac12 \sum_k \tilde a^k \right)\tr (F_2\wedge F_2)\,.
\end{aligned}\end{equation}
and the gravitational sector,
\begin{equation}
\begin{aligned}
\label{KineticCorrection3}
    & -\frac{1}{96\pi}\int \left(12 \hat a+n_aa^a - \frac{3}{2} q_a a^a \right)\tr (R\wedge R) \,.
\end{aligned}
\end{equation}
Here we have introduced the quantized constants
\begin{subequations}\label{defpa0}
\begin{align}
p_a&\equiv -\int_X\omega_a\wedge \left[\lambda(E_2)-\frac12 c_2 (X)\right] \,,\\
n_a&\equiv \frac12\int_X\omega_a\wedge c_2(X)\,.
\end{align}
\end{subequations}
After replacing $\hat a$ with $a^0$ defined as
\begin{equation}
    \hat a= a^0-\frac12 p_a a^a + \frac38 q_a a^a  + \frac12 \sum_k \tilde a^k \,,
\end{equation}
we can rewrite \eqref{KineticCorrection} and \eqref{KineticCorrection3} in the form
\begin{equation}
     -\frac{1}{8\pi}\int a^0\tr (F_1\wedge F_1)-\frac{1}{8\pi}\int \left(a^0- p_a a^a + \frac12 q_a a^a + \sum_k \tilde a^k \right)\tr (F_2\wedge F_2),
\end{equation}
and 
\begin{equation}
    -\frac{1}{96\pi}\int \left(12 a^0- 6 p_a a^a+n_a a^a + 3 q_a a^a +6\sum_k \tilde a^k \right)\tr (R\wedge R) \,,
\end{equation}
respectively. 
We finally perform another shift, this time of the M5-brane axions, redefining 
\begin{equation}\label{akred}
\tilde a^k\equiv  a^k-\frac1{2\ell^2_{\rm s}}\int_{{\cal C}^k}B_2= a^k-\frac12 m^k_a a^a\quad~~~~\text{with}\quad m^k_a\equiv D_a\cdot {\cal C}^k   \,.
\end{equation}
Note that, taking into account that
\be
q_a=\sum_km^k_a \,,
\ee
\eqref{akred} implies  that
\be
\sum_k \tilde a^k=\sum_k a^k-\frac12 q_a a^a \,.
\ee
The saxionic counterpart of this redefinition reads
\be\label{shifts0}
\tilde s^k= s^k-\frac1{2}m^k_as^a  \,.
\ee
To understand the meaning of this linear redefinition, we may observe that it can be interpreted as a shift of what we mean by  `zero-position' of the background M5-branes on the HW interval and that the chiral fields $t^k$ that contain the axions and saxions correspond to the chiral fields $-2\pi\ii\Lambda_a$ in \cite{Blumenhagen:2006ux}. Hence, by adapting their (41) we identify
\be\label{defsk2}
\tilde s^k=\lambda^k\int_{\Sigma^k}J=\lambda^km^k_a s^a\quad\text{(no sum over $k$)}\,,
\ee
where $\lambda^k$ represents the position of the M5 along the HW interval -- denoted as $\lambda_a$ in \cite{Blumenhagen:2006ux}. Now, as illustrated in Figure 2 therein, the values $\lambda^k=-\frac12$ and $\lambda^k=\frac12$ correspond to placing the $k$-th M5 on the left and right HW wall, respectively. That is, we can identify $\lambda^k=\hat y^k-\frac12$, where $\hat y^k$ is the coordinate we used on the orbifold circle, with the property of being $0$ on the left HW wall and $1$ on the right one. In combination with \eqref{defsk2}, this implies that $s^k$ introduced in \eqref{shifts0} corresponds microscopically to
\be\label{tildesk}
 s^k=\hat y^k\int_{\Sigma^k}J=\hat y^km^k_a s^a\quad\text{(no sum over $k$)} \,.
\ee

In terms of new M5 axions $ a^k$ introduced in \eqref{akred}, the axionic couplings take their final form
\begin{equation}\begin{aligned}
     &-\frac{1}{8\pi}\int a^0\tr (F_1\wedge F_1)-\frac{1}{8\pi}\int \left(a^0- p_a a^a + \sum_k {a}^k \right)\tr (F_2\wedge F_2) \,,\\&-\frac{1}{96\pi}\int \left(12 a^0- 6 p_a a^a+n_aa^a +6\sum_k {a}^k \right)\tr (R\wedge R) \,.
\end{aligned}\end{equation}
The corresponding saxionic couplings then follow by supersymmetry:
\begin{equation}\begin{aligned}
     &-\frac{1}{8\pi}\int s^0\tr (F_1\wedge *F_1)-\frac{1}{8\pi}\int \left(s^0- p_a s^a + \sum_k {s}^k \right)\tr (F_2\wedge *F_2) \,, \\&-\frac{1}{96\pi}\int \left(12 s^0- 6 p_a s^a+n_as^a +6\sum_k {s}^k \right)\tr (R\wedge *R) \,.
\end{aligned}\end{equation}

\subsection{ $SO(32)$ models} \label{app_SO32}
The $SO(32)$ heterotic string models can be treated in a similar way to the $E_8 \times E_8$ string. First, recall the form of the $I_8$ polynomial in the Green-Schwarz term needed for anomaly cancellation in this setting \cite{Green:1984sg,MR952374}. Adapting it to our conventions, this is given by
\be
I_8=\frac{1}{192(2\pi)^3} \Big[8\operatorname{tr} {\bm F}^4 + \operatorname{tr} {\bm F}^2 \operatorname{tr} {\bm R}^2+ \operatorname{tr} {\bm R}^4 + \frac{1}{4} (\operatorname{tr} {\bm R}^2)^2 \Big] \,,
\ee
We proceed by splitting ${\bm F}$ and ${\bm R}$ in their internal and external components, according to the KK ansatz. This gives us
 \be
\begin{aligned}
&\tr {\bm F}^2= \tr F^2+\tr \hat{F}^2\,,\\
&\tr {\bm R}^2=\tr R^2+\tr\hat R^2\,,\quad \tr {\bm R}^4=0   \,.
\end{aligned}
\ee
In general, contrary to the $E_8 \times E_8$ case, different breaking patterns for $SO(32)$ will result in different corrections to the 4d kinetic terms (see e.g. \cite{Blumenhagen:2005pm}). Here, for simplicity we assume an internal gauge bundle such that $\tr {\bm F}^4=0$. This choice leads to
\be
\begin{aligned}
I_8&=\frac{1}{192(2\pi)^3} \Big[ (\tr F^2+\tr \hat{F}^2 ) (\tr R^2+\tr\hat R^2 ) + \frac{1}{4} (\tr R^2+\tr\hat R^2)^2 \Big] \\
&\simeq\frac{1}{192(2\pi)^3} \left[ \tr F^2 \tr\hat R^2 -\tr R^2 \left ( -\tr \hat{F}^2 - \frac{1}{2}  \tr\hat R^2 \right ) \right] \\
&=\frac{1}{96\pi} \left[ \tr F^2 c_2(X) -\tr R^2 \left ( \lambda(E) - \frac{1}{2}  c_2(X) \right ) \right] 
\\
&\simeq\frac{1}{96\pi} \left[ \tr F^2 c_2(X) - \frac{1}{2} \tr R^2 c_2(X) \right] \,,
\end{aligned}
\ee
where we ignored the terms that will not contribute to the 4d couplings we are interested in and the last steps are to be meant cohomologically and in absence of NS5-branes, having taken into account the tadpole condition $\lambda(E)=c_2(X)$. 

Expanding the $B_2$ gauge two-form in the Green-Schwarz counterterm \eqref{hetGS}, we obtain the corrections to the axion couplings in absence of NS5-branes

\be
\begin{aligned}\label{SOcorrections}
 -\frac{1}{8\pi}\int (\hat a - \frac{1}{6} n_a a^a) \tr (F\wedge F)-\frac{1}{96\pi}\int (12 \hat a+n_a a^a)\tr (R\wedge R) \,.
\end{aligned}
\ee
If we now replace $\hat a$ with 
\be 
a^0 \equiv  \hat a - \frac{1}{6} n_a a^a \,,
\ee
we obtain the contribution
\be 
-\frac{1}{8\pi}\int a^0 \tr (F\wedge F)-\frac{1}{96\pi}\int (12 a^0+3n_a a^a)\tr (R\wedge R) \,.
\ee
By supersymmetry, the saxionic couplings are then given by
\be 
-\frac{1}{8\pi}\int s^0 \tr (F\wedge *F)-\frac{1}{96\pi}\int (12 s^0+3n_a s^a)\tr (R\wedge *R) \,.
\ee
We expect the gravitational couplings to be invariant under a transition that replaces some of gauge bundle by NS5-branes; the changes in the above computation due to the modification of the Bianchi identify should be counter-balanced by additional terms from NS5-branes. These are hard to compute directly in the heterotic frame, but are S-dual to curvature terms in the Chern-Simons actions of D5-branes in Type I string theory \cite{Blumenhagen:2005zh}.


\section{M5 instantons in $E_8\times E_8$ heterotic models}
\label{app:M5inst}

In this appendix we discuss the structure of the saxionic cone  \eqref{M5Delta} in terms of the Euclidean M5-brane instantons.

We  start by considering a process in which the first four-dimensional gauge sector forms an elementary anti-self-dual BPS instanton of unit charge, $-\frac1{16\pi^2}\int\tr(F_1\wedge F_1)=1$ -- see Footnote \ref{foot:dynkin}. This instanton contributes to the amplitudes by an exponential factor $e^{2\pi\ii t^0}=e^{-2\pi s^0}e^{2\pi\ii a^0}$. By a small instanton transition, this  gauge  instanton can be continuously deformed into an M5-brane instanton wrapping  $X$ at $\hat y=0$. Hence, by holomorphy, the M5-instanton contribution to the amplitudes should still be weighted by $e^{2\pi\ii t^0}=e^{-2\pi s^0}e^{2\pi\ii a^0}$ and then its Euclidean  action should be given by $2\pi s^0$. By repeating the same argument with the second HW wall and taking into account \eqref{hetF1F2RR}, we conclude that an M5-brane instanton at $\hat y=1$ should have Euclidean action $2\pi (s^0-p_as^a+q_as^a)$. At first sight, these results may be puzzling, and one may naively be worried that if  $(p_a-q_a)s^a>0$ the M5 instanton on the first HW wall could slip to the second wall, and vice versa if $(p_a-q_a)s^a<0$. But this would  clearly appear in tension with the assumed BPS-ness of the M5 instanton, as we would expect that  either the M5-instantons are stuck at the HW walls, or that their  on-shell action should not change as we move them  along the $y$ direction.

The key point is that in presence of a non-vanishing $G_4$ flux along $X$, the Euclidean M5-branes sitting at intermediate positions $ y_{\text{\tiny E5}}\in(0,1)$ are not isolated but must be rather connected to one of the HW walls by  open Euclidean M2-branes. Indeed recall that, in general,  if the boundary  of one or more open M2-branes contains a two-cycle $\Sigma$ supported on the M5 world-volume, they contribute  as follows to the Bianchi identity of the M5 self-dual three-form $\tilde\calh_3$: 
\be
\d\tilde\calh_3=\ell^{-3}_{\text{\tiny M}}G_4|_{\rm M5}-\delta^4(\Sigma)\,.
\ee
This implies the cohomological constraint
\be\label{WM5tadpole}
\ell^{-3}_{\text{\tiny M}}[G_4]|_{{\rm M5}}=[\Sigma]\,.
\ee
In other words, if $[G_4]_{{\rm M5}}$ is non-trivial in cohomology, then the M5-brane  must host a homologically non-trivial component $\Sigma$ of the M2 boundary, fixed by \eqref{WM5tadpole}. 

In the present setting,  \eqref{G4boundaryb} implies that the constants $p_a$ defined in \eqref{padef} can be alternatively identified with 
\be
p_a=\frac{1}{\ell_{\text{\tiny M}}^3}\lim_{y\rightarrow 1^-}\int_{\{y\} \times D_a} G_4\,,
\ee
and then measure the flux quanta of $G_4$ close to the second HW wall. As we move to the left and we cross a given subset of background M5-branes, the $G_4$ flux quanta change to
\be
\frac{1}{\ell_{\text{\tiny M}}^3}\int_{\{y\} \times D_a} G_4=p_a-\sum_{k|\hat y^k> y}m_a^k\quad\Rightarrow \quad \frac{1}{\ell_{\text{\tiny M}}^3}\lim_{y\rightarrow 0^+}\int_{\{y\} \times D_a} G_4=p_a-q_a \,.
\ee
Combined with \eqref{WM5tadpole}, this  implies that
\be\label{JG4}
\int_{\Sigma}J=\frac{1}{\ell_{\text{\tiny M}}^3}\int_{\{y_{\text{\tiny E5}}\} \times X} J\wedge G_4=p_as^a-\sum_{k|\hat y^k> y_{\text{\tiny E5}}}m_a^ks^a\,.
\ee

In our setting, BPS Euclidean M2-branes extend along the HW interval and wrap an effective (possibly reducible) curve ${\cal C}\subset X$. Considering BPS Euclidean M2-branes ending on  the M5-instanton from the left or from the right  (in  the $y$ direction), we can then identify their boundary with $\Sigma=\calc$ or $\Sigma=-\calc$, respectively. Since we necessarily have $\int_\calc J> 0$, 
\eqref{WM5tadpole} and \eqref{JG4}  imply that these open M2-branes should end on the M5-instanton from the left if  $ p_as^a-\sum_{k|\hat y^k>  y_{\text{\tiny E5}}}m_a^k s^a$ is positive, and from the right  if it is negative. Note that  $m_a^k s^a=\int_{\calc^k}J\geq 0$ and then the combination  $p_as^a-\sum_{k|\hat y^k> y_{\text{\tiny E5}}}m_a^k s^a$ increases as the M5 instanton crosses the background M5-branes from the left. 

Consider for example the case in which $(p_a-q_a)s^a>0$.  In this case an isolated M5 instanton sitting on the first HW wall can be moved away from it to a more general position $y_{\text{\tiny E5}}>0$, but this process will generate Euclidean M2-branes stretching between the HW wall, the M5 instanton and the intermediate background M5-branes. Figure \ref{fig:hetInst} in Section \ref{sec:hetscone} illustrates the case with a single background M5-brane wrapping the curve $\calc_{\text{\tiny M5}}\simeq q_a\Sigma^a$ (where $\Sigma^a\cdot D_b=\delta^a_b$) at $\hat y\leq y_{\rm E5}$, and the open Euclidean M2-branes wrapping the curves $\calc_{\text{\tiny E2}}\simeq(p_a-q_a)\Sigma^a$ and $\calc'_{\text{\tiny E2}}\simeq p_a\Sigma^a$ respectively. The total action does not change from the initial value $2\pi s^0$, but is the sum of the contributions of the different Euclidean branes.

Let us order the background M5-branes so that $\hat y^k< \hat y^{k+1}$. We can write the contribution to the total instanton action coming from the open M2-branes as  
\be
S^{\rm inst}_{\rm M2}=2\pi\Big[y_{\text{\tiny E5}}\Big(p_a-\!\!\!\!\sum_{k|\hat y^k> y_{\text{\tiny E5}}}\!\!\!\!m^k_a\Big)s^a-\!\!\!\!\sum_{k|\hat y^k< y_{\text{\tiny E5}}}\!\!\!\!s^k\Big]\,.
\ee
Hence, the contribution to the instanton action coming from the M5 at $y_{\text{\tiny E5}}$ must be given by 
\be\label{EM5action}
S^{\rm inst}_{\rm M5}=S^{\rm inst}-S^{\rm inst}_{\rm M2}=2\pi\Big[s^0-y_{\text{\tiny E5}}\Big(p_a-\!\!\!\!\sum_{k|\hat y^k> y_{\text{\tiny E5}}}\!\!\!\!m^k_a\Big)s^a+\!\!\!\!\sum_{k|\hat y^k< y_{\text{\tiny E5}}}\!\!\!\!s^k\Big]\,.
\ee
It is easy to check that $S^{\rm inst}_{\rm M2}$ and $S^{\rm inst}_{\rm M5}$ are continuous in  $y_{\text{\tiny E5}}$  and that, as expected, the latter reduces to $2\pi(s^0-p_as^a+\sum_k s^k)$  in the limit $y_{\text{\tiny E5}}\rightarrow 1$.  

Note that, on the other hand, if $(p_a-q_a)s^a>0$ an isolated M5 instanton sitting on the second  HW cannot move from it, since it would require the presence of open M2-branes ending on it from the left. If instead $p_as^a<0$ the role of the two HW walls is inverted: an isolated M5 instanton on the first HW will remain stuck on it, while an isolated M5 instanton on the second HW will be free to move away from it, forming a composite M5/M2 instanton. Indeed, this is expected from the $\mathbb{Z}_2$ symmetry \eqref{HWswap}. 

There could also be intermediate cases in which  $p_as^a>0$ and $(p_a-q_a)s^a<0$. In these cases the isolated  M5 instantons sitting in both HW walls should be trapped, while there could be composite mobile M5/M2 instantons. It would be interesting to study better the transitions between these various possibilities as we move in the K\"ahler cone, and in particular the connection with the bundle stability walls. But for  the purposes of the present paper, we just need to observe that the positivity of the M5-brane action \eqref{EM5action} is guaranteed if we impose the conditions $s^0>0$ and  $s^0-p_as^a+\sum_k s^k>0$.

\bibliographystyle{jhep}
\bibliography{references}

\end{document}